\documentclass[11pt,a4paper]{article}
\pdfoutput=1

\usepackage{dcolumn}
\usepackage{bm}
\usepackage{amsthm}
\usepackage{setspace}
\usepackage[normalem]{ulem}
\usepackage{relsize}
\usepackage{cancel}
\usepackage{verbatim}
\usepackage{enumitem}
\usepackage{mathtools}
\usepackage{empheq}

\usepackage{jheppub}

\allowdisplaybreaks[1]

\usepackage{amsthm}

\newcommand{\del}{\partial}
\newcommand{\Lie}[1]{\mathcal{L}_{#1}\,}

\newcommand{\wn}{\widetilde\nabla}

\newcommand{\tr}{\operatorname{Tr}}

\newcolumntype{I}[1]{>{\centering\arraybackslash$}m{#1}<{$}}
\newlength{\mycolwd}

\title{Lifshitz  Scale Anomalies}

\author{Igal Arav,}
\author{Shira Chapman,}
\author{Yaron Oz}
\affiliation{Raymond and Beverly Sackler School of Physics and Astronomy, Tel-Aviv University, 55 Haim Levanon street, Tel-Aviv, 69978, Israel}
\emailAdd{aravigal@post.tau.ac.il}
\emailAdd{shirator@post.tau.ac.il}
\emailAdd{yaronoz@post.tau.ac.il}

\abstract{
We analyse scale anomalies in Lifshitz field theories, formulated as the relative cohomology of the scaling operator with
respect to foliation preserving diffeomorphisms. We construct a detailed framework 
that enables us to calculate the anomalies for any number of spatial dimensions, and for any value of the dynamical exponent.
We derive selection rules, and establish the anomaly structure in diverse universal sectors.
We present the complete cohomologies for various examples in one, two and three space dimensions for several values
of the dynamical exponent.
Our calculations indicate that all the Lifshitz scale anomalies are trivial descents, called B-type 
in the terminology of conformal anomalies. However, 
not all the trivial descents are cohomologically non-trivial.
We compare the conformal anomalies to Lifshitz scale anomalies with a dynamical exponent equal to one.
}


\begin{document}

\maketitle
\flushbottom

\section{Introduction}

Lifshitz scaling is an anisotropic scaling of the time $t$ and space $x^i$ coordinates
\begin{equation}\label{intro:lifsh}
t\rightarrow \lambda^z t, \qquad x^i \rightarrow \lambda x^i,\qquad i=1,\dots,d \ .
\end{equation}
$z$ is a dynamical exponent that measures the anisotropy.
When $z=1$ the spacetime symmetry can be enhanced to include the Lorentz group, and when $z=2$ the Galilean group. 
For all other values of $z$, boost invariance is explicitly broken. 
Relativistic conformal field theories have $z=1$.

Lifshitz scaling symmetry is realized in various theoretical and experimental setups.
Scaling symmetry arises naturally in the study of quantum field theories close to a second order phase transition.
A special class of phase transitions are those that occur at zero temperature, and are driven by quantum fluctuation
rather than thermal fluctuations.
The boundary between the two phases is called a quantum critical point (QCP)  \cite{subir}.
Quantum critical points are characterized by an anisotropic scaling of space and time (\ref{intro:lifsh}).
Quantum critical points are believed to
underlie  the exotic properties of heavy fermion compounds and other materials including high $T_c$ superconductors. These materials have a 
strange metal
metallic phase, whose properties cannot be explained within the ordinary Landau-Fermi liquid theory. 
The dynamical exponent $z$ emerges as the ratio of the power law exponent of the characteristic energy scale to that of the quantum correlation length, when the tuning parameter approaches its critical value near the quantum critical point \cite{subir}.
The dynamics in the critical regime is valuable to the understanding of 
scaling properties of transport coefficients and thermodynamic quantities 
of strange metals \cite{Coleman}.
 Quantum critical systems have a hydrodynamic description with broken boost invariance, and new
 novel transports \cite{Hoyos:2013eza}.
 
  Lifshitz scaling is a property of certain covariant gravity theories that break local Lorentz invariance at the microscopic level, and 
 have been proposed 
 as potential gravitational theories with a controlled UV behaviour \cite{Horava:2009uw}.
 The black hole dynamics  in these theories exhibits 
a flux of the spin zero helicity perturbation across a universal horizon, corresponding 
to the new dissipative transport in Lifshitz field theory hydrodynamics
\cite{Eling:2014saa}.
 Lifshitz scaling shows up also in certain solutions to Einstein gravity with matter  (e.g.~\cite{Kachru:2008yh}), and in deformations
of Anti-de-Sitter \cite{Korovin:2013bua}.
 
Quantum anomalies refer to classical symmetries, that are broken at the quantum level.
Anomalous Lifshitz scaling is characterised by the weighted trace of the stress-energy
 tensor $zT^0_0+T^i_i$ obtaining a non-zero expectation value on a curved background.
Here, we will need in addition to introduce the foliation data. 
This is a generalization of
trace anomalies \cite{Deser:1993yx}. 
The latter have attracted a renewed interest recently in relation to the four-dimensional $a$ theorem \cite{Komargodski:2011vj} generalization 
of the two-dimensional $c$-theorem \cite{Zamolodchikov:1986gt}.

The aim of this paper is 
to present a general analysis of scale anomalies in Lifshitz field theories.\footnote{
Lifshitz scale anomalies have been studied in particular cases in \cite{Baggio:2011ha,Griffin:2011xs,Adam:2009gq,Gomes:2011di}.}
We will formulate the cohomological calculation as the relative cohomology of the scaling operator with
respect to foliation preserving diffeomorphisms. We will  construct a detailed framework to perform the calculation
for any number of spatial dimensions, and for any value of the dynamical exponent, and
present explicitly the complete cohomologies for various examples in one, two and three space dimensions for several values of the dynamical exponent.
Our calculations indicate that all the Lifshitz scale anomalies are trivial descents, called B-type 
in the terminology of conformal anomalies. However, 
not all the trivial descents are cohomologically non-trivial.
This is unlike the conformal case, where all Weyl invariant densities have been argued to be anomalies~\cite{Deser:1993yx}.

We end the paper with a comparison of the $z=1$ Lifshitz anomalies to the case of conformal anomalies. In both cases we consider the expectation value of the trace of the stress-energy tensor on a curved background, however, the anomalies
are different. The cohomological calculation in the conformal case involves
the relative cohomology of the scaling operator with
respect to all diffeomorphisms, and not only the foliation preserving ones as in the Lifshitz case.
Thus, for instance, there are no conformal anomalies in $2+1$ space-time dimensions but one can have $z=1$ Lifshitz anomalies.

The paper is organized as follows. In \S \ref{sec:TheCohomologicalProblem} we briefly review the cohomological calculation
of scale anomalies in conformal field theories, and define the cohomological calculation of scale anomalies
in  Lifshitz field theories. In this section we present our general prescription for finding the possible forms of the Lifshitz scale anomalies. In \S \ref{sec:1+1}, \S \ref{sec:2+1} and \S \ref{sec:3+1AnomalyResults} we present our results in $1+1$, $2+1$ and $3+1$ dimensions respectively, for various values of the dynamical exponent $z$. In \S \ref{sec:GeneralExamples} we detail some results which are valid for a general dimension and  a general value of the dynamical exponent $z$. In section \S \ref{sec:comparison} we compare our results of $z=1$ Lifshitz anomalies to the conformal anomalies.
 In \S \ref{sec:Summary} we briefly summarize the
results and discuss open problems. In appendix \S \ref{app:Notations} we detail the notations and conventions used throughout the paper. 
In appendices \S \ref{app:LifshitzWardIds}-\S \ref{app:d_1_ids} we provide proofs and derivations for various statements used throughout the text.

A note concerning  notation: throughout the paper, we will sometimes use $ A_i^{(n_T,n_S,n_\epsilon)} $ to denote the $i$-th anomaly with
 $ n_T $ time derivatives, $ n_S $ space derivatives and $ n_\epsilon $ Levi-Civita tensors in the Lifshitz cohomology with specific values of $d$ and $z$.

\section{The Cohomological Problem}\label{sec:TheCohomologicalProblem}
In this section we present a general procedure for constructing the possible forms of Lifshitz scaling anomalies allowed by the Wess-Zumino (WZ) consistency conditions in Lifshitz field theories of a given dimension $d$ and dynamical critical exponent $z$, using cohomological methods.
We begin by reviewing these methods for the case of standard (isotropic) scaling symmetry (Weyl symmetry) in conformal field theories. We then present the new tools needed for the study of the problem in the case of Lifshitz scaling symmetry. We provide a detailed recipe for constructing the possible anomalies.

\subsection{Cohomological Description of Weyl Anomalies}
    \subsubsection{Wess-Zumino Consistency Conditions and the BRST Ghost}
    \label{ssb:WZ}
    Using  the description of quantum anomalies by an effective action, one can derive the so-called Wess-Zumino 		
    consistency conditions. These are relations that must be satisfied by the anomalous Ward identities.
    Given a theory with the classical symmetries:
    \begin{equation}
    \delta_{\chi^{\alpha}} S(\{F\},\{\phi\}) = 0 ,
    \end{equation} 
    where $\chi^\alpha$ are the gauge parameters, $ \{F\} $ a set of background fields, $\{\phi\}$ the set of
     dynamic fields in		the theory, and $ S $ the classical action, and given the corresponding anomalous Ward identities:
    \begin{equation}
    \delta_{\chi^{\alpha}} W(\{F\}) = A_{\chi^{\alpha}} (\{F\}),
    \end{equation} 
    where $ W $ is the effective action (obtained after integrating out the dynamical fields), the anomalies must satisfy:
    \begin{equation}\label{review:WZ-const}
    \delta_{\chi^\alpha} A_{\chi^\beta} - \delta_{\chi^\beta} A_{\chi^\alpha} = A_{[\chi^\alpha,\chi^\beta]},
    \end{equation}
    where $[\chi^\alpha,\chi^\beta]$ are the commutation relations between the classical symmetries.
    A trivial solution to these conditions is the appropriate transformation of some functional $ G(\{F\}) $ which is local in
    the background fields:
    \begin{equation}\label{review:WZ-triv}
    A_\chi = \delta_\chi G(\{F\}) .
    \end{equation}
	Such a solution can be cancelled by an appropriate counter-term, and as such is not physically relevant. The space of
	possible physical anomalies consists of the space of all solutions to the conditions \eqref{review:WZ-const}, modulo
	the space of expressions of the form \eqref{review:WZ-triv}.    
    
    An equivalent description of the conditions, which we will use throughout the paper, is in terms of a BRST-like ghost (see e.g \cite{Bertlmann:1996xk}).
    In this description the variation parameter is replaced by a Grassmannian ghost, and its action on the fields is defined such that it becomes nilpotent:
    \begin{equation}
    (\delta_\chi)^2 =0 .
    \end{equation}
The form of the WZ consistency conditions thus becomes:
\begin{equation}
\delta_\chi A_\chi = 0 ,
\end{equation}
where $\chi$ here is the Grassmannian ghost. Similarly, the form of trivial solutions becomes:
\begin{equation}
A_\chi = \delta_\chi G(\{F\}),
\end{equation}
where $ G(\{F\} )$ is a local functional of the background fields (of zero ghost number). 
The problem of finding the physical anomalies is then mapped to the problem of finding the non-trivial terms in the cohomology
of the operator $\delta_\chi$ -- the space of $\delta_\chi$-closed terms (or cocycles) with ghost number 1, modulo the space of $\delta_\chi$-exact terms (or coboundaries).
    
\subsubsection{The Cohomology of Conformal Anomalies - a Brief Review}
\label{ssb:CohomolConfRev}

In the case of conformal theories, the relevant background field is the metric $ g_{\mu\nu} $ on the manifold on which the theory is defined, or equivalently the vielbeins $ e^a{}_\mu $.
The relevant symmetries are:
\begin{enumerate}
\item Diffeomorphism: 
\ \ \ 
$\delta_{\xi}^D g_{\mu\nu} = \nabla_\mu \xi_\nu + \nabla_\nu \xi_\mu$,\ \ \ 
$ \delta_{\xi}^D e^a{}_\mu = \xi^\nu \nabla_\nu e^a{}_\mu + \nabla_\mu \xi^\nu e^a{}_\nu$,
\item Local Lorentz\footnote{Rotations in the tangent frame.} (in the case of vielbeins):\ \ \ 
$ \delta_\alpha^L e^a{}_\mu = - \alpha^a{}_b e^b{}_\mu $ (where $ \alpha_{ab}=-\alpha_{ba} $),
\item Weyl (conformal):
\ \ \ 
$ \delta_{\sigma}^W g_{\mu\nu} = 2\sigma g_{\mu\nu} $,\ \ \ 
$ \delta_{\sigma}^W e^a{}_\mu = \sigma e^a{}_\mu $.
\end{enumerate} 

The Ward identities corresponding to these symmetries are (respectively):
\begin{enumerate}
\item From diffeomorphism invariance:\ \ \  $ \nabla_\mu T^{\mu\nu} = 0 $,
\item From local Lorentz invariance: \ \ \ $ T_{[\mu\nu]} = 0 $,
\item From Weyl invariance:\ \ \ $ T^\mu_\mu = 0 $,
\end{enumerate}
where the stress-energy tensor is defined as 
\begin{equation}
T^{\mu\nu} = \frac{2}{\sqrt{-g}} \frac{\delta S}{\delta g_{\mu\nu}} =  \frac{1}{e} e^{a\nu} \frac{\delta S}{\delta e^a{}_\mu} \, .
\end{equation}

Anomalies of the theory are the breaking of these ward identities by the effective action.
After replacing the symmetry parameters with BRST-like ghosts, the problem of finding the allowed
anomaly terms is mapped to determining the structure of the cohomology of the total operator:
\begin{equation}\label{prelims:total_cohomology_operator}
\delta \equiv \delta^D_\xi + \delta^L_\alpha + \delta^W_\sigma .
\end{equation}
When discussing the cohomological structure of an operator given as a sum of several operators, the \emph{classification theorem} due to Bonora et al.~\cite{Bonora:1984ic} allows one to split the non-trivial cocycles of the total cohomology into two sets:
\begin{enumerate}
\item Non-trivial terms in the relative cohomology w.r.t one of the symmetries --  terms which are closed but not exact under one of the symmetries, when considering only the space of terms invariant under all the rest. In this case,
the relative cohomology of the Weyl operator w.r.t diffeomorphisms and local Lorentz transformations is composed of 
terms $ A_\sigma $ that satisfy:
\begin{itemize}[label= --]
\item $ \delta^W_\sigma A_\sigma = \delta^D_\xi A_\sigma = \delta^L_\alpha A_\sigma = 0 $,
\item No local functional $ G $ (of zero ghost number) exists such that $ A_\sigma = \delta^W_\sigma G $ and $ \delta^L_\alpha G = \delta^D_\xi G =0 $.
\end{itemize} 
\item Non-trivial terms in the cohomology of one of the operators, which admit a "partner" in the cocycle space of another, such that their sum is a cocycle of the total operator. In this case, these would be terms $ A_\sigma $, $ A_\alpha $ and $ A_\xi $ which satisfy:
\begin{itemize}[label= --]
\item $ A_\xi + A_\alpha $ is a non-trivial term in the cohomology of $ \delta^D_\xi + \delta^L_\alpha $,
\item $ \delta^W_\sigma A_\sigma = 0 $,
\item $ (\delta^D_\xi + \delta^L_\alpha + \delta^W_\sigma)(A_\xi + A_\alpha + A_\sigma) = 0 $.
\end{itemize} 
\end{enumerate}

In what follows we will restrict the discussion to the anomalous structure of the relative cohomology of the Weyl operator with respect to diffeomorphisms/local Lorentz transformations.\footnote{It is known that in two space-time dimensions there are additional anomalies of the second set, which accompany gravitational anomalies \cite{Bertlmann:1996xk,Bertlmann:2000da,Ebner:1987pg}, whereas in four space-time dimensions there are none \cite{Bonora:1985cq}.} The anomalies in this relative cohomology are the ones commonly referred to as Weyl or conformal anomalies.

In order to study this structure, one has to look at diffeomorphism invariant expressions that have the right global Weyl dimension, that is all possible independent contractions of the Levi-Civita tensor, the Riemann tensor and its covariant derivatives of the correct scaling dimension. Out of these expressions one constructs the appropriate terms of ghost number 1 (multiplying by the ghost and integrating over space), and applies the Weyl operator to them. One then finds those combinations that vanish under the action of the Weyl operator, taking into consideration the Grassmannian nature of $\sigma$ -- these are the cocycles of the Weyl operator. The space of trivial terms (coboundaries) is found by applying the same operator to terms of ghost number 0 constructed from the same expressions. Finally, the anomalies are given by the quotient space of the cocycles over the coboundaries.
In two space-time dimensions for example, only one anomaly exists in the relative cohomology which is proportional to:
\begin{equation}\label{prelims:1+1d_WeylAnomaly}
\int \sigma R , 
\end{equation}
and there are no coboundary terms.
In four space-time dimensions \cite{Bonora:1985cq,Bonora:1983ff}, the possible scalar expressions of dimension 4 are: 
\begin{equation}
R^2,~~~R_{\mu\nu} R^{\mu\nu},~~~ R_{\mu\nu\rho\sigma} R^{\mu\nu\rho\sigma},~~~\epsilon^{\alpha\beta\gamma\delta} R_{\mu\nu\alpha\beta} R^{\mu\nu}{}_{\gamma\delta},~~~\Box R .
\end{equation}
The ghost number 1 combinations that vanish under the Weyl operator (the cocycles) are the Weyl tensor squared:
\begin{equation}\label{prelims:W_2}
\int \sigma  W^2 = \int \sigma [(R_{\mu\nu\rho\sigma})^2-2(R_{\mu\nu})^2+\frac{1}{3}R^2] ,
\end{equation}
 the Euler density:
 \begin{equation}\label{prelims:E_4}
\int \sigma  E_4 = \int \sigma [(R_{\mu\nu\rho\sigma})^2-4(R_{\mu\nu})^2+R^2] ,
 \end{equation}
 the Hirzebruch-Pontryagin term:
 \begin{equation}\label{prelims:P_1}
\int \sigma P_1 =  \int \sigma (\epsilon^{\alpha\beta\gamma\delta} R_{\mu\nu\alpha\beta} R^{\mu\nu\gamma\delta}) ,
 \end{equation}
  and $\int \sigma (\Box R)$. Applying the Weyl operator to zero ghost number expressions, the last term turns out to be trivial (a coboundary). Therefore one ends up with three anomalies, two in the parity even sector and one in the parity odd sector.

It has been shown \cite{Deser:1993yx,Boulanger:2007ab} that in general, in odd dimensions no Weyl anomalies exist, while in even dimensions there are in general two types of anomalies: 
the Euler density of the appropriate dimension (type A anomaly), and the various Weyl invariant scalar densities (type B anomalies).

\subsection{Lifshitz Scale Anomalies}
In Lifshitz field theories the time direction plays a major role. Since time generically scales differently than space, one has to consider the time direction separately in the construction of anomalous structures, by foliating spacetime into equal-time slices. When considering a theory defined over a general curved manifold, this structure is generalized to a codimension-1 foliation defined over the manifold.
The foliation structure over a manifold can be locally represented by a 1-form  $ t_\alpha $ defined on the manifold, the kernel of which is the tangent space to the foliation leaves (so that a vector $ V^\alpha $ is tangent to the foliation if and only if it satisfies $ t_\alpha V^\alpha = 0 $). Due to the Frobenius theorem, such a 1-form (locally) defines a codimension-1 foliation if and only if it satisfies the condition:
\begin{equation} \label{anweylan:frobcond}
t \wedge dt = 0,
\end{equation}
or equivalently in index notation:
\begin{equation}
t_{[\alpha} \del_\beta t_{\gamma]}=0.
\end{equation}

Since the 1-form $t_\alpha$ is not unique (any locally rescaled 1-form $ f t_\alpha $ represents the same foliation), the foliation is in fact represented by an equivalence class of 1-forms under rescaling. We may then choose any representative of this equivalence class. In particular, when a metric is defined, we can choose the normalized 1-form:
\begin{equation}
n_\alpha = t_\alpha /\sqrt{|g^{\beta\gamma} t_\beta t_\gamma|}  \, .
\end{equation}
Note, that the definition of $ t_\alpha $ and the foliation associated to it does not depend on the metric, while the definition of $ n_\alpha $ does.
The background fields of a Lifshitz field theory can be taken to be the metric $ g_{\mu\nu} $ and the foliation 1-form $ t_\alpha $ (in which case the action must be invariant under rescaling of $ t_\alpha $), or alternatively $ n_\alpha $. When using vielbein formalism, we often find it convenient to express the foliation 1-form in the local frame coordinates, so that the background fields will be taken to be $ e^a{}_\mu $ and $ t^a $ (or the normalized $ n^a $).

When discussing Lifshitz field theories defined on curved space, there are two symmetries that must be considered. 
The first relevant symmetry is foliation-preserving-diffeomorphisms (FPD). Since the time direction in the theory is unique, there is no longer boost invariance over flat space - only rotation invariance remains. On a curved manifold this translates into FPD invariance. That is, invariance under a coordinate transformation that preserves the foliation structure - either a coordinate transformation inside each foliation leaf, or a change in the foliation parameter. In the notation of subsection \ref{ssb:CohomolConfRev}, these are the diffeomorphisms with those $\xi$ that obey $\Lie{\xi}t_\alpha \propto t_\alpha $.
If we denote $t$ - the foliation parameter and $\mathbf{x}$ the coordinates inside each leaf, these are transformations of the form: 
\begin{equation}
t\rightarrow f(t),~~~~~~ \mathbf{x}\rightarrow \mathbf{g}(\mathbf{x},t) .
\end{equation}

However, we can easily extend this symmetry back to any $ \xi $ by having the foliation 1-form itself transform appropriately in addition to the metric (or the vielbeins). The infinitesimal form of the transformation rules will then become:
\begin{equation}\label{LifshitzAnom:FPD}
\delta^D_\xi g_{\mu\nu} = \nabla_\mu \xi_\nu + \nabla_\nu \xi_\mu,
\quad
\delta^D_\xi t_\alpha = \Lie{\xi} t_\alpha = \xi^\beta \nabla_\beta t_\alpha + \nabla_\alpha \xi^\beta t_\beta ,
\end{equation} 
or in vielbein formalism:
\begin{align}
\begin{split}
\delta^D_\xi e^a{}_\mu & = \xi^\nu \nabla_\nu e^a{}_\mu + \nabla_\mu \xi^\nu e^a{}_\nu,
\qquad
\delta^D_\xi t^a = \xi^\nu \nabla_\nu t^a,
\\
\delta^L_\alpha e^a{}_\mu & =  - \alpha^a{}_b e^b{}_\mu, 
\qquad
\delta^L_\alpha t^a = -\alpha^a{}_b t^b .
\end{split}
\end{align}

The second relevant symmetry is an anisotropic Weyl invariance.\footnote{With a slight abuse of terminology, we  will often refer to the transformation in \eqref{LifshitzAnom:AnisWeyl} as anisotropic Weyl transformation, even for $z=1$.} The Lifshitz symmetry transformation \eqref{intro:lifsh} in curved space can be generalized to an appropriate local transformation of the metric components parallel and normal to
the foliation 1-form: 
\begin{align}\label{LifshitzAnom:AnisWeyl}
\begin{split}
& \delta^W_\sigma t_\alpha    = 0 ,
\\
& \delta^W_\sigma (g^{\alpha \beta} t_\alpha t_\beta)  = -2\sigma z (g^{\alpha \beta} t_\alpha t_\beta),
\\
& \delta^W_\sigma P_{\alpha\beta}  = 2 \sigma P_{\alpha\beta},
\\
& \delta^W_\sigma n_\alpha  = z  \sigma n_\alpha, \qquad \delta^W_\sigma n^\alpha  = - z  \sigma n^\alpha,
\end{split}
\end{align}
where $P_{\mu\nu}= g_{\mu\nu} +n_\mu n_\nu$ is the projector tangent to the foliation.
Or alternatively using the vielbeins:
\begin{align}
\begin{split}
& \delta^W_\sigma (n_a e^a{}_\mu) = z \sigma n_a e^a{}_\mu,
\\
& \delta^W_\sigma (P_b^a e^b{}_\mu) = \sigma P_b^a e^b{}_\mu,
\\
& \delta^W_\sigma t^b = -z\sigma t^b,
\\
& \delta^W_\sigma n^b=0 .
\end{split}
\end{align}

Note, that when using the BRST description, one also has to define the action of $\delta^D_\xi$, $\delta^L_\alpha$ and $\delta^W_\sigma$ on the Grassmannian parameters $\xi^\mu$, $\alpha^a{}_b$ and $\sigma$ such that $\delta = \delta^D_\xi + \delta^L_\alpha + \delta^W_\sigma$ is nilpotent, as follows:
\begin{align}
\delta^D_\xi \xi^\mu &= \xi^\nu \nabla_\nu \xi^\mu,
& \delta^D_\xi \alpha^a{}_b &= \xi^\nu \nabla_\nu \alpha^a{}_b, 
& \delta^D_\xi \sigma &= \xi^\nu \nabla_\nu \sigma \notag,\\
\delta^L_\alpha \xi^\mu &= 0, 
& \delta^L_\alpha \alpha^a{}_b &= -\alpha^a{}_c \alpha^c{}_b,
& \delta^L_\alpha \sigma &= 0, \\
\delta^W_\sigma \xi^\mu &= 0,
& \delta^W_\sigma \alpha^a{}_b &= 0, 
& \delta^W_\sigma \sigma &= 0 .\notag
\end{align}

\subsubsection{Lifshitz Ward Identities}
\label{ssb:LifshitzWardIds}
In this subsection we detail the classical Ward identities corresponding to foliation preserving diffeomorphisms 
and the anisotropic Weyl scaling. The full derivation can be found in appendix \ref{app:LifshitzWardIds}. Assume a classical action depending on the metric and foliation $ S(g_{\mu\nu}, t_\alpha, \{\phi\}) $ (where $ \{\phi\} $ are the dynamic fields), or alternatively $ S(e^a{}_\mu, t^b, \{\phi\}) $.
Define the symmetric stress-energy tensor associated with the metric by:
\begin{equation}
T^{\mu\nu}_{(g)} \equiv \left. \frac{2}{\sqrt{-g}} \frac{\delta S}{\delta g_{\mu\nu}} \right|_{t_\alpha = const.}\, ,
\end{equation}
or alternatively, the stress-energy tensor associated with the vielbeins:
\begin{equation}
T_{(e)} {}^\mu {}_a \equiv \left. \frac{1}{e} \frac{\delta S}{\delta e^a{}_\mu} \right|_{t^a=const.}
\, .
\end{equation}
In addition define the variation of the action with respect to the foliation 1-form:
\begin{equation}
J^\alpha \equiv \left. \frac{1}{\sqrt{-g}}\frac{\delta S}{\delta t_\alpha} \right|_{g_{\mu\nu}=const.} =
\frac{1}{e} e^{b\alpha} \left. \frac{\delta S}{\delta t^b} \right|_{e^a{}_\mu = const.}\, .
\end{equation}
Note that $ J^\alpha $ is tangent to the foliation ($ J^\alpha t_\alpha = 0 $).
We also define a normalized version of $J^\alpha$:  
\begin{equation}\label{anweylan:J_hat_def}
\hat{J}^\alpha \equiv \sqrt{|g^{\mu\nu}t_\mu t_\nu|}J^\alpha \, .
\end{equation}
Note also that in cases where one can use either the metric or the vielbein descriptions, the following relation exists between $ T_{(g)}^{\mu\nu} $, $ T^{\mu\nu}_{(e)} \equiv T_{(e)}{}^\mu{}_a e^{a\nu} $ and $ J^\alpha $:
\begin{equation}
T_{(e)}^{\mu\nu} = T_{(g)}^{\mu\nu} + J^\mu t^\nu .
\end{equation} 
 
With these definitions, the Ward identities corresponding to invariance under the extended form of FPD \eqref{LifshitzAnom:FPD} are given by:
\begin{equation}\label{anweylan:tgwardn}
\nabla_\mu T_{(g)}^\mu{}_\nu = \hat{J}^\mu \nabla_\nu n_\mu - \nabla_\mu(\hat{J}^\mu n_\nu) = -n_\nu [\wn_\mu \hat{J}^\mu + 2\hat{J}^\mu a_\mu ],
\end{equation}
where $ a_\mu = n^\nu \nabla_\nu n_\mu $ and $ \wn $ is the covariant derivative projected on the foliation, or equivalently in terms of $T_{(e)}^{\mu\nu}$:
\begin{align}\label{anweylan:teward1n}
T_{(e)[\mu \nu]} &= \hat{J}_{[\mu}n_{\nu]} ,\\
\label{anweylan:teward2n}
\nabla_\mu T_{(e)}{}^\mu{}_\nu &= \hat{J}^\mu \nabla_\nu n_\mu .
\end{align}

The Ward identity corresponding to anisotropic Weyl symmetry is given by:
\begin{equation}\label{anweylan:tgewardn_trace}
T_{(g)}^{\mu\nu} P_{\mu\nu} - z T_{(g)}^{\mu\nu}n_\mu n_\nu
= T_{(e)}^{\mu\nu} P_{\mu\nu} - z T_{(e)}^{\mu\nu}n_\mu n_\nu = 0 .
\end{equation}
In field theories in which the above symmetries are anomalous, these Ward identities are subject to quantum corrections.

\subsubsection{Foliation Preserving Diffeomorphism Invariance }\label{ssb:fpdinvariantexpr}

As with the standard Weyl scaling, we will restrict the discussion to the anomalous structure of the relative cohomology of the anisotropic Weyl operator with respect to FPD.\footnote{These are the anomalies analogous to the standard Weyl anomalies in the Lifshitz case. Anomalies of the second set as mentioned in subsection \ref{ssb:CohomolConfRev} would accompany anomalies of FPD invariance which are analogous to gravitational anomalies in the conformal case and are outside the scope of this discussion.} That is, when studying the cohomology, we will only consider terms with Weyl ghost number one which are invariant under foliation preserving diffeomorphisms.
These are the terms which are:
\begin{itemize}[label= --]
\item closed: $\delta_\sigma^W A_\sigma  = \delta_\xi^D A_\sigma = 0$,
\item but not exact: $A_\sigma \neq \delta_\sigma^W G$ for any local $G$ that satisfies $\delta_\xi^D G = 0$,
\end{itemize}
where $\sigma(x)$ is the Grassmannian local parameter of the anisotropic Weyl transformation.

In the following we explain in detail how to build all the foliation preserving diffeomorphism (FPD) invariant expressions of a certain dimension. That is, scalars, constructed from the metric $g_{\mu\nu}$ and the foliation 1-form $t_\alpha$ which are invariant under \eqref{LifshitzAnom:FPD} (in fact it will be more convenient to work with the normalized version $n_\alpha$). We would like to pick the independent objects $\mathcal{O}$ for our construction such that they scale uniformly under anisotropic Weyl scaling transformation \eqref{LifshitzAnom:AnisWeyl} with a certain scaling dimension $d_\sigma$:
\begin{equation}\label{FPD_Objs:unif}
\delta^W_\sigma \mathcal{O} = (d_\sigma) \sigma \mathcal{O} + (\del \sigma)  ,
\end{equation}
where $\del \sigma$ represents any term proportional to derivatives of the ghost $\sigma$.
For example for $n_\alpha$, $n^\alpha$ and $P_{\alpha\beta}$, the scaling dimension $d_\sigma$ is $z$,$-z$ and $2$ respectively (see \eqref{LifshitzAnom:AnisWeyl}). 
These expressions are covariant under anisotropic Weyl transformations, i.e.\ their transformation law \eqref{FPD_Objs:unif} does not contain derivatives of the ghost.

It is clear that the terms in the cohomology (which arise as a variation of the effective action) should have a total Weyl scaling dimension of 0. They would thus be composed of scalars of dimension $ -(d+z) $ (where $d$ is the number of space dimensions), integrated over spacetime.
In order to find expressions with a uniform scaling dimension, one must decompose any tensor into components in the direction of the foliation and in the direction normal to it. For example, the metric $ g_{\mu\nu} $  or the Riemann tensor $ R_{\mu\nu\rho\sigma} $ don't have a uniform scaling dimension, while the components of the metric $ P_{\mu\nu} $ and $ n_\mu n_\nu $ do.

A tensor $ \widetilde{T}_{\alpha\beta\gamma...}$ is tangent to the foliation if 
\begin{equation}
 n^\alpha \widetilde{T}_{\alpha\beta\gamma\dots} = n^\beta \widetilde{T}_{\alpha\beta\gamma\dots} = \dots = 0 .
 \end{equation}
We claim that, in general, any scalar expression may be written as a sum of scalar expressions built by contractions of tensors that are tangent to the foliation
and have a uniform scaling dimension.
These basic tangent tensors are: 
\begin{itemize}[label= --]
\item The projector on the foliation: $ P_{\mu\nu} = g_{\mu\nu} + n_\mu n_\nu $,
\item The acceleration: $ a_\mu \equiv \Lie{n} n_\mu = n^\nu \nabla_\nu n_\mu$,
\item The extrinsic curvature: $K_{\mu\nu} \equiv \frac{1}{2}\Lie{n} P_{\mu\nu} =
P_\mu^\rho \nabla_\rho n_\nu $,
\item The intrinsic Riemann curvature of the foliation: $\widetilde R_{\mu\nu\rho\sigma}$,
\item The intrinsic Levi-Civita tensor of the foliation:  $\tilde \epsilon^{\mu\nu\rho \dots} = 
n_\alpha \epsilon^{\alpha\mu\nu\rho\dots}$,
\item The Lie derivatives (temporal derivatives) of any of the above tensors in the direction of $ n^\alpha $, for example $ \Lie{n} K_{\mu\nu} $, $ \Lie{n}\Lie{n} a_\mu $ etc.,
\item The foliation projected (spatial) covariant derivatives of any of the above tensors, for example $ \wn_\alpha K_{\mu\nu} $, $ \wn_\alpha \wn_\beta \Lie{n} a_\gamma $,
\end{itemize}
where we define the foliation projected 
covariant derivative (which we refer to as spatial derivative) of a foliation-tangent tensor as follows:
\begin{equation}
\wn_\mu \widetilde T_{\alpha\beta\dots} \equiv P^{\mu'}_\mu P^{\alpha'}_\alpha P^{\beta'}_\beta \dots \nabla_{\mu'} \widetilde T_{\alpha'\beta'\dots} \ .
\end{equation}
Note, that this is the covariant derivative that is compatible with the metric induced on the $d$-dimensional leaves of the foliation, i.e.\ $ \wn_\rho P_{\mu\nu} = 0 $. 

In order to prove this claim, we use the following statements, which can be easily established from the normalization $ n^\alpha n_\alpha = -1 $, the Frobenius condition \eqref{anweylan:frobcond} and the definitions given above for the acceleration and extrinsic curvature:
\begin{itemize}[label= --]
\item The acceleration and extrinsic curvature are both tangent to the foliation: $ n^\mu a_\mu = 0 $, $ n^\mu K_{\mu\nu} = 0 $,
\item The extrinsic curvature is symmetric: $ K_{\mu\nu} = K_{\nu\mu} $,
\item The spatial derivative of the acceleration is symmetric: $\wn_\mu a_\nu$ = $\wn_\nu a_\mu$,
\item The covariant derivative of the foliation 1-form $ n_\mu $ may be decomposed in the the following way: $ \nabla_\mu n_\nu = K_{\mu\nu} - a_\nu n_\mu $,
\item If some tensor $\widetilde T_{\alpha\beta\dots} $ is tangent to the foliation, then $ \Lie{n} \widetilde T_{\alpha\beta\dots} $ is also tangent to the foliation.
\end{itemize}
The proof of the first part of the claim regarding the ability to decompose any FPD invariant scalar in terms of the aforementioned basic tangent tensors, can be found in appendix \ref{app:claim_proof}. The last part of the claim is that these basic tangent tensors have a uniform scaling dimension. This will be proven in one of the following subsections. Here we summarize their scaling dimensions:
\begin{align}\label{FPD_Objs:scalingdims}
&d_\sigma [P_{\mu\nu}]  =2,
& &d_\sigma [P^{\mu\nu}]  =-2,
& &d_\sigma [\widetilde{R}_{\mu\nu\rho\sigma}]  =2, \notag
\\
&d_\sigma [\tilde\epsilon_{\alpha\beta\dots}] = d,
& & d_\sigma [\tilde\epsilon^{\alpha\beta\dots}] = -d, 
& &d_\sigma [K_{\mu\nu}] = 2-z,
\\
& 
d_\sigma [a_\alpha] = 0,
&& d_\sigma [\Lie{n} \widetilde T_{\alpha\beta\dots}] = d_\sigma[\widetilde T_{\alpha\beta\dots}] - z, 
& & d_\sigma [\wn_\alpha \widetilde T_{\alpha\beta\dots}] = d_\sigma [\widetilde T_{\alpha\beta\dots}], \notag
\end{align}
where $ \widetilde T_{\alpha\beta\dots} $ is any tangent tensor with uniform scaling dimension.

We conclude that in order to build all FPD invariant expressions with uniform scaling dimension, one has to form scalars from the previously defined basic tangent tensors in all possible ways.

\subsubsection{Identities for Tangent Tensors}\label{ssb:identsfortangent}

In this section we present various useful identities that relate expressions built from the basic tangent tensors. In all the formulas below $\widetilde T_{\alpha\beta\dots}$ denotes a general tensor that is tangent to the foliation, whereas $\widetilde R$, $\wn$ are Riemann tensor and the covariant derivative compatible with the induced metric $P_{\alpha\beta}$. All the other quantities are as defined above.

We begin with formulas for exchanging derivatives (these formulas can be derived from the definition of the curvature and the Gauss-Codazzi relations, see appendix \ref{app:identder}):
\begin{align}
\begin{split}\label{idents:tempspatderexchange}
\Lie{n} \wn_\mu \widetilde T_{\alpha\beta\gamma\dots} = \ &\  \wn_\mu \Lie{n} \widetilde T_{\alpha\beta\gamma\dots}
 + a_\mu \Lie{n} \widetilde T_{\alpha\beta\gamma\dots} \\
 + & \left[(\wn_\nu + a_\nu) K_{\mu\alpha} -(\wn_\alpha + a_\alpha) K_{\mu\nu} -(\wn_\mu + a_\mu) K_{\alpha\nu}  \right]  \widetilde T^\nu {}_{\beta\gamma\dots}+ \dots\ ,
\end{split}\\
\begin{split}\label{idents: spatspatderexchange}
[\wn_\mu,\wn_\nu] \widetilde T_{\lambda\delta\dots} = \ &\ \widetilde R_{\lambda \sigma \mu \nu} \widetilde T^\sigma {}_{\delta\dots} +\dots\ \ .
\end{split}
\end{align}

The cohomological calculation involves integrating by parts over the full spacetime manifold. Since we are using foliation tangent expressions, it is useful to have the following formulas for integrating by parts in terms of the foliation tangent expressions:
\begin{equation}\label{idents:integbyparts}
\begin{split}
\int \sqrt{-g}\  \nabla_\mu \widetilde J^\mu = 0 \Rightarrow  \int \sqrt{-g}\ \wn_\mu \widetilde J^\mu= -\int \sqrt{-g}\ a_\nu \widetilde J^\nu, \\
\int \sqrt{-g}\  \nabla_\mu (\phi n^\mu) = 0 \Rightarrow \int \sqrt{-g}\ \mathcal{L}_n \phi = -\int \sqrt{-g}\ K \phi ,
\end{split}
\end{equation}
where $ \widetilde{J}^\mu $ is any vector tangent to the foliation, $ \phi $ is any scalar and $ K \equiv K^\mu_\mu $. 

Other than the Bianchi identities for $ \widetilde{R}_{\lambda\sigma\mu\nu} $ and its spatial derivative, the Bianchi identities for the full $(d+1)$-dimensional Riemann curvature\footnote{Here by $d+1$ dimensional we mean the Riemann tensor of the full spacetime manifold.} also imply the following relation between $ \Lie{n} \widetilde{R}_{\lambda\sigma\mu\nu} $ and other foliation tangent tensors (this can also be derived from identities \eqref{idents:tempspatderexchange} and \eqref{idents: spatspatderexchange}, see appendix \ref{app:identder}):
\begin{align}\label{idents:second_temporal_bianchi}
\begin{split}
\Lie{n}\widetilde{R}_{\alpha\beta\mu\nu}  = \ & \widetilde{R}_{\alpha\rho\mu\nu}K^\rho_\beta - \widetilde{R}_{\beta\rho\mu\nu} K^\rho_\alpha\\ 
&+ (\wn_\mu + a_\mu) (\wn_\beta + a_\beta) K_{\nu\alpha} - (\wn_\mu + a_\mu) (\wn_\alpha + a_\alpha) K_{\nu\beta}\\
&- (\wn_\nu + a_\nu) (\wn_\beta + a_\beta) K_{\mu\alpha} + (\wn_\nu + a_\nu) (\wn_\alpha + a_\alpha) K_{\mu\beta}.
\end{split}
\end{align}

Two more useful relations give the temporal derivative of the projector $ P_{\mu\nu} $ and the foliation tangent totally anti-symmetric tensor $ \tilde{\epsilon}_{\alpha\beta\dots} $ as a direct consequence of the definition of the extrinsic curvature $ K_{\mu\nu} $:
\begin{align}\label{idents:liepsilon}
\begin{split}
\Lie{n} P_{\mu\nu} &= 2 K_{\mu\nu}, \\
\Lie{n} \tilde{\epsilon}_{\alpha\beta\dots} &= K \tilde{\epsilon}_{\alpha\beta\dots}.
\end{split}
\end{align}

Finally, we reiterate the fact that both the extrinsic curvature and the spatial derivative of the acceleration are symmetric (as a consequence of the Frobenius condition), i.e.\ $ K_{\mu\nu} = K_{\nu\mu} $  and $ \wn_\mu a_\nu = \wn_\nu a_\mu $.

\subsubsection{Anisotropic Weyl Transformation Laws}\label{ssb:weyllaws}

In this subsection we derive the anisotropic Weyl transformations of the basic tangent tensors as defined in subsection \ref{ssb:fpdinvariantexpr}.
Given some foliation tangent tensor $ \widetilde{T}_{\alpha\beta\dots} $, the Weyl transformation of its temporal derivative is given by:
\begin{align}\label{weyllaws:deltalie}
\begin{split}
\delta^W_\sigma(\Lie{n} \widetilde{T}_{\alpha\beta\dots}) &= 
\delta^W_\sigma(n^\mu \partial_\mu \widetilde{T}_{\alpha\beta\dots} + \partial_\alpha n^\mu \widetilde{T}_{\mu\beta\dots} + \partial_\beta n^\mu \widetilde{T}_{\alpha\mu\dots} + \dots) = \\
&= -z \sigma \Lie{n} \widetilde{T}_{\alpha\beta\dots} + \Lie{n} \delta^W_\sigma \widetilde{T}_{\alpha\beta\dots}\, .
\end{split}
\end{align} 
If we define:
\begin{equation}\label{weyllaws:deltatagdef}
\delta^W_\sigma \widetilde{T}_{\alpha\beta\dots} \equiv d_\sigma[\widetilde{T}] \sigma \widetilde{T}_{\alpha\beta\dots} + \delta' \widetilde{T}_{\alpha\beta\dots}\, , 
\end{equation}
where $ d_\sigma[\widetilde{T}] $ is the scaling dimension of $ \widetilde{T}_{\alpha\beta\dots} $ and $ \delta' \widetilde{T}_{\alpha\beta\dots} $ depends only on derivatives of $ \sigma $, then we obtain:
\begin{align}
\begin{split}
\delta^W_\sigma(\Lie{n} \widetilde{T}_{\alpha\beta\dots}) = 
%&= 
%-z \sigma \Lie{n} \widetilde{T}_{\alpha\beta\dots} + d_\sigma[\widetilde{T}] \sigma \Lie{n} \widetilde{T}_{\alpha\beta\dots} + d_\sigma[\widetilde{T}] (\Lie{n} \sigma) \widetilde{T}_{\alpha\beta\dots} + \Lie{n} \delta' \widetilde{T}_{\alpha\beta\dots} =\\
%& =
 (d_\sigma[\widetilde{T}] - z) \sigma \, \Lie{n}\widetilde{T}_{\alpha\beta\dots} + d_\sigma[\widetilde{T}] (\Lie{n} \sigma) \widetilde{T}_{\alpha\beta\dots} + \Lie{n} \delta' \widetilde{T}_{\alpha\beta\dots}\, ,
\end{split}
\end{align}
so that:
\begin{align}
\begin{split}
d_\sigma [\Lie{n} \widetilde{T}] &= d_\sigma[\widetilde{T}] - z, \\
\delta' \Lie{n} \widetilde{T}_{\alpha\beta\dots} &= d_\sigma[\widetilde{T}] (\Lie{n} \sigma) \widetilde{T}_{\alpha\beta\dots} + \Lie{n} \delta' \widetilde{T}_{\alpha\beta\dots}\, .
\end{split}
\end{align}

Similarly, the Weyl transformation of the spatial derivative of $ \widetilde{T}_{\alpha\beta\dots} $ is given by:
\begin{align}
\begin{split}
\delta^W_\sigma (\wn_\mu \widetilde{T}_{\alpha\beta\dots}) &=
\delta^W_\sigma (\partial_\mu \widetilde{T}_{\alpha\beta\dots} - \widetilde{\Gamma}_{\mu\alpha}^{\nu} \widetilde{T}_{\nu\beta\dots} - \dots )=\\
&=\wn_\mu (\delta^W_\sigma \widetilde{T}_{\alpha\beta\dots}) - (\delta^W_\sigma\widetilde{\Gamma}_{\mu\alpha}^{\nu}) \widetilde{T}_{\nu\beta\dots} - \dots\, ,
\end{split}
\end{align}
where $ \widetilde{\Gamma}_{\mu\alpha}^\nu $ are the Christoffel symbols associated with the foliation tangent covariant derivative $ \wn $.
Using:
\begin{equation}\label{weyllaws:christtrans}
\delta^W_\sigma \widetilde{\Gamma}_{\mu\alpha}^{\nu} = (\wn_\mu \sigma) P^\nu_\alpha + (\wn_\alpha \sigma) P^\nu_\mu - (\wn_\rho \sigma) P^{\nu\rho}P_{\mu\alpha},
\end{equation}
we get:
\begin{align}
\begin{split}
\delta^W_\sigma (\wn_\mu \widetilde{T}_{\alpha\beta\dots}) &=
\wn_\mu (\delta^W_\sigma \widetilde{T}_{\alpha\beta\dots}) - I[\widetilde{T}](\wn_\mu \sigma) \widetilde{T}_{\alpha\beta\dots} \\
&\qquad - (\wn_\alpha \sigma)\widetilde{T}_{\mu\beta\dots} + (\wn_\rho \sigma) P_{\mu\alpha} \widetilde{T}^\rho{}_{\beta\dots} \\
&\qquad - (\wn_\beta \sigma)\widetilde{T}_{\alpha\mu\dots} + (\wn_\rho \sigma) P_{\mu\beta} \widetilde{T}_{\alpha}{}^\rho{}_{\dots} - \dots\, ,
\end{split}
\end{align}
where $ I[\widetilde{T}] $ is the number of indices of $ \widetilde{T}_{\alpha\beta\dots} $. Applying the definition \eqref{weyllaws:deltatagdef} we obtain:
\begin{align}
\begin{split}
d_\sigma[\wn \widetilde{T}] &= d_\sigma[\widetilde{T}] ,\\
\delta' (\wn_\mu \widetilde{T}_{\alpha\beta\dots}) &= 
(d_\sigma[\widetilde{T}]- I[\widetilde{T}])(\wn_\mu \sigma) \widetilde{T}_{\alpha\beta\dots}
\\ & ~~~~~~
 - (\wn_\alpha \sigma)\widetilde{T}_{\mu\beta\dots} + (\wn_\rho \sigma) P_{\mu\alpha} \widetilde{T}^\rho{}_{\beta\dots} - \dots\, .
\end{split}
\end{align}

Turning to the basic tangent tensors $ P_{\mu\nu} $, $ K_{\mu\nu} $, $ a_\mu $, $ \widetilde{R}_{\mu\nu\rho\sigma} $ and $ \tilde{\epsilon}_{\alpha\beta\dots} $, we have from the definition \eqref{LifshitzAnom:AnisWeyl}:
\begin{align}
\begin{split}
\delta^W_\sigma P_{\mu\nu} &= 2\sigma P_{\mu\nu}, \\
\delta^W_\sigma P^{\mu\nu} &= -2\sigma P^{\mu\nu},\\
\delta^W_\sigma \sqrt{-g} & = (d+z)\, \sigma \sqrt{-g}.
\end{split}
\end{align}
From the definition of the extrinsic curvature along with \eqref{weyllaws:deltalie} we have:
\begin{align}
\delta^W_\sigma K_{\mu\nu}&\ = (2-z) \sigma K_{\mu\nu} + (\Lie{n}\sigma) P_{\mu\nu},
\\
\delta^W_\sigma K &\ = \delta^W_\sigma ( P^{\mu\nu} K_{\mu\nu} ) = -z\sigma K +d(\Lie{n} \sigma) .
\end{align}
From the definition of the acceleration:
\begin{equation}
\delta^W_\sigma a_\mu = \delta^W_\sigma (n^\alpha \partial_\alpha n_\mu + \partial_\mu n^\alpha n_\alpha ) = z n^\alpha n_\mu \partial_\alpha \sigma + z \partial_\mu \sigma = z P^\alpha_\mu \partial_\alpha \sigma = z \wn_\mu \sigma.
\end{equation}
From the definition of the intrinsic curvature and \eqref{weyllaws:christtrans} we get:
\begin{align}
\begin{split}
\delta^W_\sigma \widetilde R_{\mu\nu\rho\sigma}
&= 2\sigma \widetilde R_{\mu\nu\rho\sigma}
+  P_{\mu\sigma}\wn_\nu \wn_\rho\sigma
-P_{\mu\rho}\wn_\nu\wn_\sigma\sigma
+P_{\nu\rho}\wn_\mu\wn_\sigma \sigma
-P_{\nu\sigma}\wn_\mu\wn_\rho\sigma,
\\
\delta^W_\sigma  \widetilde R_{\mu\nu} &= (2-d) \wn_\mu \wn_\nu \sigma - P_{\mu\nu} \widetilde \Box \sigma,
\\
\delta^W_\sigma \widetilde R &= -2 \sigma \widetilde R - 2(d-1) \widetilde\Box\sigma.
\end{split}
\end{align}
Finally, the Weyl transformation law for the Levi-Civita tensor is given by:
\begin{equation}
\delta^W_\sigma \tilde\epsilon_{\alpha\beta\dots} = d\, \sigma \,\tilde\epsilon_{\alpha\beta\dots},
\qquad
\delta^W_\sigma \tilde\epsilon^{\alpha\beta\dots} = - d\, \sigma \, \tilde\epsilon^{\alpha\beta\dots}.
\end{equation}
The scaling dimensions \eqref{FPD_Objs:scalingdims} can easily be extracted from the above formulas.

\subsubsection{Comparison to the ADM Decomposition}

It is common in the literature to phrase the anisotropic Weyl symmetry in terms of the ADM decomposition. In this section we compare our terms and conventions with those of the ADM decomposition. In the ADM decomposition, one chooses coordinates $(t,x^i)$ such that the leaves of the foliation are given by the constant time slices $t=const$, and $ x^i $ for $ i=1,\dots,d $ are coordinates in each leaf.
Written in these coordinates the metric takes the form:
\begin{equation}
\begin{split}
g_{tt}=-N^2 + N^i N_i, \qquad g_{ti} = N_i, \qquad g_{ij} = \gamma_{ij},\\
g^{tt} = -\frac{1}{N^2}, \qquad g^{ti} = \frac{N^i}{N^2}, \qquad g^{ij} = \gamma^{ij} - \frac{N^iN^j}{N^2},
\end{split}
\end{equation}
where $ \gamma_{ij} $ is the induced metric on the foliation leaves, $ N^i $ is the shift vector and $ N $ is the lapse function.\footnote{The $i,j,\dots$ indices are raised and lowered by the metric $\gamma_{ij} $.}
The covariant volume element in these coordinates is given by:
\begin{equation}
\sqrt{-g}\,d^{(d+1)}x = N\sqrt{\gamma}\,dt\,d^d x .
\end{equation}

The timelike normal to the foliation is given by:
\begin{equation}
\begin{split}
& n_\mu = N(-1, 0) ,\\
& n^\mu = \frac{1}{N} (1,-N^i) .
\end{split}
\end{equation}
In these conventions the Frobenius condition is automatically satisfied.
Note that the foliation projector with upper indices $P^{\mu\nu}$ is nothing but the foliation induced metric $\gamma^{ij}$, therefore any foliation tangent tensors can be contracted using $\gamma^{ij}$.

The basic tangent tensors defined in subsection \ref{ssb:fpdinvariantexpr} can be written in terms of the ADM decomposition as follows. 
The spatial components
 of the extrinsic curvature are given by:\footnote{The temporal components of tangent tensors are completely determined by the spatial components according to the formula:
\begin{equation*}
\widetilde{T}_{0\beta\gamma\delta\dots} = N^i \widetilde{T}_{i\beta\gamma\delta\dots}.
\end{equation*}
For this reason we only present spatial components of tangent tensors.}
\begin{equation}
K_{ij} = \frac{1}{2N} (\del_t \gamma_{ij}-\wn_i N_j-\wn_j N_i),
\end{equation}
where $ \wn $ is the covariant derivative defined on the foliation leaves and compatible with $ \gamma_{ij} $.
The spatial components of the acceleration vector are given by:
\begin{equation}
a_i = \frac{\wn_i N}{N}.
\end{equation}
The intrinsic Riemann and Levi-Civita tensors are those associated with the metric $\gamma_{ij}$.

Given a tangent tensor $ \widetilde{T}_{ijk\dots} $, the spatial components of its temporal derivative are given by:
\begin{equation}
\mathcal{L}_n \widetilde{T}_{ijk\dots} = \frac{1}{N} \del_t \widetilde{T}_{ijk\dots} -\frac{1}{N} \mathcal{L}^{(d)}_{\vec N} \widetilde{T}_{ijk\dots},
\end{equation}
where $ \mathcal{L}^{(d)}_{\vec N} $ is the Lie derivative inside the foliation leaf in the direction of the shift vector $ N^i $.

Finally, we note that the transformation of the metric components under anisotropic Weyl is given by:
\begin{align}
\begin{split}
&\delta^W_\sigma N = z \sigma N, \\
&\delta^W_\sigma N_i = 2 \sigma N_i, \\ 
&\delta^W_\sigma \gamma_{ij} = 2\sigma \gamma_{ij} \ .
\end{split}
\end{align}

\subsubsection{Restrictions and Classification by Sectors}\label{ssb:restclasssectors}
Using the scaling dimensions \eqref{FPD_Objs:scalingdims} and some combinatorics one can derive various constraints on the possible terms in the cohomology and their properties for generic values of $z$ and $d$.
As previously mentioned, the various terms in the cohomology all have the form $ \int \sqrt{-g}\phi $, where $ \phi $ is a scalar of uniform scaling dimension $ -(d+z) $, built from contractions of the basic tangent tensors defined in subsection \ref{ssb:fpdinvariantexpr}. Suppose that $ n_K $, $n_a$, $n_R$, $n_\epsilon$, $ n_\nabla$ and $ n_\mathcal{L} $ are the number of extrinsic curvature, acceleration, intrinsic Riemann curvature, Levi-Civita tensor\footnote{It is enough to consider $ n_\epsilon = 0$ or $1$.}, spatial derivative and temporal derivative instances that appear in $\phi$ respectively, and $ n_P $ the number of induced metric instances required to contract them.     

For the scaling dimension to be correct we require:
\begin{equation}\label{rest:scalingreq}
(2-z)n_K-z n_\mathcal{L} +2n_R + d n_\epsilon -2  n_P = -d-z .
\end{equation}
For all indices in $\phi$ to be contracted in pairs we require:
\begin{equation}\label{rest:indexreq}
2 n_K + n_a + n_\nabla + 4n_R + d n_\epsilon = 2 n_P  .
\end{equation}
From requirements \eqref{rest:scalingreq} and \eqref{rest:indexreq} we obtain the conditions:
\begin{gather}\label{rest:constraint1}
z n_\mathcal{L} + z n_K + n_a + n_\nabla + 2 n_R = d+z,
\\
\label{rest:constraint2}
n_a+n_\nabla + d n_\epsilon \quad \text{is even} .
\end{gather}

Defining $ n_T \equiv n_\mathcal{L} + n_K $ as the total number of time derivatives (when writing the expression in terms of the ADM decomposition) and $ n_S \equiv n_\nabla + n_a + 2 n_R $ as the total number of spatial derivatives in the expression, we unsurprisingly get the following form for these conditions:
\begin{gather}\label{rest:constraint1b}
z n_T + n_S = d+z, \\
\label{rest:constraint2b}
n_S + d n_\epsilon \quad \text{is even} .
\end{gather}
A notable property of \eqref{rest:constraint1} is that all the coefficients are positive (assuming $z>0$), and so as expected the number of possibilities is limited.

From the transformation formulas in subsection \ref{ssb:weyllaws}, it can be easily checked that the numbers $ n_T $, $ n_S $ and $ n_\epsilon $ remain unchanged when applying the Weyl operator $ \delta^W_\sigma $ to any tangent tensor, i.e.:
\begin{align}
\begin{split}
n_S [\delta^W_\sigma \widetilde{T}_{\alpha\beta\dots} ] &= n_S [\widetilde{T}_{\alpha\beta\dots} ], \\
n_T [\delta^W_\sigma \widetilde{T}_{\alpha\beta\dots} ] &= n_T [\widetilde{T}_{\alpha\beta\dots} ], \\
n_\epsilon [\delta^W_\sigma \widetilde{T}_{\alpha\beta\dots} ] &= n_\epsilon [\widetilde{T}_{\alpha\beta\dots} ]  .
\end{split}
\end{align} 
Additionally, identities relating different tangent tensors such as the ones in subsection \ref{ssb:identsfortangent} always relate expressions with the same values of $n_S$, $n_T$ and $n_\epsilon$. Thus the linear space of expressions of the form $ \int\sqrt{-g}\phi $ that satisfy the conditions \eqref{rest:constraint1b} and \eqref{rest:constraint2b} is a direct sum of subspaces (which we refer to as \emph{sectors}), each corresponding to specific values of $ (n_S,n_T,n_\epsilon) $ that satisfy these conditions. Since expressions remain in the same sector when applying the Weyl transformation, we conclude that when studying the Lifshitz cohomology problem for given $ d $ and $z$, we may focus on each sector separately.

Two important properties that can be deduced from these numbers are concerned with
the behaviour under parity and time reversal transformations of the expressions we build. 
Under time reversal, in addition to the regular behaviour of the different tensorial expressions, the normalized foliation 1-form changes sign (assuming one defines it such that $ n^\alpha $ points "forward" in time), so that:
\begin{align}
n^\alpha &\xrightarrow{T} -n^{\alpha'} 
& n_\alpha &\xrightarrow{T} -n_{\alpha'} 
& P_{\mu\nu} &\xrightarrow{T} P_{\mu'\nu'} \notag\\
K_{\mu\nu} &\xrightarrow{T} -K_{\mu'\nu'} 
& a_{\mu} &\xrightarrow{T} a_{\mu'} 
& \widetilde{R}_{\mu\nu\rho\sigma} &\xrightarrow{T} \widetilde{R}_{\mu'\nu'\rho'\sigma'}\\
\tilde{\epsilon}_{\alpha\beta\dots} &\xrightarrow{T} \tilde{\epsilon}_{\alpha'\beta'\dots}
& \Lie{n} \widetilde{T}_{\alpha\beta\dots} &\xrightarrow{T} -T_{\widetilde{T}}\,\Lie{n} \widetilde{T}_{\alpha'\beta'\dots}
& \wn_\mu \widetilde{T}_{\alpha\beta\dots} &\xrightarrow{T} T_{\widetilde{T}}\, \wn_{\mu'} \widetilde{T}_{\alpha'\beta'\dots} \notag
\end{align}
where $ \alpha', \beta', \dots$ denote the transformed coordinate frame, and $T_{\widetilde{T}}$ denotes the sign of the tangent tensor $\widetilde{T}_{\alpha\beta\dots}$ under time reversal. Under parity, though, the foliation 1-form doesn't change sign, so we have:
\begin{align}
n^\alpha &\xrightarrow{P} n^{\alpha'} 
& n_\alpha &\xrightarrow{P} n_{\alpha'} 
& P_{\mu\nu} &\xrightarrow{P} P_{\mu'\nu'} \notag\\
K_{\mu\nu} &\xrightarrow{P} K_{\mu'\nu'} 
& a_{\mu} &\xrightarrow{P} a_{\mu'} 
& \widetilde{R}_{\mu\nu\rho\sigma} &\xrightarrow{P} \widetilde{R}_{\mu'\nu'\rho'\sigma'}\\
\tilde{\epsilon}_{\alpha\beta\dots} &\xrightarrow{P} -\tilde{\epsilon}_{\alpha'\beta'\dots}
& \Lie{n} \widetilde{T}_{\alpha\beta\dots} &\xrightarrow{P} P_{\widetilde{T}}\, \Lie{n} \widetilde{T}_{\alpha'\beta'\dots}
& \wn_\mu \widetilde{T}_{\alpha\beta\dots} &\xrightarrow{P} P_{\widetilde{T}}\, \wn_{\mu'} \widetilde{T}_{\alpha'\beta'\dots} \notag
\end{align}
where $P_{\widetilde{T}}$ is the sign of $\widetilde{T}_{\alpha\beta\dots}$ under parity.
Thus in general, the sign of a scalar built from the basic tangent tensors under time reversal and parity is given by:
\begin{equation}\label{rest:pt_props}
T = (-1)^{n_T}, \qquad P = (-1)^{n_\epsilon}.
\end{equation}

Various constraints on the possible properties of anomalies can be derived using the above considerations (\ref{rest:constraint1},  \ref{rest:constraint2}, \ref{rest:pt_props}) and known properties of the underlying field theory.
For example, if we know that the field theory is time reversal invariant it can be shown that anomalies (or in general scalar terms in the cohomology) are only possible for the following values of $z$ and $d$:
\begin{enumerate}
\item Rational (non-integer) $z=p/q$ satisfying $p\leq d$ with $q$ odd, $p$ even.
\item Rational (non-integer) $z=p/q$ satisfying $p\leq d$ with $q,\ p,\ d$ odd.
\item Even integer $z$.
\item Odd integer $z$, odd $d$. 
\end{enumerate}
Here $p/q$ should be a reduced fraction and can be greater then 1.
This can be used to show e.g.\ that for $d=2$, $z=1$ the only possible anomalous terms are time reversal breaking.  

Another interesting conclusion is, that given a specific number of space dimensions $d$, there is generally a finite number of $z>1$ values that allow for sectors with more than one time derivative. The rest allow only for sectors with $n_T=0$ (purely spatial sectors) or $n_T=1$ (universal sectors, that exist for any value of $z$).

\subsubsection{A Prescription for Finding the Anomalous Terms} \label{ssb:prescription}

In this subsection we use the previous results to give a detailed prescription for finding the anomalous terms in the relative cohomology of the anisotropic Weyl operator for any given value of $z$ and $d$.  The prescription is as follows:

\begin{enumerate}
\item As mentioned above, the various sectors of given $n_T$, $n_S$ and $n_\epsilon$ do not mix, so that it's possible to apply the prescription to each sector separately. Therefore the first step is to identify the different sectors consistent with conditions \eqref{rest:constraint1b}, \eqref{rest:constraint2b}.
\item For each of these sectors, build all possible FPD invariant expressions from the basic tangent tensors of subsection  \ref{ssb:fpdinvariantexpr} with the proper values of $n_T$, $n_S$ and $n_\epsilon$ (by contracting their indices in all possible manners). 
\item Find an independent basis of these expressions: $\phi_i,\ i=1,\dots,n_{{}_{FPD}}$, taking into account:
	\begin{itemize}[label= --]
	\item The identities of section \ref{ssb:identsfortangent}.
	\item Additional dimensionally dependent identities due to 	anti-symmetrising any set of more than $d$ indices, e.g for $d=2$ we have $\tilde{\epsilon}^{\alpha\gamma} K^\beta_\gamma - \tilde{\epsilon}^{\beta\gamma}K_\gamma^\alpha = \tilde{\epsilon}^{\alpha\beta} K$.
	\end{itemize}
\item To find the cocycles of the relative cohomology:
	\begin{itemize}[label= --]
	\item Build the integrated expressions of ghost number one: $ I_i = \int \sqrt{-g}\,\sigma \phi_i $.\footnote{Any expression with ghost number one of the form $ \int \sqrt{-g}\, (\del^k \sigma) \,\chi $ ($k>0$), where $\del^k\sigma$ involves any number of spatial or temporal derivatives of $\sigma$, can be written as a linear combination of the expressions $ I_i $ using integration by parts.}
	\item Apply the Weyl operator $\delta^W_\sigma$ to each of these terms to obtain ghost number two expressions.
	\item Use integration by parts (identities \eqref{idents:integbyparts}) and the Grassmannian nature of $\sigma$ to reduce each of them to a linear combination of independent expressions of the form:
	$L_j = \int \sqrt{-g}\, \chi_j \sigma  \del^{2k+1} \sigma,\ j=1,\dots,m $, where $\del^{2k+1} \sigma$ represents an odd number of derivatives (either temporal or spatial) applied to $\sigma$, and $\chi_j$ is any expression constructed from tangent tensors (not necessarily a scalar).\footnote{Any expression with an even number of derivatives acting on $\sigma$ can be written as a linear combination of expressions with lower odd number of derivatives, using the integration by parts identities \eqref{idents:integbyparts}, the derivative exchange formulas \eqref{idents:tempspatderexchange}, \eqref{idents: spatspatderexchange} and the Grassmannian nature of $\sigma$.} 
Suppose these linear combinations are given by: $ \delta^W_\sigma I_i = - M_{ij} L_j $.
	\item Find all linear combinations of the basic ghost number one expressions $ E = C_i I_i $ (where $C_i$ are constants) that satisfy $ \delta^W_\sigma E = 0 $, by solving the linear system of equations:
\begin{equation}\label{presc:cocycleeqs}
M_{ij} C_i = 0.
\end{equation}	
The space of solutions is the cocycle space. Let $ E_i,\ i=1,\dots,n_{cc}$ be some basis for this space, where $ n_{cc} $ is its dimension.   
	\end{itemize}
\item To find the coboundaries of the relative cohomology:
\begin{itemize}[label= --]
\item Build the integrated expressions of ghost number zero: $ G_i = \int \sqrt{-g}\phi_i $.
\item Apply the Weyl operator $ \delta^W_\sigma $ to each of these to obtain ghost number one expressions.
\item Use integration by parts to reduce each of them to a linear combination of the expressions $ I_i $.\footnote{Note that the expression $ \delta^W_\sigma G_i$ must be a linear combination of the cocycles $E_i$, since any coboundary is also a cocycle due to the nilpotence of $\delta^W_\sigma$. This can be used as a check for the calculation.}  Suppose these combinations are given by: $ \delta^W_\sigma G_i = S_{ij} I_j $. The span of these combinations is the coboundary space. Let $ F_i,\ i=1,\dots,n_{cb} $ be some basis for this space, where $n_{cb}$ is its dimension.
\end{itemize}
\item Finally, to find the anomalous terms in the cohomology, check which of the cocycles $ I_i $ are not in the span of the coboundaries $ F_i$ (or, stated differently, find the expressions that complete the basis of coboundaries into a basis of the whole cocycle space). We denote these by $ A_i,\ i=1,\dots,n_{an} $, where $ n_{an} = n_{cc} - n_{cb} $ is the number of independent anomalies. 
\end{enumerate}

Examples for the application of this prescription will be given in the following sections.\\

A useful fact that is worth mentioning is that a coboundary term must always be a total derivative. This can be explained as follows. Assume that $\mathcal{F}$ is a coboundary density. This is the case if:
\begin{equation}
\int \sqrt{-g}\,\sigma \mathcal{F} = \delta^W_\sigma \int \sqrt{-g}\, \phi,
\end{equation}
where $\int \sqrt{-g}\,\phi$ is some local functional of the background fields which must be of global scaling dimension $0$. If we then set the transformation parameter to be a constant we should find:
\begin{equation}
c \int \sqrt{-g}\, \mathcal{F} = \delta^W_c \int \sqrt{-g}\,\phi = 0,
\end{equation}
for any choice of the background fields.
$\mathcal{F}$ is therefore a total derivative.

As noted in the introduction, in the following we will sometimes use $ E_i^{(n_T,n_S,n_\epsilon)} $, $ F_i^{(n_T,n_S,n_\epsilon)} $ or $ A_i^{(n_T,n_S,n_\epsilon)} $ to denote the $i$-th cocycle, coboundary or anomaly, respectively, in the sector with  $ n_T $ time derivatives, $ n_S $ space derivatives and $ n_\epsilon $ Levi-Civita tensors in the Lifshitz cohomology with specific values of $d$ and $z$.

\subsubsection{Classification of Lifshitz Scale Anomalies}\label{ssb:prelim_classification}

In the study of the algebraic structure of Weyl anomalies in conformal theories, they have been  classified into two types (see e.g.~\cite{Deser:1993yx,Bonora:1985cq,Bonora:1983ff,Boulanger:2007ab}):
\begin{enumerate}
\item The type A anomaly, consisting of the (integrated) Euler density of the manifold,
\item The type B anomalies, consisting of strictly Weyl invariant scalar densities. Some of these are simply various contractions of products of the Weyl tensor, while others have a more complicated structure involving the Riemann tensor and its covariant derivatives. 
\end{enumerate}
This classification has been shown to be valid for any even dimension. 
The feature distinguishing between the two types of anomalies has been presented from different points of view. The authors of \cite{Deser:1993yx} gave a general argument for this structure using dimensional regularization, emphasizing the scale dependence of the effective action. In this view, type A anomalies are characterized by the scale independence of the action, which is equivalent to the vanishing of the integrated anomaly. They are, therefore, related to topological invariants (as is the case with the integrated Euler density). Type B anomalies then correspond to effective actions that contain a scale.

In \cite{Boulanger:2007ab,Boulanger:2007st}, the author presents a general cohomological argument for the aforementioned classification, in terms of descent equations. He distinguishes between the type B anomalies, that satisfy a trivial descent of equations (i.e.\ Weyl invariant densities), and type A anomalies  that have non-trivial descent. He shows that the unique anomaly with non-trivial descent is the Euler density, and as such is the counterpart of the non-abelian chiral anomaly.

For the analogous problem of classifying the Lifshitz anomalies, we choose here the latter approach, of trivial versus non-trivial descent.\footnote{The two approaches don't align in all cases. For example, if we take into account parity violating theories in the conformal case, the Pontryagin term \eqref{prelims:P_1} is a possible anomaly in the theory, and  it is both a topological term and a Weyl invariant density. Thus, it can
be considered both as type A and type B. Our results for the Lifshitz cohomology contain terms with a similar issue.} We will, therefore, be interested in the space of cocycles of the form $ E = \int \sqrt{-g}\sigma \phi $, where $\phi$ satisfies:
\begin{equation}
\delta^W_\sigma (\sqrt{-g} \phi) = 0 .
\end{equation}
We will refer to them as trivial descent cocycles. One can find them using a procedure similar to the one outlined in subsection \ref{ssb:prescription} for finding the cocycles, but without performing integration by parts in determining the independent ghost number two expressions.
One then obtains a linear system of equations, the solutions of which form the trivial descent cocycles space. We denote the basis for this space by $H_i,\ i=1,\dots,n_{td}$ where $ n_{td} $ is its dimension. We also denote by $ n_{tdcb} $ the dimension of the trivial descent coboundaries space (the intersection of the trivial descent cocycles space and the coboundaries space).

If $ A $ is an anomaly of the Lifshitz cohomology, we will refer to it as a trivial descent anomaly, or type B, if it belongs to the same cohomological class as a trivial descent cocycle, so that:
\begin{equation}\label{class:tdanomalydef}
A = H + F ,
\end{equation}
where $H$ is some trivial descent cocycle and $F$ is some coboundary term. 

Unlike the conformal case, we found in all of the examples studied of the Lifshitz cohomology, that all anomalies are trivial descent anomalies (in the sense of \eqref{class:tdanomalydef}). Equivalently, in all of the examples studied the following holds:
\begin{equation}
\Omega_{cc} = \Omega_{cb} + \Omega_{td} ,
\end{equation}
where $ \Omega_{cc} $, $ \Omega_{cb} $, $ \Omega_{td} $ denote the spaces of cocycles, coboundaries and trivial descent cocycles respectively. Our conjecture is that this may be true for the Lifshitz cohomology for any dimension and any value of the dynamical exponent $z$.

We also note in our results an additional difference compared to the conformal case: In many examples there are coboundaries which have trivial descents (i.e.\ $n_{tdcb}>0$), so that not all of the (anisotropic) Weyl invariant densities are actually anomalies. This is in contrast to the conformal case, where all Weyl invariant densities have been argued to be anomalies~\cite{Deser:1993yx}.

\section{Lifshitz Scale Anomalies in 1+1 Dimensions}\label{sec:1+1}
In this section we detail our results for Lifshitz scale anomalies in 1+1 dimensions for various values of $z$.
We begin with a detailed derivation of the $z=1$ case followed by a list of results for other values of $z$. The $z=1$ case is of special importance since it can be compared to the conformal case (which also obeys the same scaling relation). In the Lifshitz theory however, we lack the boost invariance and special conformal transformations characteristic of a completely conformal theory.

\subsection{The \texorpdfstring{$z=1$}{z=1} Case}
We follow step by step the prescription given in the previous section.
The first step is to identify the different sectors consistent with equations \eqref{rest:constraint1b}, \eqref{rest:constraint2b}. These are all values of $n_T$, $n_S$, $n_\epsilon$ consistent with:
\begin{equation}
\begin{split}
n_T + n_S = 2, \\
n_S + n_\epsilon \quad \text{even}.
\end{split}
\end{equation}
We list them below:
\begin{itemize}[label= --]
\item $n_T=2$, $n_S=0$, $n_\epsilon=0$,
\item $n_T=0$, $n_S=2$, $n_\epsilon=0$,
\item $n_T=1$, $n_S=1$, $n_\epsilon=1$.
\end{itemize}
We will now address each of these sectors separately.

\subsubsection{Parity Even Sector with Two Time Derivatives}
In the $d=1$ case the extrinsic curvature has only one independent component, let us take it to be its trace $K$. The independent FPD invariant terms are therefore:
\begin{equation}
\begin{split}
\phi_1 = K^2 , \qquad \phi_2 = \Lie{n} K  .
\end{split}
\end{equation}
The associated expressions of ghost number 1 are:
\begin{equation}
\begin{split}
I_1  = \int \sqrt{-g}\ \sigma \ K^2, \qquad
I_2  = \int \sqrt{-g}\ \sigma \ \Lie{n}K.
\end{split}
\end{equation}
The Weyl variation of each can be calculated using the rules of subsection \ref{ssb:weyllaws}:\footnote{Note that $\delta_\sigma^W$ and $\sigma$ anticommute.}
\begin{equation}
\begin{split}
& \delta^W_\sigma \ I_1= -\int \sqrt{-g}\ \sigma \ [2K \Lie{n}\sigma], \\
& \delta^W_\sigma \ I_2 = -\int \sqrt{-g}\ \sigma \ [-K \Lie{n} \sigma+\Lie{n}^2\sigma].
\end{split}
\end{equation}
We can now use \eqref{idents:integbyparts} to relate $\int \sqrt{-g}\ \sigma \ \Lie{n}^2\sigma$ to an expression with an odd number of derivatives acting on the ghost $\sigma$:
\begin{equation}
\int \sqrt{-g}\ \sigma \ \Lie{n}^2\sigma = - \int \sqrt{-g}\ \sigma \ K \Lie{n}\sigma- \int \sqrt{-g}\ (\Lie{n}\sigma)^2 = -\int \sqrt{-g}\ \sigma \ K \Lie{n}\sigma,
\end{equation}
where we used the Grassmannian nature of $\sigma$ to set $(\Lie{n}\sigma)^2=0$.
If we now define the independent ghost number 2 quantity $L_1 \equiv \int \sqrt{-g}\ \sigma \ K \Lie{n}\sigma$ we can express the Weyl variation as:
\begin{equation}
\begin{split}
& \delta^W_\sigma \ I_1= -2 L_1 ,\\
& \delta^W_\sigma \ I_2 = 2 L_1 .
\end{split}
\end{equation}
We see immediately that the cocycle space contains only one term ($n_{cc}=1$):
\begin{equation}
E_1 = I_1+I_2 = \int \sqrt{-g}\ \sigma \ [K^2+ \Lie{n}K].
\end{equation}
To find the coboundaries of the relative cohomology, we build the zero ghost expressions:
\begin{equation}
\begin{split}
G_1  = \int \sqrt{-g}\ \ K^2, \qquad
G_2  = \int \sqrt{-g}\ \ \Lie{n}K.
\end{split}
\end{equation}
Their Weyl variations are given by:
\begin{equation}
\begin{split}
& \delta^W_\sigma \ G_1= \int \sqrt{-g} \ [2K \Lie{n}\sigma], \\
& \delta^W_\sigma \ G_2 = \int \sqrt{-g} \ [-K \Lie{n} \sigma+\Lie{n}^2\sigma].
\end{split}
\end{equation}
We can integrate by parts to obtain:
\begin{equation}
\begin{split}
& \delta^W_\sigma \ G_1= - \int \sqrt{-g} \ 2\sigma [ \Lie{n}K+K^2] = -2[I_1+I_2], \\
& \delta^W_\sigma \ G_2 = \int \sqrt{-g} \ 2\sigma [\Lie{n}K+K^2] = 2[I_1+I_2].
\end{split}
\end{equation}
Therefore, the coboundary space is spanned by $F_1 = I_1+I_2$ and its dimension is $n_{cb}=1$. A useful consistency check is to make sure that all the coboundaries are in the cocycle span (which is indeed the case here).

As we can see, the cocycle space (spanned by $E_1$) is contained in the coboundary space (spanned by $F_1$) and so we have no anomalies in this sector $n_{an}=n_{cc}-n_{cb} = 1-1=0$. Note, that there are no trivial descent cocycles in this sector.

\subsubsection{Parity Even Sector with Two Space Derivatives}
We now turn to the second sector which contains two space derivatives and no $\tilde\epsilon$ tensor. In the $d=1$ case the intrinsic Riemann tensor is equal to zero. The independent FPD invariant terms are therefore:
\begin{equation}
\begin{split}
\phi_1 = a^2,\qquad
\phi_2 = \wn_\mu a^\mu.
\end{split}
\end{equation}
The integrated expressions of ghost number 1 are given by:
\begin{equation}
\begin{split}
I_1  = \int \sqrt{-g}\ \sigma \ a^2, \quad
I_2  = \int \sqrt{-g}\ \sigma \ \wn_\mu a^\mu.
\end{split}
\end{equation}
The Weyl variation of each can be calculated using the rules of subsection \ref{ssb:weyllaws}:
\begin{equation}
\begin{split}
& \delta^W_\sigma \ I_1= -\int \sqrt{-g}\ 2 \  \sigma\   a^\mu \wn_\mu \sigma, \\
& \delta^W_\sigma \ I_2 = -\int \sqrt{-g}\ \sigma \ [- a^\mu \wn_\mu \sigma + \widetilde\Box \sigma].
\end{split}
\end{equation}
where $\widetilde \Box \equiv \wn_\mu \wn^\mu$.
We can integrate by parts to express $\int \sqrt{-g}\ \sigma \ \widetilde\Box \sigma$ as an odd order derivative acting on one of the ghosts $\sigma$:
\begin{equation}
\int \sqrt{-g}\ \sigma \ \widetilde\Box\sigma = - \int \sqrt{-g}\ \sigma \ a^\mu \wn_\mu \sigma .
\end{equation}
Defining $L_1 \equiv \int \sqrt{-g}\ \sigma \ a^\mu \wn_\mu \sigma $ we can express the Weyl variation as:
\begin{equation}
\begin{split}
& \delta^W_\sigma \ I_1=- 2 L_1 ,\\
& \delta^W_\sigma \ I_2 = 2 L_1 .
\end{split}
\end{equation}
The cocycle space contains only one term ($n_{cc}=1$):
\begin{equation}
E_1 = I_1+I_2 = \int \sqrt{-g}\ \sigma \ [a^2+ \wn_\mu a^\mu].
\end{equation}
To find the coboundaries of the relative cohomology, we build the ghost number zero expressions:
\begin{equation}
\begin{split}
G_1  = \int \sqrt{-g}\ \ a^2, \quad
G_2  = \int \sqrt{-g}\ \ \wn_\mu a^\mu.
\end{split}
\end{equation}
Their Weyl variations are given by:
\begin{equation}
\begin{split}
& \delta^W_\sigma \ G_1= \int \sqrt{-g}\ 2 \   a^\mu \wn_\mu \sigma, \\
& \delta^W_\sigma \ G_2 = \int \sqrt{-g}\  [- a^\mu \wn_\mu \sigma + \widetilde\Box \sigma].
\end{split}
\end{equation}
We can integrate by parts to obtain:
\begin{equation}
\begin{split}
& \delta^W_\sigma \ G_1= - \int \sqrt{-g} \ 2\sigma\  [a^2+ \wn_\mu a^\mu] = -2[I_1+I_2], \\
& \delta^W_\sigma \ G_2 = \int \sqrt{-g} \ 2\sigma \  [a^2+ \wn_\mu a^\mu]  = 2[I_1+I_2].
\end{split}
\end{equation}
Therefore, the coboundary space is spanned by $F_1 = I_1+I_2$ and its dimension is $n_{cb}=1$.

As we can see, the cocycle space is contained in the coboundary space and so we have no anomalies in this sector $n_{an}=0$. Just as before, there are no trivial descent cocycles in this sector.

\subsubsection{Parity Odd (Universal) Sector with One Space and One Time Derivative}\label{ssb:1+1_1S1Todd_Results}
In this subsection we will detail the computation for the sector with one time derivative, one space derivative and an $\tilde\epsilon$ tensor. We will regard $z$ as a general parameter (not necessarily $z=1$). The reason for that is that this sector is \emph{universal}. That means that the same sector with the exact same terms exists for general values of $z$ in $1+1$ dimensions.

The FPD invariant terms in this sector are:
\begin{equation}
\begin{split}
\phi_1  = \ K \ \tilde  \epsilon^\mu a_\mu,\qquad
\phi_2  = \tilde \epsilon^\mu \ \wn_\mu K,\qquad
\phi_3  = \tilde \epsilon^\mu \ \mathcal{L}_n a_\mu.
\end{split}
\end{equation}
The integrated ghost number 1 expressions:
\begin{equation}
\begin{split}
I_1  = \int \sqrt{-g}\ \sigma \ K \ \tilde  \epsilon^\mu a_\mu, \qquad
I_2  = \int \sqrt{-g}\ \sigma \  \tilde \epsilon^\mu \ \wn_\mu K, \qquad
I_3  = \int \sqrt{-g}\ \sigma \  \tilde \epsilon^\mu \ \mathcal{L}_n a_\mu. \\
\end{split}
\end{equation}
The Weyl variations:
\begin{equation}
\begin{split}
& \delta^W_\sigma \ I_1= -\int \sqrt{-g}\ \  \sigma\  [\tilde  \epsilon^\mu a_\mu\ \Lie{n}\sigma 
+ z K \, \tilde  \epsilon^\mu \, \wn_\mu\sigma], \\
& \delta^W_\sigma \ I_2 = -\int \sqrt{-g}\ \sigma \ [\tilde\epsilon^\mu \, \wn_\mu \Lie{n}\sigma 
-z K \tilde\epsilon^\mu \, \wn_\mu \sigma  ],\\
& \delta^W_\sigma \ I_3 = -\int \sqrt{-g}\ \sigma z \ \tilde\epsilon^\mu \Lie{n}\wn_\mu  \sigma
.
\end{split}
\end{equation}
We can use integration by parts and equations \eqref{idents:tempspatderexchange}, \eqref{idents:liepsilon} to prove that $\int \sqrt{-g}\ \sigma \ \tilde\epsilon^\mu \Lie{n}\wn_\mu \sigma=
\int \sqrt{-g}\ \sigma \ \tilde\epsilon^\mu [\wn_\mu  \Lie{n}   \sigma +  a_\mu \Lie{n}  \sigma]=0$.
Defining:
\begin{equation}
\begin{split}
L_1  \equiv \int \sqrt{-g}\ \sigma \tilde  \epsilon^\mu a_\mu\ \Lie{n}\sigma,\qquad
L_2  \equiv \int \sqrt{-g}\ \sigma K \, \tilde  \epsilon^\mu \, \wn_\mu\sigma,
\end{split}
\end{equation}
we can express the Weyl variation as:
\begin{equation}
\begin{split}
& \delta^W_\sigma \ I_1= -(L_1+zL_2) ,\\
& \delta^W_\sigma \ I_2 = L_1+zL_2,\\
& \delta^W_\sigma \ I_3 = 0 .
\end{split}
\end{equation}
The cocycle space is the span of the two following terms ($n_{cc}=2$):
\begin{equation}
\begin{split}
E_1 & = I_1+I_2 = \int \sqrt{-g}\ \sigma \ [K \ \tilde  \epsilon^\mu a_\mu + \tilde \epsilon^\mu \ \wn_\mu K],\\
E_2 & = I_3 = \int \sqrt{-g}\ \sigma \  \tilde \epsilon^\mu \ \mathcal{L}_n a_\mu. \\
\end{split}
\end{equation}
To find the coboundaries of the relative cohomology, we build the zero ghost expressions:
\begin{equation}
\begin{split}
G_1  = \int \sqrt{-g}\ K \ \tilde  \epsilon^\mu a_\mu, \qquad
G_2  = \int \sqrt{-g}\ \tilde \epsilon^\mu \ \wn_\mu K, \qquad
G_3  = \int \sqrt{-g}\ \tilde \epsilon^\mu \ \mathcal{L}_n a_\mu. 
\end{split}
\end{equation}
Their Weyl variations are given by:
\begin{equation}
\begin{split}
& \delta^W_\sigma \ G_1= \int \sqrt{-g}\   [\tilde  \epsilon^\mu a_\mu\ \Lie{n}\sigma 
+ z K \, \tilde  \epsilon^\mu \, \wn_\mu\sigma], \\
& \delta^W_\sigma \ G_2 = \int \sqrt{-g}\ [\tilde\epsilon^\mu \, \wn_\mu \Lie{n}\sigma 
-z K \tilde\epsilon^\mu \, \wn_\mu \sigma  ],\\
& \delta^W_\sigma \ G_3 = \int \sqrt{-g}\  z \ \tilde\epsilon^\mu \Lie{n}\wn_\mu  \sigma
.
\end{split}
\end{equation}
We can integrate by parts to obtain:
\begin{equation}
\begin{split}
& \delta^W_\sigma \ G_1= - [zI_1+zI_2+I_3], \\
& \delta^W_\sigma \ G_2 = [zI_1+zI_2+I_3], \\
& \delta^W_\sigma \ G_3 = 0.
\end{split}
\end{equation}
Therefore, the coboundary space is spanned by $F_1 = z(I_1+I_2)+I_3$ and its dimension is $n_{cb}=1$.
It is useful to check that the coboundaries are in the cocycle span which is easily verified here. 

As we can see, we are left with one cocycle which is not in the coboundary span $n_{an}=n_{cc}-n_{cb}=1$. We choose it to be:
\begin{equation}\label{1d:z1_anomaly_1_1_1}
\boxed{
A_1^{(1,1,1)}  = \int \sqrt{-g}\ \sigma \  \tilde \epsilon^\mu \ \mathcal{L}_n a_\mu.
}
\end{equation}
We, therefore, see that the $z=1$ case has only one anomaly $A_1^{(1,1,1)}$.

Note, that the space of trivial descent cocycles (Weyl-invariant scalar densities) here is of dimension $ n_{td} = 1$, and is proportional to:
\begin{align}
\begin{split}
H_1 &= I_3-z(I_1+I_2).
\end{split}
\end{align}
The anomaly is, therefore, trivial descent (up to the addition of coboundary terms):
\begin{equation}
A_1^{(1,1,1)} = I_3 = \frac{1}{2}(H_1+F_1).
\end{equation}

\subsection{Other Integer Values of  \texorpdfstring{$z$}{z}}\label{1d:z_geq_1}
In this subsection we study the relative cohomology of the anisotropic Weyl operator in 1+1 dimensions for general integer $z$ greater than one. We identify only two sectors for this case:
\begin{itemize}[label= -- ]
\item Parity odd sector with one time and one space derivative: $n_T=1$, $n_S=1$, $n_\epsilon = 1$. This is the same universal sector studied in the previous subsection and it contains one anomaly: $\int \sqrt{-g}\ \sigma \  \tilde \epsilon^\mu \ \mathcal{L}_n a_\mu$.
\item Sector with $z+1$ space derivatives: $n_S=z+1$, $n_T=0$. This sector should be parity odd for even z ($n_\epsilon=1$) and parity even for odd $z$ ($n_\epsilon=0$).
\end{itemize}
Since we have already studied the first sector in the previous subsection, we will now focus on the second (purely spatial) sector. 
In 1+1 dimensions there is no need to carry indices or Levi-Civita tensors explicitly. We can write all expressions in terms of $a$ and its spatial derivatives. 
We have developed several formulas specific to the 1+1 dimensional case which can be found in appendix \ref{app:d_1_ids}.
We used a script to perform the calculation up to $z=12$ and found no anomalies in this sector.
We list in table \ref{table:1+1SpatResultsSum} the number of cocycles and coboundaries for each value of $z$ as well as the explicit form of the coboundaries for $z\leq 4$.
We conjecture that there are no anomalies in this sector for any integer $z$. We leave the proof of this statement for future work. Note also, 
 that there are no trivial descent cocycles in this sector up to $z=12$.
\begin{table}
\centering
\begin{tabular}{|c|c|c|c|c|c|}

\hline
z &  $n_{{}_{FPD}}$ & $n_{cc}$ & $n_{cb}$ & $n_{an}$ & {\rm cocycles/coboundaries}\\
\hline
& & & & & \\[-10pt]
1 & 2 & 1 & 1 & 0 & $\wn a+ a^2$ \\
2 & 3 & 1 & 1 & 0  & $a(2\wn a+ a^2)$ \\
3 & 5& 2 & 2 & 0 & $a^2(3\wn a+ a^2),\, 4a^2\wn a+4(\wn a)^2+7 a\wn^2 a +3 \wn^3 a$ \\
4 & 7 & 2 & 2 & 0 & $a^3(4\wn a+ a^2),\, 5a^3\wn a+9 a^2 \wn^2 a + 12a(\wn a)^2 +4a\wn^3 a+8\wn a$ \\
\hline
5  & 11 & 4  & 4  & 0  &\\
6  & 15 & 4  & 4  & 0  &\\
7  & 22 & 7  & 7  & 0  &\\
8  & 30 & 8  & 8  & 0  & $ a^{z-1}(a^2+z\wn a)$, $\dots$\\
9  & 42 & 12 & 12 & 0  & \\
10 & 56 & 14 & 14 & 0  &  \\
11 & 77 & 21 & 21 & 0  &  \\
12 & 101 & 24 & 24 & 0 & \\
\hline
\end{tabular}
\caption{Summary of results for the purely spatial sector in 1+1 dimensions. $z$ is the Lifshitz dynamical exponent. $n_{{}_{FPD}}$, $n_{cc}$, $n_{cb}$ and $n_{an}$ are the number of independent FPD invariant expressions, cocycles, coboundaries and anomalies for each value of $z$, respectively (see subsection \protect\ref{ssb:prescription} for more details).}
\label{table:1+1SpatResultsSum}
\end{table}

We have not identified an obvious general structure for all the cocycles and coboundaries in this sector for general values of $z$. However, it is interesting to note that $E_1 = \int\sqrt{-g}\,\sigma (\wn+a) a^z = \int \sqrt{-g} \,\sigma a^{z-1}(a^2+z\wn a)$ is both a cocycle and a coboundary term for any $z$. 
The Weyl variation of $E_1$ is given by: 
\begin{equation}
\delta'\left(a^{z-1}\left[a^2+z\wn a\right]\right) = z^2 \left[ a^z \wn \sigma + (z-1) a^{z-2} \wn a \wn \sigma +a^{z-1}\wn^2 \sigma \right],
\end{equation}
\begin{equation}
\delta_\sigma^W \left( \int \sqrt{-g}\,  \sigma a^{z-1}\left[a^2+z\wn a\right] \right) = 0,
\end{equation}
where we have used integration by parts in the last equality. We, therefore, find that $E_1$ is a cocycle in the relative cohomology. To show that it is also trivial (a coboundary term) use:
\begin{equation}
G_1 = \int \sqrt{-g} \,  a^{z+1},
\end{equation}
whose variation is given by:
\begin{equation}
\delta_\sigma^W G_1 = \int \sqrt{-g} \, z(z+1) a^{z} \wn \sigma = -z(z+1)\int \sqrt{-g} \, \sigma a^{z-1}(a^2+z\wn a)
=  -z(z+1)\, E_1.
\end{equation}

\section{Lifshitz Scale Anomalies in 2+1 Dimensions}\label{sec:2+1}

In this section we detail our results for Lifshitz scale anomalies in 2+1 dimensions for various values of $z$.
We begin with detailed results for the \emph{universal} sectors (for which we conclude there are no anomalies) and the important cases of $z=1$ (which can be compared to the conformal case) and $z=2$ (which was previously studied in \cite{Baggio:2011ha}, \cite{Griffin:2011xs}). We then present results for several other values of $z$: $ z = 2/3 $ and $z = 4 $.\footnote{The cases of $z=3$ and $z=3/2$ contain only the universal sectors.} A summary of the results can be found in table \ref{table:2+1ResultsSum}.

\subsection{The Universal Sectors}\label{sb:2+1UniversalSectors}

The universal sectors are the ones with values of $ n_T $, $ n_S $, $ n_\epsilon $ that satisfy equations \eqref{rest:constraint1b}, \eqref{rest:constraint2b} for any value of $z$. In 2+1 dimensions, these are the two following sectors:
\begin{itemize}[label= --]
\item $ n_T=1 $, $ n_S=2 $, $ n_\epsilon = 0 $,
\item $ n_T=1 $, $ n_S=2 $, $ n_\epsilon = 1 $.
\end{itemize}

\subsubsection{Parity Even Sector with Two Space and One Time Derivatives}

In $ d=2 $ spatial dimensions, the intrinsic curvature contains only one independent component, which we take to be the intrinsic Ricci scalar $ \widetilde{R} $.
Taking into account all the identities in subsection \ref{ssb:identsfortangent}, the $ n_T=1 $, $ n_S=2 $, $ n_\epsilon = 0 $ sector  contains $ n_{{}_{FPD}} = 11 $ independent FPD invariant expressions:
\begin{equation}
\begin{alignedat}{4}
\phi_1 &= K a^2,
&\qquad \phi_2 &= K_{\alpha\beta} a^\alpha a^\beta, 
&\qquad \phi_3 &= K \widetilde{R} ,
&\qquad \phi_4 &= K \wn_\alpha a^\alpha ,
 \\ 
\phi_5 &= K^{\alpha\beta} \wn_\alpha a_\beta ,
&\qquad \phi_6 &= a^\alpha \wn_\alpha K ,
&\qquad \phi_7 &= a_\alpha \wn_\beta K^{\alpha\beta} ,
&\qquad \phi_8 &= \widetilde\Box K ,\\
 \phi_9 &= \wn_\alpha \wn_\beta K^{\alpha\beta} ,
&\qquad \phi_{10} &= a^\alpha \Lie{n} a_\alpha,
&\qquad \phi_{11} &= \Lie{n} \wn_\alpha a^\alpha.
\end{alignedat}
\end{equation}
The integrated expressions of ghost number one are given by $ I_i = \int \sqrt{-g} \sigma \phi_i $.
After applying the Weyl operator to $ I_i $ and performing integration by parts, we obtain linear combinations of the following 9 independent expressions:
\begin{align}
%\begin{split}
L_1 &= \int \sqrt{-g}\, \sigma \wn_\alpha \sigma K a^\alpha ,
&L_2 &= \int \sqrt{-g}\, \sigma \wn_\beta \sigma a_\alpha K^{\alpha\beta}, 
&L_3 &= \int \sqrt{-g}\, \sigma \wn_\alpha \sigma \wn^\alpha K ,
\notag
\\
L_4 &= \int \sqrt{-g}\, \sigma \wn_\beta \sigma \wn_\alpha K^{\alpha\beta}, 
&L_5 &= \int \sqrt{-g}\, \sigma \wn^\alpha \sigma \Lie{n} a_\alpha ,
&L_6 &= \int \sqrt{-g}\, \sigma \Lie{n} \sigma a^2, \\
\notag
L_7 &= \int \sqrt{-g}\, \sigma \Lie{n} \sigma \wn_\alpha a^\alpha,
&L_8 &= \int \sqrt{-g}\, \sigma \Lie{n} \sigma \widetilde{R} ,
&L_9 &= \int \sqrt{-g}\, \sigma \Lie{n} \widetilde\Box \sigma .
%\end{split}
\end{align}
The matrix $ M_{ij} $, as defined in subsection \ref{ssb:prescription}, is given by:
\settowidth{\mycolwd}{$\,-z-2\,$} %Largest element in matrix
\begin{equation}
\left(
\begin{array}{*{9}{@{}I{\mycolwd}@{}}}
2z & 0 & 0 & 0 & 0 & 2 & 0 & 0 & 0 \\
0 & 2z & 0 & 0 & 0 & 1 & 0 & 0 & 0  \\
2 & 0 & 2 & 0 & 0 & 0 & 0 & 2 & 0 \\
-z & 0 & -z & 0 & 0 & 0 & 2 & 0 & 0 \\
1 & -z-2 & 0 & -z & 0 & 0 & 1 & 0 & 0 \\
-z-1 & 2 & z & 0 & -1 & -2 & -1 & 0 & 0 \\
-\frac{3}{2} & -z+3 & 0 & z & -\frac{1}{2} & -1 & -\frac{1}{2} & 0 & 0 \\
z & -4 & -z-2 & 0 & 2 & 2 & 0 & 0 & 2 \\
1 & -4+z & -1 & -z & 1 & 1 & 0 & 0 & 1 \\
-\frac{1}{2}z & z & 0 & 0 & \frac{1}{2}z & 0 & -\frac{1}{2}z & 0 & 0 \\
0 & 0 & 0 & 0 & 0 & 0 & -2 & 0 & z
\end{array}
\right)
\end{equation}

The solution space to the cocycle equations \eqref{presc:cocycleeqs} is of dimension $ n_{cc} = 4 $, and we choose the following basis for it:
\begin{align}
\begin{split}
E_1 &= -\frac{1}{2}z I_1 + (1-\frac{1}{2}z) I_4 -z I_6 -\frac{1}{2}z I_8 + I_{11}, \\
E_2 &= (\frac{1}{2}+\frac{1}{2}z) I_1 - I_2 + \frac{1}{2}z I_4 + \frac{1}{2}z I_6 + I_{10}, \\
E_3 &= \frac{1}{2} I_1 - I_2 +\frac{1}{2} I_4 - I_5 - \frac{1}{2} I_8 + I_9 ,\\
E_4 &= -\frac{1}{2} I_1 + I_2 - \frac{1}{2} I_4 + I_5 - \frac{1}{2} I_6 + I_7. 
\end{split}
\end{align}

The integrated expressions of ghost number zero are given by $ G_i = \int \sqrt{-g} \phi_i $. Applying the Weyl operator to these expressions and integrating by parts we obtain the coboundaries, written as linear combinations of the expressions $ I_i $. The span of these expressions is the coboundary space. It can be shown to be of dimension $ n_{cb} = 4 $ as well, and we choose the following basis for it:
\begin{align}
\begin{split}
F_1 &= \delta^W_\sigma G_1 = (-2-2z) I_1 +4 I_2 -2z I_4 -2z I_6 -4 I_{10}, \\
F_2 &= \delta^W_\sigma G_2 = - I_1 + (2-2z) I_2 - 2z I_5 - 2z I_7 - 2 I_{10}, \\
F_3 &= \delta^W_\sigma G_3 = 2 I_1 -4 I_2 +2 I_4 -4 I_5 +4 I_6 -8 I_7 +2 I_8 -4 I_9, \\
F_4 &= \delta^W_\sigma G_4 = z I_1 +(z-2) I_4 +2z I_6 + z I_8 -2 I_{11}.
\end{split}
\end{align}

It can be directly checked that all of the cocycles found are indeed linear combinations of the coboundaries:
\begin{align}
\begin{split}
E_1 &= -\frac{1}{2} F_4, \\
E_2 &= -\frac{1}{4} F_1, \\
E_3 &= -\frac{1}{4} F_3 + \frac{1}{z} F_2 - \frac{1}{2z} F_1, \\
E_4 &= -\frac{1}{2z} F_2 + \frac{1}{4z} F_1.
\end{split}
\end{align}
We conclude that there are no anomalies in this sector.

Note, that the space of trivial descent cocycles (Weyl-invariant scalar densities) here is of dimension $ n_{td} = 2$, with the basis:
\begin{align}
\begin{split}
H_1 &= E_1 + 2 E_4, \\
H_2 &= z E_3 + (2z-2) E_4.
\end{split}
\end{align}

\subsubsection{Parity Odd Sector with Two Space and One Time Derivatives}

In the $ n_T = 1 $, $ n_S = 2 $, $ n_\epsilon = 1 $ sector in $d=2$ spatial dimensions, the following dimensionally dependent identity has to be taken into account (as a result of the anti-symmetrisation of 3 indices): 
\begin{equation}
\tilde{\epsilon}^{\alpha\gamma}K_\gamma^\beta - \tilde{\epsilon}^{\beta\gamma}K_\gamma^\alpha = \tilde{\epsilon}^{\alpha\beta} K.
\end{equation}
With this in mind, the sector contains $ n_{{}_{FPD}} = 6 $ independent FPD invariant expressions, which we choose to be:
\begin{equation}
\begin{alignedat}{3}
\phi_1 &= K_S^{\alpha\beta} a_\alpha a_\beta ,
&\qquad \phi_2 &= \wn_\alpha \wn_\beta K_S^{\alpha\beta},
&\qquad \phi_3 &= \tilde{\epsilon}^{\alpha\beta} a_\alpha \Lie{n} a_\beta ,\\
\phi_4 &= K_S^{\alpha\beta} \wn_\alpha a_\beta,
&\qquad \phi_5 &= \wn_\alpha K_S^{\alpha\beta} a_\beta,
&\qquad \phi_6 &= \tilde{\epsilon}^{\alpha\beta} \wn_\alpha K a_\beta ,
\end{alignedat}
\end{equation}
where
\begin{equation}
K_S^{\alpha\beta} \equiv \frac{1}{2} ( \tilde{\epsilon}^{\alpha\gamma} K_\gamma^\beta + \tilde{\epsilon}^{\beta\gamma} K_\gamma^\alpha ) .
\end{equation}
The independent integrated ghost number one expressions are given by $ I_i = \int \sqrt{-g} \sigma \phi_i $, whereas the independent integrated ghost number two expressions are:
\begin{align}
L_1 &= \int \sqrt{-g}\, \sigma \wn_\mu \sigma K_S^{\mu\alpha} a_\alpha , \qquad
& L_2 &= \int \sqrt{-g}\, \sigma \wn_\mu \sigma \wn_\alpha K_S^{\mu\alpha} , \notag \\
L_3 &= \int \sqrt{-g}\, \sigma \wn_\mu \sigma \tilde{\epsilon}^{\mu\alpha} K a_\alpha, \qquad
&L_4 &= \int \sqrt{-g}\, \sigma \wn_\mu \sigma \tilde{\epsilon}^{\mu\alpha} \wn_\alpha K,
 \\ \notag
L_5 &= \int \sqrt{-g}\, \sigma \wn_\mu \sigma \tilde{\epsilon}^{\mu\alpha} \Lie{n} a_\alpha .
\end{align}
The matrix $ M_{ij} $ that corresponds to the cocycle equations turns out to be:
\settowidth{\mycolwd}{$\,-z-2\,$} %Largest element in matrix
\begin{equation}
\left(
\begin{array}{*{5}{@{}I{\mycolwd}@{}}}
2z & 0 & 0 & 0 & 0 \\
z-2 & -z & 0 & 0 & 0 \\
0 & 0 & 0 & 0 & \frac{3}{2}z \\
-z-2 & -z & 0 & 0 & 0 \\
2-z & z & 0 & 0 & 0 \\
0 & 0 & -z & -z & -1
\end{array}
\right)
\end{equation}

The cocycle space is of dimension $ n_{cc} = 2 $, and we choose the basis:
\begin{align}
\begin{split}
E_1 &= I_2 + I_5, \\
E_2 &= I_1 + I_4 + I_5.
\end{split}
\end{align}

The integrated expressions of ghost number zero are again $ G_i = \int\sqrt{-g}\phi_i $. The coboundary space is of dimension $ n_{cb} = 2 $, with the basis:
\begin{align}
\begin{split}
F_1 &= \delta^W_\sigma G_1 = -2z I_1 -2z I_4 -2z I_5 ,\\
F_2 &= \delta^W_\sigma G_5 = (z-2) I_1 -z I_2 +(z-2) I_4 -2 I_5 .
\end{split}
\end{align}
We can check that all the cocycles are coboundaries:
\begin{align}
\begin{split}
E_1 &= -\frac{1}{z} F_2 + \frac{1}{2z}\left(\frac{2}{z}-1 \right) F_1 , \\
E_2 &= -\frac{1}{2z} F_1.
\end{split}
\end{align}
Thus there are no anomalies in this sector.

The space of trivial descent cocycles in this sector is of dimension $ n_{td}=1 $, and consists of expressions proportional to:
\begin{equation}
H_1 = z E_1 + (z-2) E_2.
\end{equation}

\subsection{The \texorpdfstring{$z=1$}{z=1} Case}

In addition to the two universal sectors detailed in subsection \ref{sb:2+1UniversalSectors}, the $d=2$, $z=1$ case contains two additional sectors with:
\begin{itemize}[label= --]
\item $ n_T = 3$, $n_S = 0$, $n_\epsilon = 0$,
\item $ n_T = 3$, $n_S = 0$, $n_\epsilon = 1$.
\end{itemize}

\subsubsection{Parity Even Sector with Three Time Derivatives}

In this sector for $d=2$, the following identity is always satisfied:
\begin{equation}
\frac{1}{2} K^3 + \tr (K^3)  - \frac{3}{2} K\, \tr(K^2) = 0,
\end{equation}
where we define $\tr(K^2) \equiv K^{\alpha\beta}K_{\alpha\beta} $, $\tr(K^3) \equiv K_\alpha^\beta K_\beta^\gamma K_\gamma^\alpha $.

Taking this identity into account, we are left with $ n_{{}_{FPD}} = 5 $ independent, FPD invariant expressions:
\begin{equation}
\begin{alignedat}{3}
\phi_1 &= \tr(K^3), 
&\qquad \phi_2 &= K\,\tr(K^2),
&\qquad \phi_3 &= K^{\alpha\beta} \Lie{n} K_{\alpha\beta}, \\ 
\phi_4 &= K \Lie{n} K,
&\qquad  \phi_5 &= \Lie{n}^2 K.
\end{alignedat}
\end{equation}

We define the independent integrated ghost number one expressions $ I_i = \int\sqrt{-g}\sigma\phi_i $, and the independent integrated ghost number two  expressions:
\begin{align}
\begin{split}
L_1 &= \int\sqrt{-g}\,\sigma \Lie{n} \sigma\, \tr(K^2), \qquad
L_2 = \int\sqrt{-g}\,\sigma \Lie{n} \sigma K^2 , \\
L_3 &= \int\sqrt{-g}\,\sigma \Lie{n} \sigma \Lie{n} K ,\qquad
\ \ \ \, L_4 = \int\sqrt{-g}\,\sigma \Lie{n}^3 \sigma .
\end{split}
\end{align}
The matrix $M_{ij}$ is given by:
\begin{equation}
\begin{pmatrix}
3 & 0 & 0 & 0 \\
2 & 2 & 0 & 0 \\
5 & -1 & 0 & 0 \\
0 & -3 & 0 & 0 \\
0 & 1 & -2 & 2
\end{pmatrix}
\end{equation}
Solving the cocycle equations, we obtain a cocycle space of dimension $ n_{cc} = 2 $, with the basis:
\begin{align}
\begin{split}
E_1 &= -2 I_1 + \frac{1}{2} I_2 + I_3, \\
E_2 &= -I_1 + \frac{3}{2} I_2 + I_4 .
\end{split}
\end{align}

We define the integrated expressions of ghost number zero $ G_i = \int\sqrt{-g}\phi_i $, and obtain a coboundary space of dimension $ n_{cb} = 2 $, with the basis:
\begin{align}
\begin{split}
F_1 &= \delta^W_\sigma G_1 = 12 I_1 - 3 I_2 - 6 I_3, \\
F_2 &= \delta^W_\sigma G_2 = 12 I_1 - 8 I_2 - 4 I_3 - 4 I_4 .
\end{split}
\end{align}
All of the cocycles are indeed coboundaries:
\begin{align}
\begin{split}
E_1 &= -\frac{1}{6} F_1, \\
E_2 &= -\frac{1}{4} F_2 + \frac{1}{6} F_1,
\end{split}
\end{align}
and therefore there are no anomalies in this sector.
The space of trivial descent cocycles in this sector is of dimension $ n_{td} = 1 $, and consists of expressions proportional to:
\begin{equation}
H_1 = -2 E_1 + E_2 .
\end{equation}

\subsubsection{Parity Odd Sector with Three Time Derivatives}

Interestingly, due to the symmetry of the extrinsic curvature, there is only one possible FPD invariant expression in this sector (that isn't identically zero):
\begin{equation}
\phi_1 = \tilde{\epsilon}^{\alpha\beta} K_\alpha^\gamma \Lie{n} K_{\gamma\beta} .
\end{equation}
The only integrated ghost number one expression is then $ I_1 = \int \sqrt{-g} \sigma \phi_1 $, and there are no ghost number two expressions. The cocycle space is thus one dimensional ($ n_{cc} = 1 $), with $ I_1 $ the only independent (and trivial descent) cocycle. Since there are no coboundaries, it's an anomaly as well:
\begin{equation}\label{2d:z1_anomaly_3_0_1}
\boxed{
A_1^{(3,0,1)} = I_1 = \int \sqrt{-g}\, \sigma \tilde{\epsilon}^{\alpha\beta} K_\alpha^\gamma \Lie{n} K_{\gamma\beta} .
}
\end{equation}
We conclude that there is only one anomaly in 2+1 dimensions for $z=1$ which is given in equation \eqref{2d:z1_anomaly_3_0_1}.

\subsection{The \texorpdfstring{$z=2$}{z=2} Case}

The $ d=2 $, $ z=2 $ case is of particular importance, as certain condensed matter systems have been shown to exhibit a Lifshitz scaling symmetry with this value of the dynamical critical exponent \cite{Ardonne:2003wa}. The parity even sector for this case has been extensively studied in the literature (see e.g.~\cite{Baggio:2011ha}, \cite{Griffin:2011xs}). Here we repeat the cohomological analysis of this case in our terms for comparison. We also show that in the parity odd sectors there are no possible anomalies.

In addition to the universal sectors detailed in subsection \ref{sb:2+1UniversalSectors}, this case contains 4 additional sectors:
\begin{itemize}[label=--]
\item $ n_T = 2 $, $ n_S = 0 $, $ n_\epsilon = 0 $,
\item $ n_T = 2 $, $ n_S = 0 $, $ n_\epsilon = 1 $,
\item $ n_T = 0 $, $ n_S = 4 $, $ n_\epsilon = 0 $,
\item $ n_T = 0 $, $ n_S = 4 $, $ n_\epsilon = 1 $.
\end{itemize}
However, the second sector above ($ n_T = 2 $, $ n_S = 0 $, $ n_\epsilon = 1 $) is clearly empty and contains no FPD invariant expressions at all, due to the symmetry of the extrinsic curvature. We are, therefore, left with only 3 sectors. 

\subsubsection{Parity Even Sector with Two Time Derivatives}

This sector contains the following FPD invariant expressions ($ n_{{}_{FPD}} = 3 $):
\begin{align}
\begin{split}
\phi_1 = \tr(K^2), \qquad
\phi_2 = K^2, \qquad
\phi_3 = \Lie{n} K ,
\end{split}
\end{align}
where $ \tr(K^2) \equiv K^{\alpha\beta} K_{\alpha\beta} $.
The integrated ghost number one expressions are $ I_i = \int \sqrt{-g} \sigma \phi_i $.
The only independent ghost number two expression is
\begin{equation}
L_1 = \int \sqrt{-g}\, \sigma \Lie{n}\sigma K .
\end{equation}
The matrix $ M_{ij} $ is then given by:
\begin{equation}
\begin{pmatrix}
2 \\
4 \\
-4
\end{pmatrix}
\end{equation}

The cocycle space is 2 dimensional ($ n_{cc} = 2 $), with the basis:
\begin{align}
\begin{split}
E_1 &= I_1 - \frac{1}{2} I_2, \\
E_2 &= I_2 + I_3 .
\end{split}
\end{align}
Defining the integrated ghost number zero expressions $ G_i = \int \sqrt{-g} \phi_i $, the coboundary space is of dimension $ n_{cb} = 1 $, with the basis:
\begin{equation}
F_1 = \delta_\sigma^W G_1 = -2 I_2 - 2 I_3 .
\end{equation}
The second cocycle is a coboundary:
\begin{equation}
E_2 = - \frac{1}{2} F_1,
\end{equation}
and we are left with 1 anomaly ($ n_{an} = 1 $) in this sector, given by:
\begin{equation}\label{2d:z2_anomaly_2_0_0}
\boxed{
A_1^{(2,0,0)} = E_1 = \int \sqrt{-g}\,\sigma \left[ \tr(K^2) - \frac{1}{2} K^2 \right].
}
\end{equation} 
Note, that the trivial descent cocycle space here is also of dimension $ n_{td} = 1 $, and spanned by the anomaly $ A_1^{(2,0,0)} $.
This is the same result previously obtained for this sector, e.g.\ in \cite{Baggio:2011ha}.

\subsubsection{Parity Even Sector with Four Spatial Derivatives}

There are 12 independent, FPD invariant expressions in this sector ($ n_{{}_{FPD}} = 12 $):
\begin{equation}
\begin{alignedat}{4}
\phi_1 &= \widetilde{R}^2 ,
&\qquad \phi_2 &= \widetilde{R} a^2 ,
&\qquad \phi_3 &= a^\alpha \wn_\alpha \widetilde{R} ,
&\qquad \phi_4 &= \widetilde{R} \wn_\alpha a^\alpha ,
\\
\phi_5 &= \widetilde{\Box} \widetilde{R} ,
&\qquad\phi_6 &= a^4 ,
&\qquad\phi_7 &= a^\alpha \wn_\alpha (a^2) ,
&\qquad\phi_8 &= a^2 \wn_\alpha a^\alpha ,
\\
\phi_9 &= (\wn_\alpha a^\alpha)^2 ,
&\qquad\phi_{10} &= \wn_\alpha a_\beta \wn^\alpha a^\beta ,
&\qquad\phi_{11} &= a^\alpha \wn_\alpha \wn_\beta a^\beta ,
&\qquad\phi_{12} &= \widetilde{\Box} \wn_\alpha a^\alpha .
\end{alignedat}
\end{equation}
The integrated ghost number one expressions are then $ I_i = \int \sqrt{-g} \sigma \phi_i $.
We choose the independent, integrated ghost number two expressions to be:
\begin{equation}
\begin{alignedat}{2}
L_1 &= \int\sqrt{-g}\, \sigma \wn_\alpha\sigma \widetilde{R}a^\alpha ,
&\qquad L_2 &= \int\sqrt{-g}\, \sigma \wn_\alpha\sigma \wn_\alpha \widetilde{R},\\ 
L_3 &= \int\sqrt{-g}\, \sigma \wn_\alpha\sigma a^\alpha \wn_\beta a^\beta ,
&\qquad L_4 &= \int\sqrt{-g}\, \sigma \wn_\alpha\sigma a^\alpha a^2, \\
 L_5 &= \int\sqrt{-g}\, \sigma \wn_\alpha\sigma \wn^\alpha (a^2) ,
&\qquad L_6 &= \int\sqrt{-g}\, \sigma \wn^\alpha\sigma \wn_\alpha \wn_\beta a^\beta ,\\
L_7 &= \int\sqrt{-g}\, \sigma \widetilde{\Box}\wn_\alpha\sigma a^\alpha .
\end{alignedat}
\end{equation}
The matrix $ M_{ij} $ is given by:
\begin{equation}
\begin{pmatrix}
4 & 4 & 0 & 0 & 0 & 0 & 0 \\
4 & 0 & 0 & 2 & 2 & 0 & 0 \\
-1 & 2 & 0 & 0 & 0 & 0 & -2 \\
-2 & -2 & 2 & 0 & 0 & 2 & 0 \\
0 & -2 & -2 & -2 & -2 & -2 & 4 \\
0 & 0 & 0 & 8 & 0 & 0 & 0 \\
0 & 0 & -4 & -6 & 2 & 0 & 0 \\
0 & 0 & 4 & -2 & -2 & 0 & 0 \\
0 & 0 & -4 & 0 & 0 & -4 & 0 \\
-2 & 0 & 2 & 0 & -4 & -4 & 0 \\
-1 & 0 & -2 & 0 & 0 & 2 & 2 \\
2 & 0 & 4 & 2 & 2 & 0 & -4
\end{pmatrix}
\end{equation}
The cocycle space has dimension $ n_{cc} = 6 $, and we choose the basis:
\begin{align}
\begin{split}
E_1 &= I_2 + 2 I_3 + I_4 + I_5 ,\\
E_2 &= I_6 + I_7 + I_8 ,\\
E_3 &= I_1 + 2 I_4 + I_9, \\
E_4 &= I_1 + \frac{1}{2} I_2 + 2 I_4 + I_6 + \frac{3}{2} I_7 + I_{10} ,\\
E_5 &= - I_1 + I_2 + I_3 - I_4 - I_6 - I_7 + I_{11} ,\\
E_6 &= I_1 - 2 I_2 - 2 I_3 + I_6 + I_7 + I_{12} .
\end{split}
\end{align}
The coboundary space in this sector has dimension $ n_{cb} = 5 $, and we choose the basis:
\begin{align}
\begin{split}
F_1 &= \delta^W_\sigma G_1 = -4 I_2 - 8 I_3 - 4 I_4 - 4 I_5, \\
F_2 &= \delta^W_\sigma G_2 = -6 I_2 -4 I_3 -4 I_4 -2 I_6 -4 I_7 -2 I_8 -4 I_{10} -4 I_{11} ,\\
F_3 &= \delta^W_\sigma G_3 = 4 I_2 +2 I_4 -2 I_5 +2 I_6 +4 I_7 +4 I_8 +2 I_9 +4 I_{10} +8 I_{11} +2 I_{12}, \\
F_4 &= \delta^W_\sigma G_6 = -8 I_6 -8 I_7 -8 I_8, \\
F_5 &= \delta^W_\sigma G_7 = -2 I_2 +6 I_6 +4 I_7 +10 I_8 +4 I_9 -4 I_{10} ,
\end{split}
\end{align}
where $ G_i = \int \sqrt{-g} \phi_i $.
We conclude that there is $ n_{an} = 1 $ anomaly in this sector, which is given by (up to coboundary terms):
\begin{equation}\label{2d:z2_anomaly_0_4_0}
\boxed{
A_1^{(0,4,0)} = E_3 = \int \sqrt{-g}\, \sigma \left( \widetilde{R} + \wn_\alpha a^\alpha \right)^2.
}
\end{equation}
We can check that all other cocycles are indeed linear combinations of the anomaly term $ A_1^{(0,4,0)} $ (which we abbreviate here as $A_1$) and the coboundaries $ F_i $:
\begin{align}
\begin{split}
E_1 &= -\frac{1}{4} F_1, \\
E_2 &= -\frac{1}{8} F_4, \\
E_4 &= A_1 - \frac{1}{4} F_5 - \frac{5}{16} F_4 ,\\
E_5 &= -A_1 + \frac{1}{4} F_5 + \frac{3}{8} F_4 - \frac{1}{4} F_2 ,\\
E_6 &= A_1 -\frac{1}{2} F_5 -\frac{5}{8} F_4 +\frac{1}{2} F_3 + F_2 -\frac{1}{4} F_1 .
\end{split}
\end{align}
This is the same result previously obtained for this sector in e.g.~\cite{Baggio:2011ha}.

We note that the space of trivial descent cocycles here has dimension $ n_{td} = 2 $, with the basis:
\begin{align}
\begin{split}
H_1 &= E_3 = A_1, \\
H_2 &= E_1 + E_2 +2 E_5 + E_6 .
\end{split}
\end{align}
The anomaly in this sector is thus also a trivial descent (up to coboundary terms), and there is a one dimensional space of trivial descent coboundaries, proportional to $ H_1 + H_2 $.

\subsubsection{Parity Odd Sector with Four Spatial Derivatives}

In this sector, we note the following identity (that comes into play when considering the independent ghost number two expressions):
\begin{equation}
\tilde{\epsilon}^{\alpha\gamma}\wn_\gamma a^\beta - \tilde{\epsilon}^{\beta\gamma}\wn_\gamma a^\alpha = \tilde{\epsilon}^{\alpha\beta} \wn_\gamma a^\gamma .
\end{equation}
There are 3 independent FPD invariant expressions ($ n_{{}_{FPD}} = 3 $):
\begin{align}
\begin{split}
\phi_1 = \tilde{\epsilon}^{\alpha\beta} a_\alpha a^\gamma \wn_\beta a_\gamma ,\qquad
\phi_2 = \tilde{\epsilon}^{\alpha\beta} a_\alpha \wn_\beta \widetilde{R} ,\qquad
\phi_3 = \tilde{\epsilon}^{\alpha\beta} a_\alpha \wn_\beta \wn_\gamma a^\gamma .
\end{split}
\end{align}
The integrated expressions of ghost number one are again defined as $ I_i = \int \sqrt{-g} \sigma \phi_i $. Taking into account the previously mentioned identity, the independent integrated ghost number two expressions are:
\begin{equation}
\begin{alignedat}{2}
L_1 &= \int\sqrt{-g}\, \sigma\wn_\alpha\sigma\, \tilde{\epsilon}^{\alpha\beta} a_\beta a^2 ,
&\qquad L_2 &= \int\sqrt{-g}\, \sigma\wn_\alpha\sigma\, \tilde{\epsilon}^{\alpha\beta} a_\beta \wn_\gamma a^\gamma ,\\
L_3 &= \int\sqrt{-g}\, \sigma\wn_\alpha\sigma\, \tilde{\epsilon}^{\alpha\gamma} a_\beta \wn_\gamma a^\beta ,
&\qquad L_4 &= \int\sqrt{-g}\, \sigma\wn_\alpha\sigma\, \tilde{\epsilon}^{\alpha\beta} \wn_\beta \wn_\gamma a^\gamma ,\\
L_5 &= \int\sqrt{-g}\, \sigma\wn_\alpha\sigma\, \tilde{\epsilon}^{\alpha\beta} a_\beta \widetilde{R} ,
&\qquad L_6 &= \int\sqrt{-g}\, \sigma\wn_\alpha\sigma\, \tilde{\epsilon}^{\alpha\beta} \wn_\beta \widetilde{R} ,\\
L_7 &= \int\sqrt{-g}\, \sigma\wn_\alpha\widetilde\Box \sigma\, \tilde{\epsilon}^{\alpha\beta} a_\beta .
\end{alignedat}
\end{equation}
The matrix $M_{ij}$ corresponding to the cocycle equations is:
\begin{equation}
\begin{pmatrix}
2 & 0 & 4 & 0 & 0 & 0 & 0 \\
0 & 0 & 0 & 0 & 2 & 2 & 2 \\
0 & 2 & 0 & 2 & 0 & 0 & -2
\end{pmatrix}
\end{equation}
The only solution to the cocycle equations in this case is $0$. Therefore, there are no cocycles in this sector and the cohomology is empty.

\subsection{Several Other Values of \texorpdfstring{$z$}{z}}

In this subsection we present our results for the Lifshitz cohomology for several other values of the dynamical critical exponent $z$. These calculations were performed using a script that implements the prescription outlined in subsection \ref{ssb:prescription}.

We begin with $ z = 2/3 $. Field theories with $z<1$ may not be realized in low energy physical systems. Nevertheless, studying their corresponding cohomologies can
give valuable mathematical insight into the possible structure of Lifshitz cohomologies. 
Aside from the universal sectors studied in subsection \ref{sb:2+1UniversalSectors}, this case contains two other sectors:
\begin{itemize}[label=--]
\item $ n_T = 4 $, $ n_S = 0 $, $ n_\epsilon = 0 $: This sector contains $ n_{{}_{FPD}} = 11 $ independent FPD invariant expressions.  There are $ n_{cc} = 6 $ independent cocycles, $ n_{cb} = 4 $ coboundaries and $ n_{td} = 3 $ trivial descent cocycles. We are left with $ n_{an} = 2 $ independent anomalous terms, both of which are trivial descents (up to addition of coboundaries), given by:
\begin{align}
A_1^{(4,0,0)} = \int \sqrt{-g}\, \sigma & \left[ \frac{2}{3} \tr(K^4) - \frac{4}{3} K \tr(K^3) + \tr(K^2)\tr(K^2) \right], 
\label{2+1Other:z2/3AnomStart}
\\
A_2^{(4,0,0)} = \int \sqrt{-g}\, \sigma & \left[ -\frac{152}{27} \tr(K^4) + \frac{4}{27} K \tr(K^3) - 2 \Lie{n} K_{\alpha\beta} \Lie{n} K^{\alpha\beta} \right.\notag
\\
& \ \ \left. - \frac{4}{3} K_\alpha^\gamma K^{\alpha\beta} \Lie{n} K_{\beta\gamma} + (\Lie{n} K)^2 + \frac{2}{3} \tr(K^2) \Lie{n} K \right] ,
\label{2+1Other:z2/3AnomEnd}
\end{align}
where $ \tr(K^2) \equiv K^{\alpha\beta} K_{\alpha\beta} $, $ \tr(K^3) \equiv K^\alpha_\beta K^\beta_\gamma K^\gamma_\alpha $ and $ \tr(K^4) \equiv K^\alpha_\beta K^\beta_\gamma K^\gamma_\delta K^\delta_\alpha $.

\item $ n_T = 4 $, $ n_S = 0 $, $ n_\epsilon = 1 $: This sector contains $ n_{{}_{FPD}} = 2 $ independent FPD invariant expressions. There is only $ n_{cc} = 1 $ independent cocycle, which is also a trivial descent and a coboundary. Therefore, no anomalies appear in this sector.

\end{itemize}

The case of $ z=4 $ contains two sectors other than the universal ones:
\begin{itemize}[label=--]
\item $ n_T = 0 $, $ n_S = 6 $, $ n_\epsilon = 0 $: There are $ n_{{}_{FPD}} = 44 $ independent FPD invariant expressions in this sector. The cocycle space turns out to have dimension $ n_{cc} = 18 $, with $ n_{cb} = 16 $ independent coboundaries, and $ n_{td} = 4 $ trivial descent cocycles. We find $ n_{an} = 2 $ anomalies, both of which are trivial descents:
\begin{align}
A_1^{(0,6,0)} = \int\sqrt{-g}\,\sigma & \left[ 2\widetilde{R} + \wn_\alpha a^\alpha \right]^3,
\label{2+1Other:z4AnomStart}
\\
A_2^{(0,6,0)} = \int\sqrt{-g}\,\sigma & \left[ a^2 \widetilde{R}^2 + 4 a^\alpha \widetilde{R} \wn_\alpha \widetilde{R} + \frac{3}{2} a^\alpha \widetilde{R} (\wn_\alpha \wn_\beta a^\beta) + 4 \wn_\alpha \widetilde{R} \wn^\alpha \widetilde{R} \right.
\notag
\\ 
& \ \ + 4 (\wn_\alpha \wn_\beta a^\beta )(\wn^\alpha \widetilde{R}) + a^2 \widetilde{R} (\wn_\beta a^\beta)
\notag
\\
& \ \ + 2 (a^\alpha \wn_\alpha \widetilde{R})(\wn_\beta a^\beta) 
+ a^\alpha (\wn_\alpha \wn_\gamma a^\gamma) (\wn_\beta a^\beta)\notag
\\
& \ \ \left.+ \frac{1}{4} a^2 (\wn_\alpha a^\alpha)^2 
 + (\wn_\beta \wn_\alpha a^\alpha) (\wn_\gamma \wn^\beta a^\gamma)  \right].\label{2+1Other:z4AnomEnd}
\end{align}

\item $ n_T = 0 $, $ n_S = 6 $, $ n_\epsilon = 1 $: This sector contains $ n_{{}_{FPD}} = 20 $ independent FPD invariant expressions, $ n_{cc} = 3 $ independent cocycles, $ n_{cb} = 3 $ independent coboundaries and no trivial descent cocycles. We therefore find no anomalies in this sector.

\end{itemize}

We finally note that the $ z = 3/2 $ and $ z = 3 $ cases contain only the universal sectors, and thus no anomalies. All Our results in $2+1$ dimensions are summarized in table \ref{table:2+1ResultsSum}.

\begin{table}
\centering
\begin{tabular}{|c|c|c|c|c|c|c|c|c|c|c|}
\hline
$z$ & $ n_T $ & $ n_S $ & $ n_\epsilon $ & $ n_{{}_{FPD}} $ & $ n_{cc} $ & $ n_{cb} $ & $ n_{td} $ & $ n_{tdcb} $ & $ n_{an} $ & Anomalies \\
\hline 
Universal & 1 & 2 & 0 & 11 & 4 & 4 & 2 & 2 & 0 & --- \\
	      & 1 & 2 & 1 & 6 & 2 & 2 & 1 & 1 & 0 & --- \\
\hline
1         & 3 & 0 & 0 & 5 & 2 & 2 & 1 & 1 & 0 & --- \\	      
          & 3 & 0 & 1 & 1 & 1 & 0 & 1 & 0 & 1 & $\tilde{\epsilon}^{\alpha\beta} K_\alpha^\gamma \Lie{n} K_{\gamma\beta}$ \\
\hline
2         & 2 & 0 & 0 & 3 & 2 & 1 & 1 & 0 & 1 & $\tr(K^2) - \frac{1}{2} K^2$ \\
		  & 0 & 4 & 0 & 12 & 6 & 5 & 2 & 1 & 1 & $\left( \widetilde{R} + \wn_\alpha a^\alpha \right)^2$ \\
		  & 0 & 4 & 1 & 3 & 0 & 0 & 0 & 0 & 0 & --- \\
\hline
2/3		  & 4 & 0 & 0 & 11 & 6 & 4 & 3 & 1 & 2 & See \eqref{2+1Other:z2/3AnomStart}-\eqref{2+1Other:z2/3AnomEnd} \\
		  & 4 & 0 & 1 & 2 & 1 & 1 & 1 & 1 & 0 & --- \\
\hline
4		  & 0 & 6 & 0 & 44 & 18 & 16 & 4 & 2 & 2 & See \eqref{2+1Other:z4AnomStart}-\eqref{2+1Other:z4AnomEnd} \\
		  & 0 & 6 & 1 & 20 & 3 & 3 & 0 & 0 & 0 & --- \\
\hline
\end{tabular}
\caption{Summary of results for the Lifshitz cohomology in 2+1 dimensions. $z$ is the Lifshitz dynamical exponent. $n_T$, $n_S$ and $n_\epsilon$ are the number of time derivatives, space derivatives and Levi-Civita tensors in each sector, respectively (see subsection \protect\ref{ssb:restclasssectors} for more details). $n_{{}_{FPD}}$, $n_{cc}$, $n_{cb}$, $n_{td}$, $n_{tdcb}$ and $n_{an}$ are the number of independent FPD invariant expressions, cocycles, coboundaries, trivial descent cocycles, trivial descent coboundaries and anomalies in each sector, respectively (see subsection \protect\ref{ssb:prescription} for more details). Note that the $z=3/2$ and $z=3$ cases contain only the universal sector and thus no anomalies.}
\label{table:2+1ResultsSum}
\end{table}

\section{Lifshitz Scale Anomalies in 3+1 Dimensions}\label{sec:3+1AnomalyResults}
In this section we detail our results for the Lifshitz scale anomalies in 3+1 dimensions for several values of $z$ i.e.\ $z=1$, $z=2$, 
 $z=3$ and $z=3/2$. The calculation turned out to be quite involved and we used a script to perform it. In this section we describe our results with less detail than those of the previous sections. For each value of $z$ we list the various sectors. For each sector we list the number of independent FPD invariant terms, cocycles, coboundaries, anomalies and trivial descent cocycles. We then list the explicit expressions for the anomalies only. The results are summarized in table \ref{table:3+1ResultsSum}. We begin with the \emph{universal} sector which is common to all values of $z$. 

\subsection{Universal Sector}
This sector which is common to all values of $z$ has $n_T=1$ time derivative, $n_S=3$ space derivatives and $n_\epsilon=1$. This sector contains $ n_{{}_{FPD}} =7  $ independent FPD invariant expressions.  There are $ n_{cc} = 3 $ independent cocycles, $ n_{cb} = 2 $ coboundaries and $ n_{td} = 2 $ trivial descent cocycles. We are left with $ n_{an} = 1 $ independent anomalous term which is trivial descent (up to addition of coboundaries), and is given by:\footnote{In this section we specify anomaly densities $\mathcal{A}_i$ which are related to the anomalies by $A_i = \int \sqrt{-g} \sigma \mathcal{A}_i$.}
\begin{align}\label{3d:universal_1T3S}
\mathcal{A}_1^{(1,3,1)} = \ 
&
(1 -  z) a^{\alpha} \tilde{\epsilon}_{\alpha \gamma \delta} K^{\beta \gamma} \widetilde{R}_{\beta}{}^{\delta} + \frac{z-1}{z} a^{\alpha} \tilde{\epsilon}_{\alpha \gamma \delta} K_{\beta}{}^{\delta} \wn^{\gamma}a^{\beta}\notag\\
& + \frac{1}{z} a^{\alpha} a^{\beta} \tilde{\epsilon}_{\beta \gamma \delta} \wn^{\delta}K_{\alpha}{}^{\gamma} + z \tilde{\epsilon}_{\beta \gamma \delta} \widetilde{R}^{\alpha \beta} \wn^{\delta}K_{\alpha}{}^{\gamma} + \tilde{\epsilon}_{\beta \gamma \delta} \wn^{\beta}a^{\alpha} \wn^{\delta}K_{\alpha}{}^{\gamma}.
\end{align}

\subsection{The \texorpdfstring{$z=1$}{z=1} Case}
In this subsection we specify all the non-universal sectors in the calculation of the $z=1$ Lifshitz cohomology.

\subsubsection*{Parity Even Sector with Four Time Derivatives}
This sector has $n_T=4$ time derivatives, $n_S=0$ space derivatives and $n_\epsilon=0$. This sector contains $ n_{{}_{FPD}} =13  $ independent FPD invariant expressions.  There are $ n_{cc} = 7 $ independent cocycles, $ n_{cb} = 5 $ coboundaries and $ n_{td} = 4 $ trivial descent cocycles. We are left with $ n_{an} = 2 $ independent anomalous terms, both of which are trivial descents (up to addition of coboundaries), and read:
\begin{align}
\label{3d:z1_even4T_start}
\mathcal{A}_1^{(4,0,0)} = \ 
&
\tr(K^4) -  \frac{4}{3} K \tr(K^3) + \tr(K^2) \tr(K^2),
\\
\label{3d:z1_even4T_end}
\mathcal{A}_2^{(4,0,0)} = \ 
&
-7 \tr{K^4} + \frac{1}{3} K \tr(K^3) + \tr{K^2} \Lie{n} K\notag\\
& + (\Lie{n} K)^2 - 3 \Lie{n} K_{\alpha \beta} \Lie{n} K^{\alpha \beta} - 3 K_{\alpha}{}^{\gamma} K^{\alpha \beta} \Lie{n} K_{\beta \gamma}.
\end{align}

\subsubsection*{Parity Even Sector with Two Space and Two Time Derivatives}
This sector has $n_T=2$ time derivatives, $n_S=2$ space derivatives and $n_\epsilon=0$. This sector contains $ n_{{}_{FPD}} =47  $ independent FPD invariant expressions.  There are $ n_{cc} = 23 $ independent cocycles, $ n_{cb} = 17 $ coboundaries and $ n_{td} = 13 $ trivial descent cocycles. We are left with $ n_{an} = 6 $ independent anomalous terms, all of which are trivial descents (up to addition of coboundaries), and are given by:
\begin{align}\label{3d:z1_even2S2T_start}
\mathcal{A}_1^{(2,2,0)} = \ 
& \left(K^2-3\tr\left(K^2\right)\right) \left(\wn_{\alpha}a^{\alpha} - \frac{1}{2} a^2 + \frac{1}{4} \widetilde{R} \right),
\\
\mathcal{A}_2^{(2,2,0)} = \ 
&
- \frac{1}{2} a^2 K^2 + K a^{\alpha} a^{\beta} K_{\alpha \beta} -  \frac{3}{2} a^{\alpha} a^{\beta} K_{\alpha}{}^{\gamma} K_{\beta \gamma} + a^2 \tr(K^2) \notag\\
& -  \frac{1}{4} K^2 \widetilde{R} + \frac{1}{4} \tr(K^2) \widetilde{R} -  \frac{3}{2} K_{\alpha}{}^{\gamma} K^{\alpha \beta} \widetilde{R}_{\beta \gamma} + K K^{\beta \gamma} \widetilde{R}_{\beta \gamma}\notag\\
& -  \frac{1}{2} \tr(K^2) \wn_{\alpha}a^{\alpha} + K K_{\alpha \beta} \wn^{\beta}a^{\alpha} -  \frac{3}{2} K_{\alpha}{}^{\gamma} K_{\beta \gamma} \wn^{\beta}a^{\alpha},
\\
\mathcal{A}_3^{(2,2,0)} = \ 
&
- \frac{2}{3} a^2 K^2 + 4 K a^{\alpha} a^{\beta} K_{\alpha \beta} - 6 \
a^{\alpha} a^{\beta} K_{\alpha}{}^{\gamma} K_{\beta \gamma} + \
\frac{2}{3} K a^{\alpha} \wn_{\alpha}K\notag\\
& - 2 a^{\alpha} K_{\alpha}{}^{\beta} \wn_{\beta}K -  \frac{1}{6} \
\wn_{\beta}K \wn^{\beta}K - 2 K a^{\alpha} \wn_{\gamma}K_{\alpha}{}^{\
\gamma}\notag\\
& + 6 a^{\alpha} K_{\alpha}{}^{\beta} \
\wn_{\gamma}K_{\beta}{}^{\gamma} -  \frac{3}{2} \wn_{\alpha}K^{\alpha \
\beta} \wn_{\gamma}K_{\beta}{}^{\gamma} + \wn^{\beta}K \wn_{\gamma}K_{\beta}{}^{\gamma},
\\
\mathcal{A}_4^{(2,2,0)} = \ 
&
- \frac{1}{3} a^2 K^2 - 4 K a^{\alpha} a^{\beta} K_{\alpha \beta} + 6 a^{\alpha} a^{\beta} K_{\alpha}{}^{\gamma} K_{\beta \gamma} + 3 a^2 \tr{K^2} \notag\\
& -  \frac{2}{3} K a^{\alpha} \wn_{\alpha}K + 2 a^{\alpha} K^{\beta \gamma} \wn_{\alpha}K_{\beta \gamma} -  \frac{1}{3} \wn_{\beta}K \wn^{\beta}K \notag\\
& + 4 a^{\alpha} K^{\beta \gamma} \wn_{\gamma}K_{\alpha \beta} - 4 a^{\alpha} K_{\alpha}{}^{\beta} \wn_{\gamma}K_{\beta}{}^{\gamma} + \wn_{\gamma}K_{\alpha \beta} \wn^{\gamma}K^{\alpha \beta},
\\
\mathcal{A}_5^{(2,2,0)} = \ 
&
- \frac{4}{9} a^2 K^2 + 2 K a^{\alpha} a^{\beta} K_{\alpha \beta} - 6 a^{\alpha} a^{\beta} K_{\alpha}{}^{\gamma} K_{\beta \gamma} + \frac{2}{3} K a^{\alpha} \Lie{n} a_{\alpha}\notag\\
& - 2 a^{\alpha} K_{\alpha \beta} \Lie{n} a^{\beta} + \frac{1}{9} K a^{\alpha} \wn_{\alpha}K -  \frac{1}{3} \Lie{n} a^{\alpha} \wn_{\alpha}K \notag\\
& -  \frac{2}{3} a^{\alpha} K_{\alpha}{}^{\beta} \wn_{\beta}K + \Lie{n} a^{\alpha} \wn_{\beta}K_{\alpha}{}^{\beta} + \frac{1}{18} \wn_{\beta}K \wn^{\beta}K \notag\\
&-  K a^{\alpha} \wn_{\gamma}K_{\alpha}{}^{\gamma} + 4 a^{\alpha} K_{\alpha}{}^{\beta} \wn_{\gamma}K_{\beta}{}^{\gamma} -  \frac{1}{2} \wn_{\alpha}K^{\alpha \beta} \wn_{\gamma}K_{\beta}{}^{\gamma},
\\
\label{3d:z1_even2S2T_end}
\mathcal{A}_6^{(2,2,0)} = \ 
&
\frac{1}{9} a^2 K^2 + 4 a^{\alpha} a^{\beta} K_{\alpha}{}^{\gamma} K_{\beta \gamma} -  \frac{2}{3} K a^{\alpha} \Lie{n} a_{\alpha}\notag\\
& + \Lie{n} a_{\alpha} \Lie{n} a^{\alpha} + 2 a^{\alpha} K_{\alpha \beta} \Lie{n} a^{\beta} + \frac{2}{9} K a^{\alpha} \wn_{\alpha}K\notag\\
& -  \frac{2}{3} \Lie{n} a^{\alpha} \wn_{\alpha}K -  \frac{4}{3} a^{\alpha} K_{\alpha}{}^{\beta} \wn_{\beta}K + \frac{1}{9} \wn_{\beta}K \wn^{\beta}K.
\end{align}

\subsubsection*{Parity Even Sector with Four Space Derivatives}
This sector has $n_T=0$ time derivatives, $n_S=4$ space derivatives and $n_\epsilon=0$. This sector contains $ n_{{}_{FPD}} =15  $ independent FPD invariant expressions.  There are $ n_{cc} = 8 $ independent cocycles, $ n_{cb} = 6 $ coboundaries and $ n_{td} = 3 $ trivial descent cocycles. We are left with $ n_{an} = 2 $ independent anomalous terms, both of which are trivial descents (up to addition of coboundaries), and are given by:
\begin{align}\label{3d:z1_even4S_start} 
\mathcal{A}_1^{(0,4,0)} = \ 
& \left(\nabla_\alpha a^\alpha - \frac{1}{2} a^2 + \frac{1}{4} \widetilde R\right)^2 ,
\\
\label{3d:z1_even4S_end}
\mathcal{A}_2^{(0,4,0)} = \ 
&
\frac{3}{4} a^4 -  \frac{3}{4} a^2 \widetilde{R} -  \frac{5}{16} \widetilde{R}^2 + 2 a^{\alpha} a^{\beta} \widetilde{R}_{\alpha \beta} + \widetilde{R}_{\alpha \beta} \widetilde{R}^{\alpha \beta} -  \frac{1}{2} \widetilde{R} \wn_{\alpha}a^{\alpha} \notag\\
&  + 2 a^{\alpha} a^{\beta} \wn_{\beta}a_{\alpha} -  a^2 \wn_{\beta}a^{\beta} + 2 \widetilde{R}_{\alpha \beta} \wn^{\beta}a^{\alpha}+ \wn_{\alpha}a_{\beta} \wn^{\beta}a^{\alpha}.
\end{align}

\subsubsection*{Parity Odd Sector with One Space and Three Time Derivatives}
This sector has $n_T=3$ time derivatives, $n_S=1$ space derivative and $n_\epsilon=1$. This sector contains $ n_{{}_{FPD}} =5  $ independent FPD invariant expressions.  There are $ n_{cc} = 3 $ independent cocycles, $ n_{cb} = 2 $ coboundaries and $ n_{td} = 3 $ trivial descent cocycles. We are left with $ n_{an} = 1 $ independent anomalous term which is trivial descent (up to addition of coboundaries), and is given by:
\begin{align}\label{3d:z1_Odd1S3T}
\mathcal{A}_1^{(3,1,1)} = \ 
&
2 \tilde{\epsilon}_{\beta \delta \mu} K^{\alpha \beta} K^{\gamma \delta} \wn_{\gamma}K_{\alpha}{}^{\mu} + \tilde{\epsilon}_{\gamma \delta \mu} K_{\alpha}{}^{\gamma} K^{\alpha \beta} \wn^{\mu}K_{\beta}{}^{\delta}.
\end{align}

\subsection{The \texorpdfstring{$z=2$}{z=2} Case}
In this subsection we specify all the non-universal sectors in the calculation of the $z=2$ Lifshitz cohomology.

\subsubsection*{Parity Odd Sector with One Space and Two Time Derivatives}

This is the sector with $n_T = 2$ time derivatives, $n_S = 1$ space derivative and $n_\epsilon = 1$. It contains only $ n_{{}_{FPD}} = 1 $ FPD invariant expression, which results in $n_{cc}=1$ cocycle, that is trivial descent ($n_{td}=1$) and no coboundaries. There is thus $n_{an}=1$ anomaly in this sector, which is trivial descent and given by:
\begin{equation}\label{3d:z2_odd1S2T}
\mathcal{A}_1^{(2,1,1)} = \tilde{\epsilon}_{\beta \gamma \delta} K^{\alpha \beta} \wn^{\delta}K_{\alpha}{}^{\gamma}.
\end{equation}

\subsubsection*{Parity Odd Sector with Five Space Derivatives}

This sector has $n_T=0$ time derivatives, $n_S=5$ space derivatives and $n_\epsilon=1$. There are $n_{{}_{FPD}}=4$ independent FPD invariant expressions in this sector. We find $n_{cc}=2$ independent cocycles, $n_{cb}= 1$ independent coboundary and $n_{td}=1$ trivial descent cocycle. We then have $n_{an}=1$ anomaly in this sector, which is trivial descent:
\begin{equation}\label{3d:z2_odd5S}
\mathcal{A}_1^{(0,5,1)} = \frac{1}{2} a^{\alpha} a^{\beta} \tilde{\epsilon}_{\beta \gamma \delta} \wn^{\delta}\widetilde{R}_{\alpha}{}^{\gamma} + 2 \tilde{\epsilon}_{\beta \gamma \delta} \widetilde{R}^{\alpha \beta} \wn^{\delta}\widetilde{R}_{\alpha}{}^{\gamma} + \tilde{\epsilon}_{\beta \gamma \delta} \wn^{\beta}a^{\alpha} \wn^{\delta}\widetilde{R}_{\alpha}{}^{\gamma} .
\end{equation}

\subsection{The \texorpdfstring{$z=3$}{z=3} Case}
In this subsection we specify all the non-universal sectors in the calculation of the $z=3$ Lifshitz cohomology.

\subsubsection*{Parity Even Sector with Two time Derivatives}

This sector has $n_T=2$ time derivatives, $n_S=0$ space derivatives and $n_\epsilon=0$. There are $n_{{}_{FPD}}=3$ independent FPD invariant expressions in this sector. There are $n_{cc}=2$ independent cocycles, $n_{cb}=1$ coboundary and $n_{td}=1$ trivial descent cocycle. We conclude that there is only $n_{an}=1$ anomaly in this sector, which is trivial descent (up to addition of coboundaries), given by:
\begin{equation}\label{3d:z3_even2T}
\mathcal{A}_1^{(2,0,0)} = K^2 - 3 \tr(K^2) .
\end{equation}

\subsubsection*{Parity Even Sector with Six Space Derivatives}

This is the sector with $n_T=0$ time derivatives, $n_S=6$ space derivatives and $n_\epsilon=0$. This sector contains $n_{{}_{FPD}}=76$ independent FPD invariant expressions. We find $n_{cc} = 33$ independent cocycles, $n_{cb}=28$ independent coboundaries and $n_{td}=10$ trivial descent cocycles. We are left with $n_{an}=5$  independent anomalous terms, all of which are trivial descents, that are given by:
\begin{align}\label{3d:z3_even6S_start}
\mathcal{A}_1^{(0,6,0)} = \
& \left( \wn_{\alpha}a^{\alpha} - \frac{1}{6} a^2 + \frac{3}{4} \widetilde{R} \right)^3 ,
\\
\mathcal{A}_2^{(0,6,0)} = \ 
& \frac{1}{81} a^6 -  \frac{1}{9} a^4 \widetilde{R} + \frac{1}{4} a^2 \widetilde{R}^2 + \frac{3}{4} a^{\alpha} \widetilde{R} \wn_{\alpha}\widetilde{R} + a^{\alpha} \widetilde{R} \wn_{\alpha}\wn_{\beta}a^{\beta}\notag\\ 
& + \frac{9}{16} \wn_{\alpha}\widetilde{R} \wn^{\alpha}\widetilde{R} -  \frac{1}{3} a^{\alpha} a^{\beta} \widetilde{R} \wn_{\beta}a_{\alpha} + \frac{2}{3} a^2 \widetilde{R} \wn_{\beta}a^{\beta}\notag\\ 
& + a^{\alpha} \wn_{\alpha}\widetilde{R} \wn_{\beta}a^{\beta} + \frac{4}{3} a^{\alpha} \wn_{\alpha}\wn_{\gamma}a^{\gamma} \wn_{\beta}a^{\beta} + \frac{1}{9} a^{\alpha} a^{\beta} \wn_{\alpha}a^{\gamma} \wn_{\beta}a_{\gamma}\notag\\ 
& -  \frac{1}{6} a^2 a^{\beta} \wn_{\beta}\widetilde{R} -  \frac{3}{2} a^{\alpha} \widetilde{R}_{\alpha}{}^{\beta} \wn_{\beta}\widetilde{R} -  \frac{1}{2} a^{\alpha} \wn_{\alpha}a^{\beta} \wn_{\beta}\widetilde{R}\notag\\ 
& -  \frac{2}{9} a^2 a^{\beta} \wn_{\beta}\wn_{\gamma}a^{\gamma} -  \frac{2}{3} a^{\alpha} \wn_{\alpha}a^{\beta} \wn_{\beta}\wn_{\gamma}a^{\gamma} + \frac{2}{27} a^2 a^{\beta} a^{\gamma} \wn_{\gamma}a_{\beta}\notag\\ 
& -  \frac{4}{27} a^4 \wn_{\gamma}a^{\gamma} -  \frac{4}{9} a^{\alpha} a^{\beta} \wn_{\beta}a_{\alpha} \wn_{\gamma}a^{\gamma} + \frac{4}{9} a^2 \wn_{\beta}a^{\beta} \wn_{\gamma}a^{\gamma}\notag\\ 
& + 3 \wn_{\alpha}\wn^{\beta}a^{\alpha} \wn_{\gamma}\widetilde{R}_{\beta}{}^{\gamma} -  a^{\alpha} \widetilde{R}_{\alpha}{}^{\gamma} \wn_{\gamma}\wn_{\beta}a^{\beta}\notag\\ 
& + \wn_{\beta}\wn_{\alpha}a^{\alpha} \wn_{\gamma}\wn^{\beta}a^{\gamma} ,
\\
\mathcal{A}_3^{(0,6,0)} = \ 
& - \frac{1}{216} a^6 + \frac{1}{16} a^4 \widetilde{R} + \frac{63}{32} a^2 \widetilde{R}^2 + \frac{243}{64} \widetilde{R}^3 -  \frac{9}{2} a^{\alpha} a^{\beta} \widetilde{R} \widetilde{R}_{\alpha \beta}\notag\\ 
& -  \frac{81}{4} \widetilde{R} \widetilde{R}_{\alpha \beta} \widetilde{R}^{\alpha \beta} + 9 a^{\alpha} a^{\beta} \widetilde{R}_{\alpha}{}^{\gamma} \widetilde{R}_{\beta \gamma} + 27 \widetilde{R}_{\alpha}{}^{\gamma} \widetilde{R}^{\alpha \beta} \widetilde{R}_{\beta \gamma}\notag\\ 
& -  \frac{9}{2} a^2 \widetilde{R}_{\beta \gamma} \widetilde{R}^{\beta \gamma} + \frac{27}{16} \widetilde{R}^2 \wn_{\alpha}a^{\alpha} + 6 a^{\alpha} a^{\beta} \widetilde{R}_{\beta \gamma} \wn_{\alpha}a^{\gamma}\notag\\ 
& -  \frac{3}{2} a^{\alpha} a^{\beta} \widetilde{R} \wn_{\beta}a_{\alpha} + \frac{3}{4} a^2 \widetilde{R} \wn_{\beta}a^{\beta} + a^{\alpha} a^{\beta} \wn_{\alpha}a^{\gamma} \wn_{\beta}a_{\gamma}\notag\\ 
& -  \frac{27}{2} \widetilde{R} \widetilde{R}_{\alpha \beta} \wn^{\beta}a^{\alpha} + 27 \widetilde{R}_{\alpha}{}^{\gamma} \widetilde{R}_{\beta \gamma} \wn^{\beta}a^{\alpha} -  \frac{9}{4} \widetilde{R} \wn_{\alpha}a_{\beta} \wn^{\beta}a^{\alpha}\notag\\ 
& + 9 \widetilde{R}_{\alpha \gamma} \wn_{\beta}a^{\gamma} \wn^{\beta}a^{\alpha} + \frac{1}{12} a^4 \wn_{\gamma}a^{\gamma} + \wn^{\beta}a^{\alpha} \wn_{\gamma}a_{\beta} \wn^{\gamma}a_{\alpha}\notag\\ 
& - 3 a^2 \widetilde{R}_{\beta \gamma} \wn^{\gamma}a^{\beta} -  \frac{1}{2} a^2 \wn_{\beta}a_{\gamma} \wn^{\gamma}a^{\beta} ,
\\
\mathcal{A}_4^{(0,6,0)} = \ 
& - \frac{1}{8} \wn_{\alpha}\widetilde{R} \wn^{\alpha}\widetilde{R} -  \wn_{\beta}\widetilde{R}_{\alpha \gamma} \wn^{\gamma}\widetilde{R}^{\alpha \beta} + \wn_{\gamma}\widetilde{R}_{\alpha \beta} \wn^{\gamma}\widetilde{R}^{\alpha \beta} ,
\\
\label{3d:z3_even6S_end}
\mathcal{A}_5^{(0,6,0)} = \ 
& \frac{7}{81} a^6 -  \frac{7}{9} a^4 \widetilde{R} -  \frac{15}{4} a^2 \widetilde{R}^2 + 3 a^{\alpha} a^{\beta} \widetilde{R} \widetilde{R}_{\alpha \beta} + \frac{16}{9} a^2 a^{\beta} a^{\gamma} \widetilde{R}_{\beta \gamma}\notag\\ 
& - 7 a^{\alpha} a^{\beta} \widetilde{R}_{\alpha}{}^{\gamma} \widetilde{R}_{\beta \gamma} + 11 a^2 \widetilde{R}_{\beta \gamma} \widetilde{R}^{\beta \gamma} -  \frac{9}{4} a^{\alpha} \widetilde{R} \wn_{\alpha}\widetilde{R}\notag\\ 
& + 6 a^{\alpha} \widetilde{R}^{\beta \gamma} \wn_{\alpha}\widetilde{R}_{\beta \gamma} -  a^{\alpha} \widetilde{R} \wn_{\alpha}\wn_{\beta}a^{\beta} + 8 a^{\alpha} \widetilde{R}^{\beta \gamma} \wn_{\alpha}\wn_{\gamma}a_{\beta}\notag\\ 
& -  \frac{27}{16} \wn_{\alpha}\widetilde{R} \wn^{\alpha}\widetilde{R} -  a^{\alpha} a^{\beta} \widetilde{R} \wn_{\beta}a_{\alpha} - 2 a^2 \widetilde{R} \wn_{\beta}a^{\beta} -  a^{\alpha} \wn_{\alpha}\widetilde{R} \wn_{\beta}a^{\beta}\notag\\ 
& + \frac{5}{3} a^{\alpha} a^{\beta} \wn_{\alpha}a^{\gamma} \wn_{\beta}a_{\gamma} -  \frac{7}{6} a^2 a^{\beta} \wn_{\beta}\widetilde{R} -  \frac{15}{2} a^{\alpha} \widetilde{R}_{\alpha}{}^{\beta} \wn_{\beta}\widetilde{R}\notag\\ 
& -  \frac{7}{2} a^{\alpha} \wn_{\alpha}a^{\beta} \wn_{\beta}\widetilde{R} -  \frac{2}{3} a^2 a^{\beta} \wn_{\beta}\wn_{\gamma}a^{\gamma} - 2 a^{\alpha} \wn_{\alpha}a^{\beta} \wn_{\beta}\wn_{\gamma}a^{\gamma}\notag\\ 
& + \frac{10}{9} a^2 a^{\beta} a^{\gamma} \wn_{\gamma}a_{\beta} -  \frac{4}{9} a^4 \wn_{\gamma}a^{\gamma} -  \frac{4}{3} a^{\alpha} a^{\beta} \wn_{\beta}a_{\alpha} \wn_{\gamma}a^{\gamma}\notag\\ 
& + \frac{8}{3} a^{\alpha} a^{\beta} a^{\gamma} \wn_{\gamma}\widetilde{R}_{\alpha \beta} + 18 a^{\alpha} \widetilde{R}^{\beta \gamma} \wn_{\gamma}\widetilde{R}_{\alpha \beta}\notag\\ 
& - 3 \wn_{\alpha}\wn^{\beta}a^{\alpha} \wn_{\gamma}\widetilde{R}_{\beta}{}^{\gamma} + \frac{8}{9} a^{\alpha} a^{\beta} a^{\gamma} \wn_{\gamma}\wn_{\beta}a_{\alpha} - 4 a^{\alpha} \widetilde{R}_{\alpha}{}^{\gamma} \wn_{\gamma}\wn_{\beta}a^{\beta}\notag\\ 
& + 8 a^2 \widetilde{R}_{\beta \gamma} \wn^{\gamma}a^{\beta} + 4 a^{\alpha} \wn_{\alpha}\widetilde{R}_{\beta \gamma} \wn^{\gamma}a^{\beta} + 4 a^{\alpha} \wn_{\alpha}\wn_{\beta}a_{\gamma} \wn^{\gamma}a^{\beta}\notag\\ 
& + \frac{4}{3} a^2 \wn_{\beta}a_{\gamma} \wn^{\gamma}a^{\beta} + 8 a^{\alpha} \wn_{\beta}\widetilde{R}_{\alpha \gamma} \wn^{\gamma}a^{\beta} + 9 \wn_{\beta}\widetilde{R}_{\alpha \gamma} \wn^{\gamma}\widetilde{R}^{\alpha \beta}\notag\\ 
& + 6 \wn_{\alpha}\widetilde{R}_{\beta \gamma} \wn^{\gamma}\wn^{\beta}a^{\alpha} + \wn_{\alpha}\wn_{\beta}a_{\gamma} \wn^{\gamma}\wn^{\beta}a^{\alpha} .
\end{align}

\subsection{The \texorpdfstring{$z=3/2$}{z=3/2} Case}
In this subsection we specify the non-universal sector in the calculation of the $z=3/2$ Lifshitz cohomology.

\subsubsection*{Parity Even Sector with Three time Derivatives}
This sector has $n_T=3$ time derivatives, $n_S=0$ space derivatives and $n_\epsilon=0$. There are $n_{{}_{FPD}}=6$ independent FPD invariant expressions in this sector. It has $n_{cc}=3$ independent cocycles, $n_{cb}=2$ coboundaries and $n_{td}=2$ trivial descent cocycles. There is then $n_{an}=1$ anomaly in this sector, which is trivial descent and given by:
\begin{equation}\label{3d:z3/2_even3T}
\mathcal{A}_1^{(3,0,0)} = K^3  -  \frac{9}{2} K \tr(K^2) + \frac{9}{2} \tr(K^3) .
\end{equation}

\begin{table}
\centering
\begin{tabular}{|c|c|c|c|c|c|c|c|c|c|c|}
\hline
$z$ & $ n_T $ & $ n_S $ & $ n_\epsilon $ & $ n_{{}_{FPD}} $ & $ n_{cc} $ & $ n_{cb} $ & $ n_{td} $ & $ n_{tdcb} $ & $ n_{an} $ & Anomalies \\
\hline 
Universal   & 1 & 3 & 1 & 7 & 3 & 2 & 2 & 1 & 1 & See \eqref{3d:universal_1T3S} \\
\hline
1           & 4 & 0 & 0 & 13 & 7 & 5 & 4 & 2 & 2 & See \eqref{3d:z1_even4T_start}-\eqref{3d:z1_even4T_end} \\
		    & 2 & 2 & 0 & 47 & 23 & 17 & 13 & 7 & 6 & See \eqref{3d:z1_even2S2T_start}-\eqref{3d:z1_even2S2T_end} \\
            & 0 & 4 & 0 & 15 & 8 & 6 & 3 & 1 & 2 & See \eqref{3d:z1_even4S_start}-\eqref{3d:z1_even4S_end} \\
		    & 3 & 1 & 1 & 5 & 3 & 2 & 3 & 2 & 1 & See \eqref{3d:z1_Odd1S3T} \\
\hline
2  		  & 2 & 1 & 1 & 1 & 1 & 0 & 1 & 0 & 1 & See \eqref{3d:z2_odd1S2T} \\
		  & 0 & 5 & 1 & 4 & 2 & 1 & 1 & 0 & 1 & See \eqref{3d:z2_odd5S} \\
\hline
3		  & 2 & 0 & 0 & 3 & 2 & 1 & 1 & 0 & 1 & See \eqref{3d:z3_even2T} \\
		  & 0 & 6 & 0 & 76 & 33 & 28 & 10 & 5 & 5 & See \eqref{3d:z3_even6S_start}-\eqref{3d:z3_even6S_end} \\
\hline
3/2		  & 3 & 0 & 0 & 6 & 3 & 2 & 2 & 1 & 1 & See \eqref{3d:z3/2_even3T} \\
\hline
\end{tabular}
\caption{Summary of results for the Lifshitz cohomology in 3+1 dimensions. $z$ is the Lifshitz dynamical exponent. $n_T$, $n_S$ and $n_\epsilon$ are the number of time derivatives, space derivatives and Levi-Civita tensors in each sector, respectively (see subsection \protect\ref{ssb:restclasssectors} for more details). $n_{{}_{FPD}}$, $n_{cc}$, $n_{cb}$, $n_{td}$, $n_{tdcb}$ and $n_{an}$ are the number of independent FPD invariant expressions, cocycles, coboundaries, trivial descent cocycles, trivial descent coboundaries and anomalies in each sector, respectively (see subsection \protect\ref{ssb:prescription} for more details).}
\label{table:3+1ResultsSum}
\end{table}

\section{General Examples}\label{sec:GeneralExamples}
In this subsection we detail some results that are valid for general values of $d$ and $z$. The examples we present do not consist of a full analysis of the anomalies in the cases under consideration. Rather, it is a study of particular sectors that could be either fully solved, or that we have something general to say about their anomalous structure.

\subsection{The \texorpdfstring{$z=d$}{z=d} Purely Temporal Sector}
When the number of space dimensions $d$ equals the value of the dynamical exponent $z$ we have a universal sector whose structure does not depend on the value of $d$ (or $z$). This sector contains 2 time derivatives, zero space derivatives and is parity even.

We have the following FPD invariant terms:
\begin{equation}
\begin{split}
\phi_1 = \tr{(K^2)}, \qquad \phi_2 = K^2 , \qquad \phi_3 = \Lie{n} K ,.
\end{split}
\end{equation}
The associated expressions of ghost number 1 are $I_i = \int \sqrt {-g} \sigma \phi_i$.
The Weyl variation of each can be calculated using the rules of subsection \ref{ssb:weyllaws}:
\begin{equation}
\begin{split}
& \delta^W_\sigma \ I_1= -\int \sqrt{-g}\ \sigma \ [2K \Lie{n}\sigma], \\
& \delta^W_\sigma \ I_2= -\int \sqrt{-g}\ \sigma \ [2d K \Lie{n}\sigma], \\
& \delta^W_\sigma \ I_3 = -\int \sqrt{-g}\ \sigma \ [-zK \Lie{n} \sigma+d\Lie{n}^2\sigma].
\end{split}
\end{equation}
Defining $L_1 \equiv \int \sqrt{-g}\ \sigma \ K \Lie{n}\sigma$ we get after integrating by parts (remember that $d=z$):
\begin{equation}
\begin{split}
& \delta^W_\sigma \ I_1= -2 L_1 ,\\
& \delta^W_\sigma \ I_2= -2z L_1 ,\\
& \delta^W_\sigma \ I_3 = 2z L_1 .
\end{split}
\end{equation}
There are two independent cocycles ($n_{cc}=2$):
\begin{equation}
\begin{split}
E_1 =\ & I_1-\frac{1}{z}I_2 = \int \sqrt{-g}\ \sigma \ [\tr{(K^2)}-\frac{1}{z}K^2],\\
E_2 =\ & I_2+I_3 = \int \sqrt{-g}\ \sigma \ [K^2+ \Lie{n}K].
\end{split}
\end{equation}
The second of them turns out to be a coboundary term ($n_{cb}=1$). Notice that for the special case of $z=1$ the first cocycle identically vanish and then we have no anomalies in this sector.
There is one (trivial descent) anomaly for $z\neq 1$ ($n_{an}=1$):
\begin{equation}\label{general:d_z_temporal}
\begin{split}
\boxed{
A^{(2,0,0)}_1 = \int \sqrt{-g}\ \sigma \ [\tr{(K^2)}-\frac{1}{z}K^2].}
\end{split}
\end{equation}
Note that this structure is consistent with what we got in equations: \eqref{2d:z2_anomaly_2_0_0}, \eqref{3d:z3_even2T},
for the specific cases of $d=z=2$ and $d=z=3$ respectively.
A similar analysis could be performed for any $d=nz$ where $n$ is some integer. Similar conclusions can then be drawn regarding the sector with $n+1$ temporal derivatives.

\subsection{General Purely Spatial Anomaly for Even \texorpdfstring{$d+z$}{d+z}}
For general even $d+z$ (where $d\geq 2$), a parity even purely spatial sector always exists in the cohomology (with $n_T = 0$, $n_S = d+z$ and $ n_\epsilon = 0 $). While a full analysis of the cohomology in this sector seems to require solving for each $d$ and $z$ individually, we noted that the anomalies in this sector always contain the following expression:
\begin{equation}\label{general:d_z_spatial}
\boxed{A^{(0,z+d,0)}_1 = \int \sqrt{-g} \sigma \left(\wn_\mu a^\mu - \frac{d-2}{2z} a^2 + \frac{z}{2(d-1)} \widetilde R\right)^{(d+z)/2}.}
\end{equation}
This is the case of the anomalies in equations:
\eqref{2d:z2_anomaly_0_4_0}, \eqref{2+1Other:z4AnomStart}, \eqref{3d:z1_even4S_start}, \eqref{3d:z3_even6S_start}.

It can be easily checked using the transformation rules of subsection \ref{ssb:weyllaws} that the Weyl variation of the density associated with the above expression vanishes. It is therefore a trivial descent cocycle. Since this scalar density is not a total derivative, it cannot be a coboundary in the cohomology, and must represent an anomaly. 

We believe that additional general structures can be identified in the other sectors, which we leave for future work.

\section{Comparison to the Conformal Case}\label{sec:comparison}
In this section we compare the Lifshitz scale anomalies with $z=1$, to the well known cases of conformal anomalies in $1+1$, $2+1$ (no anomalies)
and $3+1$ dimensions. We find that the Euler density, which always represents an anomaly in the conformal case in even dimensions, becomes a coboundary term in the Lifshitz $z=1$ case, and thus no longer an anomaly. We also explain the relation between the parity odd Lifshitz anomaly in $1+1$ dimensions and gravitational anomalies.

\subsection{Comparison of the Cohomologies}
Since conformal theories obey a scaling symmetry of the type \eqref{intro:lifsh} with $z=1$, it is possible to regard them as Lifshitz theories with $z=1$, with the addition of an arbitrary foliation of spacetime (where the classical action is independent of this arbitrary foliation). We can then compare their cohomologies.

Since in the Lifshitz case we do not require full diffeomorphism invariance but only FPD invariance, there are many more allowed terms in the cohomology. It is clear however that any diffeomorphism invariant term that may appear in the conformal cohomology may also appear in the Lifshitz one.\footnote{More precisely, assuming we restrict the discussion to the relative cohomology of the Weyl operator with respect to diffeomorphisms, the diffeomorphism invariant effective action of the conformal picture $W_{\text{conf}}[e^a{}_\mu]$ may also function as an FPD invariant effective action in the Lifshitz picture $ W_{\text{Lif}}[e^a{}_\mu, t^a] \equiv W_{\text{conf}}[e^a{}_\mu]$.}
In addition we expect the conformal cocycles and coboundaries to be contained in the Lifshitz cocycles and coboundaries respectively. 
In particular, since the Weyl operator does not mix between Lifshitz sectors, the projection of each conformal cocycle on each of the sectors should also be a cocycle of the Lifshitz cohomology.
While any conformal coboundary is also a coboundary in the Lifshitz cohomology, the opposite is not necessarily true. Therefore anomalies of the conformal case may now become coboundaries in the Lifshitz cohomology. This is due to the fact that the counterterms needed to cancel them may not be fully diffeomorphism invariant but merely FPD invariant.
We show below how this is manifested in the different dimensions.

\paragraph{In $1+1$ dimensions} we have in the conformal case one anomaly of the form \eqref{prelims:1+1d_WeylAnomaly}. In foliation projected terms it reads:
\begin{equation}
\int \sqrt{-g}\sigma R =\int \sqrt{-g} \sigma \left[ 2(K^2+\Lie{n}K)-2(a^2+\wn_\mu a^\mu) \right] = 2 E_1^{(2,0,0)} - 2 E_1^{(0,2,0)},
\end{equation}
where $ E_i^{(n_T,n_S,n_\epsilon)} $ represents the $i$-th cocycle in the sector that corresponds to $n_T$ time derivatives, $n_S$ space derivatives and $n_\epsilon$ Levi-Civita tensors. Thus we observe that the conformal anomaly is indeed a cocycle in the Lifshitz cohomology, but unlike the conformal case, it is now a coboundary. It can be removed by adding an appropriate counterterm to the action:
\begin{equation}
W_{c.t.} = \int \sqrt{-g} (K^2-a^2) .
\end{equation}
Note, that adding this counterterm to the action also breaks the diffeomorphism invariance (while FPD invariance is preserved), thus ``shifting'' the Weyl anomaly into a diffeomorphism anomaly.

Recall that we also found in the $1+1$, $z=1$ Lifshitz cohomology an apparently new anomaly of the form \eqref{1d:z1_anomaly_1_1_1}.
However, in the next subsection we show that this anomaly is actually related to the Weyl ``partner'' of the gravitational anomaly in the $1+1$ conformal case.\footnote{Weyl ``partner''	as defined in the classification theorem mentioned in subsection \ref{ssb:CohomolConfRev}.}

\paragraph{In $2+1$ dimensions} in the conformal picture there are no Weyl anomalies since no invariant of the right dimension can be constructed. In the $z=1$ Lifshitz case we have one new anomaly \eqref{2d:z1_anomaly_3_0_1} in the parity odd sector.

\paragraph{In $3+1$ dimensions} in the conformal picture there are two parity even anomalies \eqref{prelims:W_2}, \eqref{prelims:E_4} and one parity odd anomaly \eqref{prelims:P_1}. By writing these expressions in foliation projected terms we verified that they are indeed linear combinations of cocycles in the appropriate sectors of the $3+1$, $z=1$ Lifshitz cohomology. The Euler density and the Weyl tensor squared are a linear combination of cocycles of the parity even sectors $(n_T,n_S,n_\epsilon) = (4,0,0) $, $ (2,2,0) $ and $(0,4,0) $, whereas the Pontryagin term is a linear combination of cocycles from the parity odd sectors $(3,1,1)$ and $(1,3,1)$.
Both the Weyl tensor squared and the Pontryagin term are anomalies in the Lifshitz cohomology as well, and can be decomposed as a combination of the anomaly expressions that were specified in section \ref{sec:3+1AnomalyResults} as follows:
\begin{align}
\begin{split}
W^2 =\  &  
\frac{2}{3} \mathcal{A}_1^{(4,0,0)} 
-\frac{2}{3} \mathcal{A}_2^{(4,0,0)}
+\frac{32}{3} \mathcal{A}_2^{(2,2,0)}
-\frac{4}{3} \mathcal{A}_3^{(2,2,0)}
\\
& -4 \mathcal{A}_4^{(2,2,0)}
-4 \mathcal{A}_5^{(2,2,0)}
-\frac{2}{3} \mathcal{A}_1^{(0,4,0)}
+2 \mathcal{A}_2^{(0,4,0)} + \dots,
\end{split}\\
\begin{split}
P_1 =\  &  
16 \mathcal{A}_1^{(1,3,1)}+\dots,
\end{split}
\end{align}
where ``$\dots$'' stands for coboundary terms.
Note that the Pontryagin term consists only of the universal sector anomaly (plus coboundaries from both parity odd sectors). 
The Euler density, however, again turns out to be a coboundary in the Lifshitz cohomology. It is therefore not an anomaly of the $z=1$ Lifshitz cohomology, and can be removed by adding the following counterterm to the action:
\begin{align}
W_{c.t.} = - \int \sqrt{-g}
& \left[
\frac{401}{42} a^2 \Lie{n} K -  \frac{22}{63} K_{\alpha \beta} K^{\alpha \beta} \Lie{n} K -  \frac{6}{7} \widetilde{R} \Lie{n} K + \frac{16}{63} (\Lie{n} K)^2
\right.
\notag\\ 
&\  -  \frac{76}{21} \Lie{n} a_{\alpha} \Lie{n} a^{\alpha} -  \frac{5}{7} a^{\alpha} a^{\beta} \Lie{n} K_{\alpha \beta} -  \frac{4}{7} \widetilde{R}^{\alpha \beta} \Lie{n} K_{\alpha \beta}\notag\\ 
&\  + \frac{32}{63} \Lie{n} K_{\alpha \beta} \Lie{n} K^{\alpha \beta} + \frac{221}{42} a^{\alpha} \Lie{n} \wn_{\alpha}K -  \frac{20}{7} \Lie{n} \wn_{\alpha}\wn_{\beta}K^{\alpha \beta}\notag\\ 
&\  -  \frac{332}{21} a^{\alpha} \Lie{n} \wn_{\beta}K_{\alpha}{}^{\beta} -  \frac{260}{21} K_{\alpha \beta} \Lie{n} \wn^{\beta}a^{\alpha} -  \frac{1}{63} \Lie{n} \wn^{\beta}\wn_{\beta}K\notag\\ 
&\  -  \frac{391}{21} a^{\alpha} \Lie{n}^{2} a_{\alpha} + \frac{68}{63} K^{\alpha \beta} \Lie{n}^{2} K_{\alpha \beta} -  \frac{115}{18} \Lie{n}^{2} \wn_{\alpha}a^{\alpha}\notag\\ 
&\  -  \frac{4}{63} \Lie{n}^{3} K -  \frac{28}{9} \Lie{n} a^{\alpha} \wn_{\alpha}K + 2 \widetilde{R} \wn_{\alpha}a^{\alpha} - 2 a^{\alpha} \wn_{\alpha}\wn_{\beta}a^{\beta}\notag\\ 
&\  - 4 \wn_{\alpha}\wn_{\beta}\widetilde{R}^{\alpha \beta} - 4 \wn_{\alpha}\wn^{\beta}\wn_{\beta}a^{\alpha} + 2 \wn_{\alpha}a^{\alpha} \wn_{\beta}a^{\beta}\notag\\ 
&\  + \frac{220}{21} \Lie{n} a^{\alpha} \wn_{\beta}K_{\alpha}{}^{\beta} -  \frac{512}{21} \Lie{n} K_{\alpha \beta} \wn^{\beta}a^{\alpha} \notag\\ 
&\ \left. + \frac{94}{7} \wn_{\gamma}K_{\alpha \beta} \wn^{\gamma}K^{\alpha \beta} + \frac{68}{7} K^{\alpha \beta} \wn^{\gamma}\wn_{\gamma}K_{\alpha \beta} \right].
\end{align}
We also note that in the Lifshitz case we have many anomalies that are independent from the projections of the Weyl tensor squared and Pontryagin term to the various sectors (7 in the parity even and 1 in the parity odd sectors). 

In conclusion, in the examples studied, the Euler densities turned out to consist of coboundary terms only, while the rest of the conformal anomalies are still present as anomalies in the Lifshitz cohomology. We propose that this might be the case for all even spacetime dimensions. We also found additional anomalies in the Lifshitz cohomology that are not diffeomorphism invariant.

\subsection{Relation to Gravitational Anomalies in 1+1 Dimensions}

When the conformal theory contains gravitational anomalies, its description in terms of a $z=1$ Lifshitz theory will not in general be FPD invariant and is therefore outside the scope of our discussion. 
In $1+1$ dimensional theories, however, we can always extend the conformal effective action to an FPD invariant Lifshitz one.
This is done by ``shifting'' the anomalies into a foliation dependence. Assume that the conformal (non diffeomorphism invariant) action takes the form $W_{\text{conf}}[e^a{}_\mu] =W_{\text{conf}}[e^0{}_\mu,e^1{}_\mu]$, where $W_{\text{conf}}$ is invariant under spacetime coordinates transformations but not under local Lorentz transformations.\footnote{By adding an appropriate counterterm one may always shift the gravitational anomaly to a pure Lorentz anomaly \cite{Bertlmann:1996xk,Bardeen:1984pm}.}
We can define the Lifshitz effective action as follows:
\begin{equation}\label{compartoconf:lifactiondef}
W_{\text{Lif}}[e^a{}_\mu,t^a] \equiv W_{\text{conf}} [-e^a{}_\mu n_a, e^a{}_\mu \widetilde{n}_a] = W_{\text{conf}}[-n_\mu, \widetilde{n}_\mu],
\end{equation}
where $ \widetilde{n}^a \equiv  \epsilon^{ab} n_b $. This effective action is clearly FPD invariant, and coincides with the original $ W_{\text{conf}} $ for the ``flat'' foliation $ n^a = (1,0) $. It represents a choice of a local Lorentz frame which is aligned with the arbitrary foliation.\footnote{Since the theory is classically invariant under local Lorentz transformations, $ W_{\text{Lif}} $ differs from $ W_{\text{conf}} $ by a local term and therefore the actions are equivalent.} Indeed, in terms of this Lifshitz action, the Lorentz non invariance has been converted to a foliation dependence.\footnote{Note that there is only one anomalous gravitational degree of freedom in $1+1$ dimensional field theories, which is known to be associated with area preserving diffeomorphisms (i.e.\ $\del_\mu \xi^\mu = 0$). This is consistent with the anomalous degree of freedom of the foliation dependence in the Lifshitz field theories.}
The anomalous Ward identities that correspond to the Lorentz anomaly in conformal theories are well known, and given by (see e.g.~\cite{Bertlmann:1996xk,Bertlmann:2000da,Ebner:1987pg}): 
\begin{align}\label{compartoconf:confgravanomward1}
T_{(e)[\mu\nu]} = &\  a R \epsilon_{\mu\nu},\\
\nabla_\mu T^{\mu\nu}_{(e)} = &\  a R \epsilon^{ab} \omega^\nu{}_{ab}, \\
\label{compartoconf:confgravanomward3}
T^\mu_{(e)\mu} = &\ -2 a \epsilon^{ab} \nabla_\mu \omega^\mu{}_{ab},
\end{align}
where $a$ is a model dependent anomaly coefficient.
Note, that this anomaly belongs to the second set in the classification theorem mentioned in subsection \ref{ssb:CohomolConfRev}. Identity \eqref{compartoconf:confgravanomward3} represents the Weyl ``partner'' of the Lorentz anomaly.

Written in terms of the previously defined Lifshitz effective action, the corresponding anomalous Ward identities are:
\begin{align}\label{compartoconf:lifgravanomward1}
\hat{J}_\mu &= -2a R \widetilde{n}_\mu , \\
\label{compartoconf:lifgravanomward2}
T^\mu_{(e)\mu} &= 4a \tilde \epsilon^\rho (\Lie{n} a_\rho- a_\rho K - \wn_\rho K) ,
\end{align}
along with the (non anomalous) Ward identities \eqref{anweylan:teward1n} and \eqref{anweylan:teward2n}. The identity \eqref{compartoconf:lifgravanomward1} corresponds to the anomalous independence of the action on the foliation. It can be derived from \eqref{compartoconf:confgravanomward1} and \eqref{anweylan:teward1n}. Identity \eqref{compartoconf:lifgravanomward2} is the Weyl ``partner'' of this foliation independence anomaly, and can be derived from \eqref{compartoconf:confgravanomward3} by replacing the vielbeins $ e^0{}_\mu $, $ e^1{}_\mu $ by $- n_\mu $ and $ \widetilde{n}_\mu $ respectively (according to \eqref{compartoconf:lifactiondef}), as follows:
\begin{align}
\begin{split}
T^\mu_{(e)\mu} 
= &\ -2 a \epsilon^{ab} \nabla_\mu \omega^\mu{}_{ab}
%= -2 a \epsilon^{ab} \nabla_\mu \left[e_{a\nu} \nabla^\mu e_b{}^\nu\right]
 =  
-4 a \nabla_\mu \left[e_{0\nu} \nabla^\mu e_1{}^\nu\right] \rightarrow
\\
&
-4 a \nabla_\mu \left[n_\nu \nabla^\mu \widetilde n^\nu\right] 
%= -4 a \epsilon^{\nu\rho} \nabla_\mu \left[n_\nu \nabla^\mu  n_\rho\right]
%= -4a \epsilon^{\nu\rho} n_\nu  \nabla_\mu\left[K^\mu_\rho - a_\rho n^\mu\right] =
= -4 a \epsilon^{\nu\rho} n_\nu  \nabla_\mu\left[K P^\mu_\rho - a_\rho n^\mu\right] 
\\
&  = 4a \tilde \epsilon^\rho (\Lie{n} a_\rho- (a_\rho K + \wn_\rho K)).
%\\
%& = 4a \left(2A^{(2,0,1)}_1-F^{(2,0,1)}_1\right)
\end{split}
\end{align}
It can be easily checked that expression \eqref{compartoconf:lifgravanomward2} is a linear combination of the anomaly found in the parity odd sector of the $1+1$, $z=1$ Lifshitz cohomology in subsection \ref{ssb:1+1_1S1Todd_Results}, and a coboundary term:
\begin{equation}
\int \sqrt{-g} \sigma T^\mu_{(e)\mu} =  4a \left(2A^{(2,0,1)}_1-F^{(2,0,1)}_1\right).
\end{equation}
We therefore see that in the case of a conformal theory, the anomaly we found in the $1+1$ Lifshitz cohomology functions as the Weyl part of the gravitational anomaly in the theory. 
We do not expect this to be the case for higher dimensions, since the anomalous degrees of freedom of diffeomorphism invariance would in general contain not only the foliation degrees of freedom, but those of FPD invariance as well. One would therefore need to consider non FPD invariant expressions in order to account for all possible gravitational anomalies.

\section{Summary and Outlook}\label{sec:Summary}

In this work we presented a detailed prescription for finding the general anomalous structures associated with scaling symmetry in Lifshitz field theories. 
One has to consider a foliation one form as a background field in addition to the spacetime metric. It is then possible to check, which non-trivial foliation preserving  invariants can be built that are consistent with the Wess-Zumino consistency conditions.

We performed the analysis for various values of $z$ and $d$. Our results are summarised in table \ref{summary_table} below. 
In general we found that all the anomalies are type B in the sense explained in subsection \ref{ssb:prelim_classification}.
We suspect that this might be the case for general $d$ and $z$. We leave this for future study. We also noted that as opposed to the conformal case not every Weyl invariant scalar density is an anomaly.

For $z=1$ we compared our results to the conformal Weyl anomalies. We showed that the Euler densities in 1+1 and 3+1 dimensions become trivial. This is due to the variety of new counterterms, which are FPD invariant but not fully diffeomorphism invariant. We suggest that this may be the case for any number of dimensions.
In 1+1 we related the single parity odd anomaly to gravitational anomalies.
In $2+1$ dimensions, as opposed to the conformal case, we found various anomalies. This is again due to the rich structure induced by the foliation. In $3+1$ we found a larger number of anomalies than in the conformal case. 

We found two specific examples of anomaly expressions that are valid for general $d$ and $z$. 
This suggests the possibility that more general structures like these may be found, or even a full cohomological analysis that would be valid for any dimension and any value of the dynamical exponent. In particular, since we noted that in $1+1$ dimensions, other than the anomaly in the universal sector, there are no further anomalies in any of the integer values of $z$ up to 12, it would be interesting to see if this can be proven for any $z$.

We worked out the relative cohomology w.r.t foliation preserving diffeomorphisms. Another interesting extension of this work would be to study the full cohomology, including anomalies of FPD invariance. Among other things, it may enable us to find a more general relation between the known gravitational anomalies in the conformal case and anomalies in Lifshitz field theories. 

The anomaly terms allowed by the WZ consistency conditions in $ 2+1 $ dimensions with $ z=2 $ have  been shown to appear both in field theory \cite{Baggio:2011ha} and holographic \cite{Baggio:2011ha,Griffin:2011xs}  calculations. It would be interesting to similarly reproduce the other anomaly terms that we found in the Lifshitz cohomology for $1+1$, $2+1$ and $3+1$ dimensions in field theory or holographic models, and calculate the associated coefficients. 

Two other interesting research directions follow from our work.
Studying  the behaviour of the coefficients associated with the Lifshitz scale anomalies along RG flows 
can shed light on RG flows in non-Lorentz invariant field theories.
A derivation of the contribution of Lifshitz  anomalies to the entanglement entropy in Lifshitz field theories, as has been done in the conformal case in \cite{Fursaev:2013fta}, can also give valuable insights to non-Lorentz invariant field theories. We leave these for future work.

\begin{table}
\centering
\begin{tabular}{|c|c|c|c|c|}
\hline
$d$ &  $z$ & $n_{an}$ & Anomalies \\
\hline
1+1 & 1-12 & 1 & See \eqref{1d:z1_anomaly_1_1_1}\\
\hline
2+1 & 1 & 1 & See \eqref{2d:z1_anomaly_3_0_1}\\
\hline
 & 2 & 2 & See \eqref{2d:z2_anomaly_2_0_0}, \eqref{2d:z2_anomaly_0_4_0} \\
\hline
 & 3 & 0 & -- \\
\hline
 & 4 & 2 & See \eqref{2+1Other:z4AnomStart} - \eqref{2+1Other:z4AnomEnd} \\
\hline
 & 2/3 & 2 & See \eqref{2+1Other:z2/3AnomStart} - \eqref{2+1Other:z2/3AnomEnd} \\
\hline
 & 3/2 & 0 & -- \\
\hline
3+1 & 1 & 12 & See \eqref{3d:universal_1T3S} - \eqref{3d:z1_Odd1S3T}
 \\
\hline
 & 2 & 3 &  See \eqref{3d:universal_1T3S}, \eqref{3d:z2_odd1S2T} - \eqref{3d:z2_odd5S} \\
\hline
 & 3 & 7 &  See \eqref{3d:universal_1T3S},
\eqref{3d:z3_even2T} -  \eqref{3d:z3_even6S_end} 
 \\
\hline
 & 3/2 & 2 &  See \eqref{3d:universal_1T3S}, \eqref{3d:z3/2_even3T} \\
\hline 
general & -- & -- & See \eqref{general:d_z_temporal} - \eqref{general:d_z_spatial} \\
\hline
\end{tabular}
\caption{Summary of the results. $d$, $z$ and $n_{an}$ are the space dimension, dynamical exponent and number of anomalies respectively.}
\label{summary_table}
\end{table}

\acknowledgments
We would like to thank Adam Schwimmer for a valuable email correspondence. 
A large number of the calculations in this paper were
performed using xAct \cite{Garcia} and xTras \cite{Nutma:2013zea}, tensor computer algebra packages for \emph{Mathematica}.
This work is supported in part by the Israeli Science Foundation Center of Excellence, BSF, GIF and the I-CORE program of Planning and Budgeting Committee and the Israel Science Foundation (grant number 1937/12). The work of S.C is partially supported by the Israel Ministry of Science and Technology. S.C would like to thank the hospitality of the physics department at Princeton University during the completion of this manuscript.

%\hrulefill

\appendix

\section{Notations and Conventions}\label{app:Notations}

This appendix serves as a quick reference for our notations, conventions and definitions.
We work with a $d+1$ dimensional spacetime manifold, where $d$ is the number of space dimensions. The manifold is foliated into $d$ dimensional leaves. We use Greek indices $\alpha,\beta,\dots$ for the $d+1$ dimensional spacetime coordinates. We use the Latin indices $i,j,k,\dots$ for coordinates on a $d$ dimensional foliation leaf. We use $a,b,c,\dots$ as indices in the local Lorentz frame of the spacetime manifold (when using vielbeins). 

We use a metric $g_{\mu\nu}$ with a Lorentzian signature of $\{-1,1,1,\dots\}$. The vielbeins $e^a{}_\mu$ then satisfy $ g^{\mu\nu} e^a{}_\mu e^b{}_\nu = \eta^{ab} $, where $ \eta^{00}=-1$ and $\eta^{11}=\eta^{22}=\dots=1$. We assume that the foliation leaves are spacelike and the foliation 1-form is timelike, so that the normalized foliation 1-form $n_\mu$ satisfies: $ n_\mu n^\mu = -1 $.

For the Levi-Civita tensor (of the full spacetime) we use the definition: 
\begin{equation}
\epsilon^{\alpha\beta\dots} \equiv \frac{1}{\sqrt{|g|}} \hat{\epsilon}^{\alpha\beta\dots},
\end{equation}
where $ \hat{\epsilon}^{\alpha\beta\dots} $ is the totally antisymmetric symbol, defined such that $ \hat{\epsilon}^{012\dots} = 1 $.

We use the standard torsionless metric compatible connection on the spacetime manifold, with the following convention for the associated Riemann curvature:
\begin{equation}
[\nabla_\mu,\nabla_\nu] V^\alpha = R^\alpha{}_{\rho\mu\nu} V^\rho,
\end{equation} 
where $V^\alpha$ is some vector, and the following conventions for the Ricci tensor and scalar:
\begin{equation}
R_{\rho\nu} \equiv R^\alpha{}_{\rho\alpha\nu}, \qquad R \equiv g^{\rho\nu} R_{\rho\nu} .
\end{equation}
When working with vielbeins we use the following convention for the spin connection:
\begin{equation}
\omega_\mu{}^a{}_b = - e_b{}^\nu \nabla_\mu e^a{}_\nu,
\end{equation} 
where the covariant derivative here operates only on the spacetime indices.
Also note that we define the temporal derivative of a tensor $ T^{\alpha_1\alpha_2\dots}_{\beta_1\beta_2\dots} $ to be its Lie derivative in the direction of the normalized foliation $n^\mu$:
\begin{equation}
\Lie{n} T^{\alpha_1\alpha_2\dots}_{\beta_1\beta_2\dots} = n^\mu \del_\mu T^{\alpha_1\alpha_2\dots}_{\beta_1\beta_2\dots} - \del_\mu n^{\alpha_1}\, T^{\mu\alpha_2\dots}_{\beta_1\beta_2\dots} -\dots +\del_{\beta_1} n^{\mu}\, T^{\alpha_1\alpha_2\dots}_{\mu\beta_2\dots} +\dots\, .
\end{equation}

We end this appendix with table \ref{table:notations}, summarizing the various notations used throughout this work.

\begin{table}
\centering
\begin{tabular}{p{0.19\textwidth}p{0.81\textwidth}}
\hline\hline
$d$ & Number of space dimensions \\
$z$ & Lifshitz dynamical exponent \\
$t_\mu$, $n_\mu$ & Non-normalized/normalized foliation 1-form \\
$\xi_\mu$, $\alpha^a{}_b$, $\sigma$ & Diffeomorphism/local Lorentz/Weyl transformations parameter/ghost \\
$\delta^D_\xi$, $\delta^L_\alpha$, $\delta^W_\sigma$ & Diffeomorphism/local Lorentz/Weyl operator \\
$T^{\mu\nu}_{(g)}$, $T_{(e)} {}^\mu {}_a$ & Stress-energy tensor defined by variation w.r.t metric/vielbeins\\
$J^\alpha$, $\hat{J}^\alpha$ & Non-normalized/normalized ``foliation current'' (action variation w.r.t $t_\alpha$)\\
$d_\sigma$ & Global Lifshitz scaling dimension \\
$ P_{\mu\nu} $ & Foliation projector $ P_{\mu\nu} \equiv g_{\mu\nu} + n_\mu n_\nu $ \\
$ a_\mu $ & Acceleration associated with the foliation $ a_\mu \equiv \Lie{n} n_\mu = n^\nu \nabla_\nu n_\mu$\\
$K_{\mu\nu}$ & Extrinsic curvature of the foliation $K_{\mu\nu} \equiv \frac{1}{2}\Lie{n} P_{\mu\nu} = P_\mu^\rho \nabla_\rho n_\nu $\\
$ K $ & Trace of the foliation extrinsic curvature, $K \equiv K^\mu_\mu$\\
$ \tr(K^n) $ & Trace of the product of $n$ extrinsic curvatures, $\tr(K^n)=K^\alpha_\beta K^\beta_\gamma \dots K^\mu_\alpha$\\
$ \widetilde{R}_{\mu\nu\rho\sigma} $, $\widetilde{R}_{\mu\nu}$, $\widetilde{R}$ & Intrinsic Riemann tensor/Ricci tensor/Ricci scalar of the foliation\\
$\tilde \epsilon^{\mu\nu\rho \dots}$ & Intrinsic Levi-Civita tensor of the foliation, i.e.\ $\tilde \epsilon^{\mu\nu\rho \dots} = n_\alpha \epsilon^{\alpha\mu\nu\rho\dots}$\\
$\wn_\mu$ & Spatial (foliation projected) covariant derivative \\
$ \widetilde{T}_{\alpha\beta\dots} $ & Any tensor tangent to the foliation\\
$n_T$, $n_S$, $n_\epsilon$ & Total number of time derivatives/space derivatives/Levi-Civita tensors in an expression or in a sector of the Lifshitz cohomology\\
$n_{{}_{FPD}}$ & Number of independent FPD invariant expressions in a given sector\\
$n_{cc}$, $n_{cb}$, $n_{an}$ & Number of independent cocycles/coboundaries/anomalies in a sector\\
$n_{td}$, $n_{tdcb}$ & Number of independent trivial descent cocycles/coboundaries in a sector\\
$\phi_i$ & The $i$-th independent FPD invariant expression in a given sector\\
$I_i$ & The $i$-th independent integrated expression of ghost number one in a given sector $ I_i = \int \sqrt{-g} \sigma \phi_i $\\
$L_i$ & The $i$-th independent integrated expression of ghost number two in a given sector\\
$G_i$ & The $i$-th integrated expression of ghost number zero in a given sector $ G_i = \int \sqrt{-g}\phi_i $\\
$E_i^{(n_T,n_S,n_\epsilon)}$ or $E_i$ & The $i$-th independent cocycle in the sector corresponding to the values $(n_T,n_S,n_\epsilon)$\\
$F_i^{(n_T,n_S,n_\epsilon)}$ or $F_i$ & The $i$-th independent coboundary in the sector corresponding to the values $(n_T,n_S,n_\epsilon)$\\
$A_i^{(n_T,n_S,n_\epsilon)}$ or $A_i$ & The $i$-th independent anomaly in the sector corresponding to the values $(n_T,n_S,n_\epsilon)$\\
$H_i$ & The $i$-th independent trivial descent cocycle in a given sector\\
$\mathcal{A}_i^{(n_T,n_S,n_\epsilon)}$ or $\mathcal{A}_i$ & The $i$-th independent anomaly density in the sector corresponding to the values $(n_T,n_S,n_\epsilon)$, related to the $i$-th anomaly by $A_i = \int \sqrt{-g} \sigma \mathcal{A}_i$\\
\hline\hline
\end{tabular}
\caption{Notations and definitions.}
\label{table:notations}
\end{table}

\section{Derivation of the Lifshitz Ward Identities}\label{app:LifshitzWardIds}

In this appendix we derive the form of the classical Ward identities corresponding to foliation preserving diffeomorphisms 
and the anisotropic Weyl scaling as presented in subsection \ref{ssb:LifshitzWardIds}.

Assume a classical action depending on the metric and foliation $ S(g_{\mu\nu}, t_\alpha, \{\phi\}) $ (where $ \{\phi\} $ are the dynamic fields), or alternatively $ S(e^a{}_\mu, t^b, \{\phi\}) $, along with the definitions of subsection \ref{ssb:LifshitzWardIds}. 
Since the foliation 1-form is defined only up to rescaling, the action must be invariant under rescaling
of the 1-form:
\begin{equation}
S(g_{\mu\nu},t_\alpha,\{\phi\}) = S(g_{\mu\nu},f t_\alpha,\{\phi\}) .
\end{equation}
Applying an infinitesimal rescaling, we obtain for any $\delta f$:
\begin{equation}
\delta S = \int \frac{\delta S}{\delta t_\alpha} \delta t_\alpha = \int \sqrt{-g} J^\alpha t_\alpha \delta f = 0 ,
\end{equation}
so that $ J^\alpha $ is tangent to the foliation:
\begin{equation}
J^\alpha t_\alpha = 0.
\end{equation}

Starting with invariance under the extended form of FPD \eqref{LifshitzAnom:FPD}, we apply the transformation operator to the action (using the metric formalism):\footnote{We assume here that the dynamic fields $ \{\phi\} $ satisfy the E.O.M $ \frac{\delta S}{\delta \phi^i} = 0 $ and don't contribute to the variation.}
\begin{align}
\delta^D_\xi S &=
\int \frac{\delta S}{\delta g_{\mu\nu}} \delta^D_\xi g_{\mu\nu} + \frac{\delta S}{\delta t_\alpha} \delta^D_\xi t_\alpha = \notag\\
&= \int \sqrt{-g} \left[ \frac{1}{2} T_{(g)}^{\mu\nu} (\nabla_\mu \xi_\nu + \nabla_\nu \xi_\mu) + J^\mu (\xi^\nu \nabla_\nu t_\mu + \nabla_\mu \xi^\nu t_\nu ) \right] = \notag\\
&= \int \sqrt{-g}\ \xi^\nu \left[ -\nabla_\mu T_{(g)}^\mu{}_\nu + J^\mu \nabla_\nu t_\mu - \nabla_\mu(J^\mu t_\nu) \right] .
\end{align}
Requiring that $ \delta^D_\xi S = 0 $ for any $\xi$ we obtain the Ward identity:
\begin{equation}
\nabla_\mu T_{(g)}^\mu{}_\nu = J^\mu \nabla_\nu t_\mu - \nabla_\mu(J^\mu t_\nu).
\end{equation}
Using $\hat J^\alpha$ as defined in equation \eqref{anweylan:J_hat_def}, this identity can be written as in equation \eqref{anweylan:tgwardn}:
\begin{equation}
\boxed{
\nabla_\mu T_{(g)}^\mu{}_\nu = \hat{J}^\mu \nabla_\nu n_\mu - \nabla_\mu(\hat{J}^\mu n_\nu) = -n_\nu [\wn_\mu \hat{J}^\mu + 2\hat{J}^\mu a_\mu ],
}
\end{equation}
where $ a_\mu = n^\nu \nabla_\nu n_\mu $ and $ \wn $ is the covariant derivative projected on the foliation.
Alternatively, using the vielbein formalism, we apply first the local Lorentz transformation operator, giving:
\begin{align}
\delta^L_\alpha S &= \int \frac{\delta S}{\delta e^a{}_\mu} \delta^L_\alpha e^a{}_\mu + \frac{\delta S}{\delta t^a} \delta^L_\alpha t^a = \notag\\
&= \int e \left[ - T_{(e)} {}^\mu {}_a \alpha^a{}_b e^b{}_\mu - J_a \alpha^a{}_b t^b \right] = \notag\\
&= \int  e\ \alpha^{ab} \left[ - T_{(e)ba} - J_a t_b \right].
\end{align}
Requiring $ \delta^L_\alpha S = 0 $ for all $ \alpha^{ab} $ such that $ \alpha^{ab} = -\alpha^{ba} $, we get the Ward identity:
\begin{equation} \label{appendixward:teward1}
T_{(e)[\mu \nu]} = J_{[\mu}t_{\nu]}.
\end{equation}
We then apply the diffeomorphism operator:
\begin{align}
\delta^D_\xi S &= \int \frac{\delta S}{\delta e^a{}_\mu} \delta^D_\xi e^a{}_\mu + \frac{\delta S}{\delta t^a} \delta^D_\xi t^a = \notag\\
&= \int e \left[ T_{(e)} {}^\mu {}_a (\xi^\nu \nabla_\nu e^a{}_\mu + \nabla_\mu \xi^\nu e^a{}_\nu) + J_a \xi^\nu \nabla_\nu t^a \right] = \notag\\
&= \int e\ \xi^\nu \left[ -\nabla_\mu T_{(e)}{}^\mu{}_\nu + T_{(e)}{}^\mu{}_a \nabla_\nu e^a{}_\mu + J_a \nabla_\nu t^a \right].
\end{align}
By requiring $ \delta^D_\xi S = 0 $ for any $\xi$, we obtain the identity:
\begin{align}\label{appendixward:teward2}
\nabla_\mu T_{(e)}{}^\mu{}_\nu &=  T_{(e)}{}^\mu{}_a \nabla_\nu e^a{}_\mu + J_a \nabla_\nu t^a
= T_{(e)}^{ab} \omega_{\nu ab} + J_a \nabla_\nu t^a = \notag\\
&= J^a t^b \omega_{\nu ab} + J_a \nabla_\nu t^a
= J_a D_\nu t^a = J^\mu \nabla_\nu t_\mu,
\end{align}
where we used \eqref{appendixward:teward1}, and $D_\nu$ represents the covariant derivative with respect to local Lorentz transformations.
Using the normalized foliation 1-form, identities \eqref{appendixward:teward1} and \eqref{appendixward:teward2} can be written in the form of equations \eqref{anweylan:teward1n} -- \eqref{anweylan:teward2n}:
\begin{empheq}[box=\fbox]{align}\label{appendixward:teward1n}
T_{(e)[\mu \nu]} &= \hat{J}_{[\mu}n_{\nu]} ,\\
\label{appendixward:teward2n}
\nabla_\mu T_{(e)}{}^\mu{}_\nu &= \hat{J}^\mu \nabla_\nu n_\mu \ .
\end{empheq}
Turning to the anisotropic Weyl symmetry, applying the Weyl operator to the action we have:
\begin{align}
\delta^W_\sigma S &=
\int \frac{\delta S}{\delta g_{\mu\nu}} \delta^W_\sigma g_{\mu\nu} + \frac{\delta S}{\delta t_\alpha} \delta^W_\sigma t_\alpha = \int \sqrt{-g} 2 \sigma T_{(g)}^{\mu\nu} (P_{\mu\nu} - z n_\mu n_\nu),
\end{align}
or, using veilbein formalism:
\begin{align}
\delta^W_\sigma S &= \int \frac{\delta S}{\delta e^a{}_\mu} \delta^W_\sigma e^a{}_\mu + \frac{\delta S}{\delta t^a} \delta^W_\sigma t^a
= \int e\ \sigma T_{(e)}{}^\mu{}_a ( P^a_b - z n^a n_b ) e^b{}_\mu \ .
\end{align}
Requiring that $ \delta^W_\sigma S = 0 $ for any $ \sigma $, we obtain the Ward identity equation \eqref{anweylan:tgewardn_trace}:
\begin{equation}
\boxed{
T_{(g)}^{\mu\nu} P_{\mu\nu} - z T_{(g)}^{\mu\nu}n_\mu n_\nu
= T_{(e)}^{\mu\nu} P_{\mu\nu} - z T_{(e)}^{\mu\nu}n_\mu n_\nu = 0 \ .
}
\end{equation}

\section{Decomposition to Basic Tangent Tensors}\label{app:claim_proof}
In this appendix we present a proof for the statement of subsection \ref{ssb:fpdinvariantexpr}, that any FPD invariant scalar expression may be written as a sum of scalar expressions built by contracting the basic tangent tensors: $P_{\mu\nu}$, $a_\mu$, $K_{\mu\nu}$, $\widetilde R_{\mu\nu\rho\sigma}$, $\tilde \epsilon_{\mu\nu\rho\dots}$ and their temporal derivatives $\Lie{n}$ and spatial derivatives $\wn_\mu$ (as defined in subsection \ref{ssb:fpdinvariantexpr}).

As a first step in the proof, we note that any tensorial expression $ T_{\alpha\beta\dots} $ may be decomposed as a sum of terms of the form
\begin{equation}
T_{\alpha\beta\dots\mu\nu\dots} = \sum \widetilde{T}_{\alpha\beta\dots} n_\mu n_\nu \dots
\label{deco}
\end{equation}
where $ \widetilde{T}_{\alpha\beta\dots} $ are either tensors tangent to the foliation, or scalars. We refer to these tensors as the foliation tangent components of $ T_{\alpha\beta\dots} $. 
It can be easily seen that any scalar built by contracting all the indices of any number of tensors can thus be written as a sum of scalars built by contracting the indices of the tangent components of these tensors. Therefore, in order to prove our claim it is sufficient to show that the tangent components of any tensor built from the metric and foliation 1-form are polynomials in the previously mentioned basic tangent tensors.

The metric, the foliation 1-form and the Levi-Civita tensor clearly satisfy this condition, with the following decompositions:
\begin{align}
\begin{split}
n_\alpha &= n_\alpha ,
\\
g_{\alpha\beta} &= P_{\alpha\beta}-n_\alpha n_\beta,
\\
\epsilon_{\alpha\beta\gamma\delta\dots} &= - n_{\alpha} \tilde{\epsilon}_{\beta\gamma\delta\dots} +  n_{\beta} \tilde{\epsilon}_{\alpha\gamma\delta\dots} - n_{\gamma} \tilde{\epsilon}_{\alpha\beta\delta\dots} + \dots\, .
\end{split}
\end{align}
As previously noted, the covariant derivative of $ n_\alpha $ also satisfies it:
\begin{equation}
\nabla_\mu n_\nu = K_{\mu\nu} - a_\nu n_\mu .
\end{equation}
The Riemann tensor satisfies it due to the Gauss-Codazzi relations:
\begin{align}\label{Gauss-Codazzi}
\begin{split}
P_\mu^{\mu'} P_\nu^{\nu'} P_\rho^{\rho'} P_\sigma^{\sigma'}  R_{\mu'\nu'\rho'\sigma'}& = \widetilde R_{\mu\nu\rho\sigma}  + K_{\mu\rho} K_{\nu\sigma} - K_{\mu\sigma} K_{\nu\rho},
\\
n^{\mu'} P_\nu^{\nu'} P_\rho^{\rho'} P_\sigma^{\sigma'}  R_{\mu'\nu'\rho'\sigma'}& = \wn_\sigma K_{\nu\rho} - \wn_\rho K_{\nu\sigma} ,
\\
n^{\mu'} P_\nu^{\nu'} n^{\rho'} P_\sigma^{\sigma'}  R_{\mu'\nu'\rho'\sigma'}& = -\Lie{n} K_{\nu\sigma} + K_{\nu\alpha}K^\alpha_\sigma +\wn_\sigma a_\nu +a_\nu a_\sigma .
\end{split}
\end{align}

It remains to show that given a tensor that satisfies the above condition (all its tangent components are polynomials in the basic tangent tensors), its covariant derivative also satisfies it. Suppose then that $ T_{\alpha\beta\dots} $ is such a tensor, with the decomposition (\ref{deco}).
Then:
\begin{equation}
\nabla_\rho T_{\alpha\beta\dots\mu\nu\dots} = \sum (\nabla_\rho\widetilde{T}_{\alpha\beta\dots} )n_\mu n_\nu \dots + \widetilde{T}_{\alpha\beta\dots} (\nabla_\rho n_\mu) n_\nu \dots + \dots\, .
\end{equation} 
Since $ \widetilde{T}_{\alpha\beta\dots} $ and $ \nabla_\rho n_\mu $ both satisfy the condition, it is enough to focus on the expression $ \nabla_\rho\widetilde{T}_{\alpha\beta\dots} $. In general we can always write:
\begin{equation}
\begin{split}
\nabla_\rho\widetilde{T}_{\alpha\beta\dots} & =
(P_\rho^{\rho'} - n_\rho n^{\rho'})
(P_\alpha^{\alpha'} - n_\alpha n^{\alpha'})
(P_\beta^{\beta'} - n_\beta n^{\beta'})
\dots
\nabla_{\rho'}\widetilde{T}_{\alpha'\beta'\dots} 
\\
& = P_\rho^{\rho'}P_\alpha^{\alpha'}P_\beta^{\beta'} \dots 
\nabla_{\rho'} \widetilde{T}_{\alpha'\beta'\dots} 
\\
& \ \ \ +
n_\alpha n^{\alpha'} (\dots)^{\rho'\beta'\dots}_{\rho\beta \dots} \nabla_{\rho'}\widetilde{T}_{\alpha'\beta'\dots}
+
n_\beta n^{\beta'} (\dots)^{\rho'\alpha'\dots}_{\rho\alpha \dots} \nabla_{\rho'}\widetilde{T}_{\alpha'\beta'\dots} +\dots
\\
& \ \ \ 
- n_\rho n^{\rho'} P_\alpha^{\alpha'} P_\beta^{\beta'}
\dots \nabla_{\rho'}\widetilde{T}_{\alpha'\beta'\dots}
\\
& = \wn_{\rho} \widetilde{T}_{\alpha\beta\dots} \
 -
n_\alpha \nabla_{\rho'} n^{\alpha'} (\dots)^{\rho'\beta'\dots}_{\rho\beta \dots} \widetilde{T}_{\alpha'\beta'\dots}
+
\dots
\\
& \ \ \ 
- n_\rho P_\alpha^{\alpha'} P_\beta^{\beta'}
\dots 
(\Lie{n} \widetilde{T}_{\alpha'\beta'\dots}
+\nabla_{\alpha'} n^{\alpha''} \widetilde{T}_{\alpha''\beta'\dots}
+\nabla_{\beta'} n^{\beta''} \widetilde{T}_{\alpha'\beta''\dots} \dots),
\end{split}
\end{equation}
where the terms of the form $(\dots)^{\rho'\alpha'\dots}_{\rho\alpha \dots}$ represent various products of the foliation projector $P_\alpha^{\alpha'}$ and the normalized foliation 1-form $n_\alpha$.
The last expression is clearly a polynomial in $ P_{\alpha\beta}$,  $ n_\alpha $, $\nabla_\alpha n_\beta $, $\widetilde{T}_{\alpha\beta\dots} $, $\wn_\rho \widetilde{T}_{\alpha\beta\dots} $ and $ \Lie{n} \widetilde{T}_{\alpha\beta\dots} $, and therefore satisfies the condition that all its tangent components are polynomials in the basic tangent tensors. 
Thus our statement is proven.

\section{Derivation of Identities for Tangent Tensors}\label{app:identder}

In this appendix we detail the derivation of some of the identities of subsection  \ref{ssb:identsfortangent}.

The derivation of the temporal and spatial derivative exchange formula for tangent tensor $\widetilde T_{\alpha\beta\gamma}$ (equation \eqref{idents:tempspatderexchange}) is as follows:
\begin{align*}
& \Lie{n} \wn_\mu \widetilde T_{\alpha\beta\gamma\dots} = 
\Lie{n} \left[P_\mu^{\mu'} 
P_\alpha^{\alpha'}P_\beta^{\beta'}
\dots \nabla_{\mu'} \widetilde T_{\alpha'\beta'\gamma'\dots} \right]
\\ &
=
\Lie{n} \left(P_\mu^{\mu'} \right) 
P_\alpha^{\alpha'}P_\beta^{\beta'}
\dots \nabla_{\mu'} \widetilde T_{\alpha'\beta'\gamma'\dots}
+
P_\mu^{\mu'}
\Lie{n} \left(P_\alpha^{\alpha'}\right)
P_\beta^{\beta'}
\dots \nabla_{\mu'} \widetilde T_{\alpha'\beta'\gamma'\dots}
+ \dots \\
&~~~~
+
P_\mu^{\mu'} 
P_\alpha^{\alpha'}P_\beta^{\beta'}
\dots \Lie{n} \nabla_{\mu'} 
 \widetilde T_{\alpha'\beta'\gamma'\dots}
\\
&
= a_\mu  
P_\alpha^{\alpha'}P_\beta^{\beta'}
\dots n^{\mu'}   \nabla_{\mu'}\widetilde T_{\alpha'\beta'\gamma'\dots}
+
a_\alpha P_\mu^{\mu'}
P_\beta^{\beta'}
\dots n^{\alpha'} \nabla_{\mu'} \widetilde T_{\alpha'\beta'\gamma'\dots}
+ \dots \\
&~~~~
+
P_\mu^{\mu'} 
P_\alpha^{\alpha'}P_\beta^{\beta'}
\dots  \left[ 
n^\nu \nabla_\nu \nabla_{\mu'}\widetilde T_{\alpha'\beta'\gamma'\dots}
+(\nabla_{\mu'} n^\nu) \nabla_{\nu} \widetilde T_{\alpha'\beta'\gamma'\dots}
+(\nabla_{\alpha'} n^\nu)  \nabla_{\mu'} \widetilde T_{\nu\beta'\gamma'\dots} +\dots
\right]
\\
&  
= a_\mu  
P_\alpha^{\alpha'}P_\beta^{\beta'}
\dots n^{\mu'}   \nabla_{\mu'}\widetilde T_{\alpha'\beta'\gamma'\dots}
-
a_\alpha P_\mu^{\mu'}
P_\beta^{\beta'}
\dots (\nabla_{\mu'} n^{\alpha'})  \widetilde T_{\alpha'\beta'\gamma'\dots}
- \dots \\
&~~~~
+
P_\mu^{\mu'} 
P_\alpha^{\alpha'}P_\beta^{\beta'}
\dots  \left[ 
n^\nu  \nabla_{\mu'}\nabla_\nu\widetilde T_{\alpha'\beta'\gamma'\dots}
+ n^\nu R_{\nu\mu'\alpha'\rho}\, \widetilde T^\rho{}_{\beta'\gamma'\dots}
 + \dots \right] \\ 
& ~~~~
+P_\mu^{\mu'} 
P_\alpha^{\alpha'}P_\beta^{\beta'}
\dots  (\nabla_{\mu'} n^\nu) \nabla_{\nu} \widetilde T_{\alpha'\beta'\gamma'\dots}
+P_\mu^{\mu'} 
P_\alpha^{\alpha'}P_\beta^{\beta'}
\dots  K_{\alpha'}^\nu  \nabla_{\mu'} \widetilde T_{\nu\beta'\gamma'\dots} +\dots
\\
& 
= a_\mu  
P_\alpha^{\alpha'}P_\beta^{\beta'}
\dots \left[ \Lie{n}\widetilde T_{\alpha'\beta'\gamma'\dots}
- K_{\alpha'}^\nu \widetilde T_{\nu\beta'\gamma'\dots}
- \dots\right]
-
a_\alpha P_\mu^{\mu'}
P_\beta^{\beta'}
\dots K_{\mu'}^{\alpha'}  \widetilde T_{\alpha'\beta'\gamma'\dots}
- \dots \\
&~~~~
+
P_\mu^{\mu'} 
P_\alpha^{\alpha'}P_\beta^{\beta'}
\dots  \left[ \nabla_{\mu'} ( n^\nu \nabla_{\nu} \widetilde T_{\alpha'\beta'\gamma'\dots})+ (\wn_\rho K_{\mu'\alpha'} -
\wn_{\alpha'} K_{\mu'\rho})
\, \widetilde T^\rho{}_{\beta'\gamma'\dots}
 + \dots \right] \\ 
& ~~~~
+P_\mu^{\mu'} 
P_\alpha^{\alpha'}P_\beta^{\beta'}
\dots  K_{\alpha'}^\nu  \nabla_{\mu'} \widetilde T_{\nu\beta'\gamma'\dots} +\dots\\
%%%%%%%%%%%%%%%%%%%%%%%%%%%%%%
&
= a_\mu  
 \left[ \Lie{n}\widetilde T_{\alpha\beta\gamma\dots}
- K_{\alpha}^\nu \widetilde T_{\nu\beta\gamma\dots}
- \dots\right]
-
a_\alpha 
K_{\mu}^{\nu}  \widetilde T_{\nu\beta\gamma\dots}
- \dots \\
&~~~~
+ P_\mu^{\mu'} 
P_\alpha^{\alpha'}P_\beta^{\beta'}
\dots \nabla_{\mu'} ( \Lie{n} \widetilde T_{\alpha'\beta'\gamma'\dots} - (K_{\alpha'}^\nu - a^\nu n_{\alpha'}) \widetilde T_{\nu\beta'\gamma'}-\dots) 
\\ & ~~~~ 
 +  (\wn_\rho K_{\mu\alpha} -
\wn_{\alpha} K_{\mu\rho})
\, \widetilde T^\rho{}_{\beta\gamma\dots}
 + \dots + K_{\alpha}^\nu  \wn_{\mu} \widetilde T_{\nu\beta\gamma\dots} +\dots
 \\
 %%%%%%%%%%%%%%%%%%%%
 & = 
 a_\mu \Lie{n}\widetilde T_{\alpha\beta\gamma\dots}
- a_\mu K_{\alpha}^\nu \widetilde T_{\nu\beta\gamma\dots}
- \dots
-
a_\alpha 
K_{\mu}^{\nu}  \widetilde T_{\nu\beta\gamma\dots}
- \dots \\
&~~~~
+ \wn_{\mu}  \Lie{n} \widetilde T_{\alpha\beta\gamma\dots}
- \wn_{\mu} (K_{\alpha}^\nu\, \widetilde T_{\nu\beta\gamma}) - \dots
+  K_{\mu\alpha} a^\nu \widetilde T_{\nu\beta\gamma}
+\dots 
\\ & ~~~~
 +  (\wn_\rho K_{\mu\alpha} -
\wn_{\alpha} K_{\mu\rho})
\, \widetilde T^\rho{}_{\beta\gamma\dots}
 + \dots + K_{\alpha}^\nu  \wn_{\mu} \widetilde T_{\nu\beta\gamma\dots} +\dots
 \\
%%%%%%%%%%%%%%%%%%%%%%%%%%%%%%%
 & = \wn_{\mu}  (\Lie{n} \widetilde T_{\alpha\beta\gamma\dots})+
 a_\mu \Lie{n}\widetilde T_{\alpha\beta\gamma\dots}
+ \left[ (\wn_\nu+a_\nu) K_{\mu\alpha}
-
(\wn_{\alpha}+a_\alpha) K_{\mu\nu}
\, 
- (\wn_{\mu}+a_\mu) K_{\alpha\nu} \right]
\widetilde T^\nu{}_{\beta\gamma\dots},
\end{align*}
where we used the Gauss-Codazzi relations \eqref{Gauss-Codazzi} and the fact the the Lie derivative of a tangent tensor is also tangent to the foliation.

The derivation of the temporal derivative of the Riemann curvature equation \eqref{idents:second_temporal_bianchi} is as follows. Use the commutation relation for two space derivatives:
\begin{equation}\label{app:second_bianchi_step1}
\left[\wn_\mu, \wn_\nu \right] \widetilde V_\alpha = \widetilde R_{\mu\nu\alpha\rho}\widetilde V^\rho,
\end{equation}
and apply a temporal (Lie) derivative to both sides. Exchanging the temporal derivative with both the spatial derivatives on the l.h.s of \eqref{app:second_bianchi_step1} using equation \eqref{idents:tempspatderexchange} gives:
\begin{align*}
\Lie{n} \wn_{[\mu} \wn_{\nu]} \widetilde V_\alpha
 = & \widetilde R_\alpha{}^\rho{}_{\mu\nu} \Lie{n} \widetilde V_\rho 
 - \widetilde R_\alpha{}^\rho{}_{\mu\nu} K_{\rho\beta} \widetilde V^\beta
 - \widetilde R_\beta{}^\rho{}_{\mu\nu} K_{\rho\alpha} \widetilde V^\beta \\
& +\left[
(\wn_\mu+a_\mu)(\wn_\rho+a_\rho)K_{\nu\alpha}
-(\wn_\mu+a_\mu)(\wn_\alpha+a_\alpha)K_{\nu\rho}
-(\mu\leftrightarrow\nu)
\right] \widetilde V^\rho .
\end{align*}
Applying the temporal derivative to the r.h.s of \eqref{app:second_bianchi_step1} we obtain:
\begin{equation*}
\Lie{n} (\widetilde R_{\mu\nu\alpha}{}^{\rho} \widetilde V_\rho) = 
\Lie{n}\widetilde R_{\mu\nu\alpha}{}^{\rho} \cdot \widetilde V_\rho
+\widetilde R_{\mu\nu\alpha}{}^{\rho} \cdot \Lie{n} \widetilde V_\rho.
\end{equation*}
We therefore end up with:
\begin{align}
\begin{split}
\Lie{n}\widetilde{R}_{\alpha\beta\mu\nu}  = \ & \widetilde{R}_{\alpha\rho\mu\nu}K^\rho_\beta - \widetilde{R}_{\beta\rho\mu\nu} K^\rho_\alpha\\ 
&+ (\wn_\mu + a_\mu) (\wn_\beta + a_\beta) K_{\nu\alpha} - (\wn_\mu + a_\mu) (\wn_\alpha + a_\alpha) K_{\nu\beta}\\
&- (\wn_\nu + a_\nu) (\wn_\beta + a_\beta) K_{\mu\alpha} + (\wn_\nu + a_\nu) (\wn_\alpha + a_\alpha) K_{\mu\beta},
\end{split}
\end{align}
which is precisely \eqref{idents:second_temporal_bianchi} 

One can use this to derive similar identities for the temporal derivatives of the Ricci tensor:
\begin{align}
\begin{split}
\Lie{n}\widetilde{R}_{\alpha\mu}  = \ & 
- \widetilde{R}_{\alpha\beta\mu\rho}K^{\beta\rho}
+ \widetilde{R}_{\beta\mu} K^\beta_\alpha\\ 
&+ (\wn_\mu + a_\mu) (\wn_\rho+ a_\rho) K^\rho_\alpha - (\wn_\mu + a_\mu) (\wn_\alpha + a_\alpha) K\\
&- (\wn_\rho + a_\rho) (\wn^\rho + a^\rho) K_{\mu\alpha} + (\wn_\rho + a_\rho) (\wn_\alpha + a_\alpha) K_{\mu}^\rho,
\end{split}
\end{align}
 and the Ricci scalar:
\begin{align}
\begin{split}
\Lie{n}\widetilde{R} = \ &
-2K^{\alpha\mu}\widetilde R_{\alpha\mu} 
+ 2(\wn_\mu + a_\mu) (\wn_\rho+ a_\rho) K^{\rho\mu}- 2(\wn_\rho + a_\rho) (\wn^\rho + a^\rho) K .
\end{split}
\end{align}

\section{Useful Formulas for the Spatial Sector in 1+1 Dimensions}\label{app:d_1_ids}
In this appendix we present two useful formulas relevant to the purely spatial sector in $1+1$ dimensions for a general integer value of the dynamical exponent $z$.
We suppress indices in the formulas below since they are not needed in 1+1 dimensions as explained in subsection \ref{1d:z_geq_1}. 

The variation of any number of lower indexed derivatives acting on the acceleration vector is given by:
\begin{equation}
\begin{split}
\delta^W_\sigma (\wn^n a) & = \wn (\delta^W_\sigma  \wn^{n-1} a) -n \wn \sigma \cdot \wn^{n-1} a
%\\
%\delta^W_\sigma (\wn^n a) &
 = z \wn^{n+1}\sigma - \sum_{k=0}^{n-1} \left({n+1 \atop k}\right) \wn^k a \wn^{n-k} \sigma.
\end{split}
\end{equation}
For integration by parts of an expression of ghost number two with an even number of derivatives acting on one of the ghosts, we have the following identity:
\begin{equation}
\begin{split}
\int \sqrt{-g}\, f\sigma \wn^{2n} \sigma
= & - \sum_{k=1}^n \left(n \atop k \right) \int \sqrt{-g}\,  (\wn+a)^k f \cdot \sigma \wn^{2n-k} \sigma.
\end{split}
\end{equation}

\end{document}